\documentclass[11pt,a4paper]{article}
\pdfoutput=1
\usepackage{graphicx}
\usepackage{jheppub}

\usepackage{xcolor}      
\usepackage{xspace}
\usepackage{slashed}

\usepackage[font=small]{caption}    
\usepackage[shortlabels]{enumitem}  
\usepackage{tensor}    
\usepackage{cancel}    
\usepackage[text,italic]{esdiff}    

\usepackage{epstopdf, color}
\graphicspath{{figures/}}
\usepackage{mathrsfs}
\usepackage{braket}
\usepackage{array}

\allowdisplaybreaks

\newcommand{\eq}[1]{eq.~\eqref{eq:#1}}
\newcommand{\eqs}[2]{eqs.~\eqref{eq:#1} and \eqref{eq:#2}}

\renewcommand{\sec}[1]{section~\ref{sec:#1}}

\newcommand{\ssec}[1]{section~\ref{subsec:#1}}

\newcommand{\app}[1]{Appendix~\ref{app:#1}}

\newcommand{\fig}[1]{figure~\ref{fig:#1}}
\newcommand{\Fig}[1]{Figure~\ref{fig:#1}}
\newcommand{\tab}[1]{table~\ref{tab:#1}}

\newcommand{\ie}{i.e.}
\newcommand{\eg}{e.g.}

\newcommand{\aS}{\ensuremath{\alpha_S}\xspace}
\newcommand{\bmT}{\ensuremath{\bar{m}_T}\xspace}
\newcommand{\pT}{\ensuremath{p_T}\xspace}
\newcommand{\Kfac}{\ensuremath{K}\xspace}
\newcommand{\ord}{\ensuremath{O}}
\newcommand{\GeV}{\text{ GeV}}
\newcommand{\TeV}{\text{ TeV}}
\newcommand{\LO}{\text{LO}}

\newcommand{\QED}{\text{QED}}
\newcommand{\A}{\mathcal{A}}
\newcommand{\wwSing}{ww_{\bf1}}
\newcommand{\wwTrip}{ww_{\bf3}}
\newcommand{\qqSing}{q\bar{q}_{\bf1}}
\newcommand{\qqTrip}{q\bar{q}_{\bf3}}

\newcommand{\sh}{\hat{s}}
\renewcommand{\th}{\hat{t}}
\newcommand{\uh}{\hat{u}}

\title{Precision Diboson Observables for the LHC}

\author[a]{Christopher Frye,}
\author[a,b]{Marat Freytsis,}
\author[a]{Jakub Scholtz}
\author[a]{and Matthew J.~Strassler}

\affiliation[a]{Department of Physics, Harvard University, Cambridge, MA 02138}
\affiliation[b]{Institute for Theoretical Science, University of Oregon, Eugene, OR 97403}

\emailAdd{frye@physics.harvard.edu}
\emailAdd{freytsis@uoregon.edu}
\emailAdd{jscholtz@physics.harvard.edu}
\emailAdd{strassler@physics.harvard.edu}

\abstract{Motivated by the restoration of $SU(2)\times U(1)$ at high energy, we suggest that certain ratios of diboson differential cross sections can be used as high-precision observables at the LHC.  We rewrite leading-order diboson partonic cross sections in a form that makes their $SU(2)\times U(1)$ and custodial $SU(2)$ structure more explicit than in previous literature, and identify important aspects of this structure that survive even in hadronic cross sections.  We then focus on higher-order corrections to ratios of $\gamma\gamma$, $Z\gamma$ and $ZZ$ processes, including full next-to-leading-order corrections and $gg$ initial-state contributions, and argue that these ratios can likely be predicted to better than 5\%, which should make them useful in searches for new phenomena. The ratio of $Z\gamma$ to $\gamma\gamma$ is especially promising in the near term, due to large rates and to exceptional cancellations of QCD-related uncertainties.  We argue that electroweak corrections are moderate in size, have small uncertainties, and can potentially be observed in these ratios in the long run.
}


\begin{document}
\maketitle

\section{Introduction}
\label{sec:Intro}

With no signs as yet of physics beyond the Standard Model (SM), it is essential that measurements at the Large Hadron Collider (LHC) become increasingly precise in the coming years, allowing tests of new SM effects and leading to greater sensitivity to subtle non-SM phenomena.  In many cases the limiting factor is a lack of confidence in theoretical calculations, so it is particularly important to find more examples of measurable quantities that are widely agreed to have small theoretical uncertainties.

In this paper we consider production of pairs of electroweak (EW) bosons, collectively referred to as ``diboson processes'' or $pp \to V_1 V_2$, where $V_i = \gamma, W^\pm, Z$, which have by now been an object of study for almost four decades~\cite{Brown:1978mq,Brown:1979ux,Mikaelian:1979nr,Combridge:1980sx,Glover:1988rg,Smith:1989xz,Ohnemus:1990za,Mele:1990bq,Ohnemus:1991gb,Ohnemus:1991kk,Frixione:1992pj,Bailey:1992br,Ohnemus:1992jn,Frixione:1993yp,Dixon:1998py,Campbell:1999ah,Campbell:2011bn}.  These processes have been measured individually by the ATLAS and CMS collaborations~\cite{Chatrchyan:2011rr,Chatrchyan:2011qt,Aad:2012tba,Aad:2013izg,ATLAS:2013gma,ATLAS:2013fma,CMS:2013qea,ATLAS:2014xea,CMS:2014xja,Khachatryan:2015kea,Khachatryan:2015sga,ATL-PHYS-PUB-2015-20}.  Our goal here is to consider combinations of these measurements.

In the SM the EW bosons originate from a triplet and singlet of $SU(2) \times U(1)$, becoming massive and mixing after EW symmetry breaking. But at the high energies accessible to the LHC, the symmetry breaking effects are moderated, and one might imagine the underlying $SU(2) \times U(1)$ structure might more directly relate diboson processes to one another.  It turns out that although this naive expectation is not automatically satisfied, there are nevertheless some elegant and interesting relations.

In this paper we identify numerous independent ratios of diboson measurements that are special at tree level and that offer moderate to excellent potential for both high-precision predictions and high-precision measurements.  These ratios, in contrast to the differential cross sections themselves, are flat or slowly-varying as functions of \pT (and other kinematic variables), making them stable against certain experimental problems.  Moreover, we expect that many of them receive controllable QCD corrections, especially at high \pT.  Electroweak corrections are expected to be important at the 10--20\% level, and may be visible in these ratios, without clutter from large QCD uncertainties.  Since the uncertainties on these EW corrections will be small after ongoing calculations are completed, the ratios potentially also offer sensitivity to high-energy beyond-the-Standard-Model (BSM) phenomena.  These would include BSM corrections to triple-gauge-boson vertices and broad diboson resonances, though we do not investigate this issue carefully here.

To illustrate these features, we will perform a detailed study of three related ratios, each of which has a different pattern of uncertainties, though only two of the central values are independent.  We will show that their special properties survive to higher order, though with an interesting array of subtleties.  Specifically we will consider $d\sigma/d\bmT$ for $\gamma\gamma$, $Z\gamma$ and $ZZ$ at next-to-leading order (NLO), where \bmT is the average transverse mass of the two vector bosons:
\begin{equation}
  \bmT = \frac{1}{2}\left(\sqrt{p_{T,1}^2 + m_{1}^2} + \sqrt{p_{T,2}^2 + m_{2}^2}  \, \right)\,.
\end{equation}
We will discuss issues arising at NNLO, and include the $gg$-initiated loop contribution explicitly. We will give evidence that a number of uncertainties are reduced by taking the various ratios of these three processes, and also argue that experimental technicalities do not interfere with the measurements.  The effect of higher-order corrections on our other observables will be studied elsewhere.

The use of ratios of measurements to reduce theoretical and experimental errors has a long history, with perhaps the most famous and successful in particle physics involving the measurements of $R_\text{had} = \sigma(e^+e^- \to \text{hadrons})/\sigma(e^+e^- \to \mu^+\mu^-)$ in the early 1970s. In the study of hadronic decay processes, ratios have long been used to reduce systematic uncertainties from higher-order and non-perturbative corrections (see ref.~\cite{Charles:2004jd} and references therein).  These methods have seen continuing use at the LHC, and similar approaches have been extended to the study of Higgs decays in order to better constrain its properties~\cite{Djouadi:2012rh,Goertz:2013eka,Banerjee:2015bla}.

Ratios of production cross sections at hadron colliders have seen more limited use due to the more complex initial state.   At the LHC in particular, the use of $d\sigma(\gamma + nj)$ to calibrate the process $d\sigma(Z + nj)$, an irreducible background for many BSM searches, has been investigated at leading order (LO) for $n = 1$~\cite{Ask:2011xf} and NLO for $n = 2$ and 3~\cite{Bern:2011pa,Bern:2012vx}, and implemented in an analysis by the CMS collaboration~\cite{Chatrchyan:2014lfa}.  Similar studies have been carried out for ratios of $Z$ and $W^\pm$ processes~\cite{Malik:2013kba}.  Moreover, data comparing $Z$ to $\gamma$  production has recently been shown to be in good agreement with theoretical predictions~\cite{Khachatryan:2015ira}, and ratios of single-boson production cross sections have been measured~\cite{Aad:2011dm,ATLAS-CONF-2015-039}, primarily to aid with fits for parton distribution functions.  Searches for new colored states in ratios of multijet processes have been proposed in ref.~\cite{Becciolini:2014lya}, while the gradual ramp-up of beam energies at the LHC has also motivated looking at total cross sections of individual processes across a range of energies~\cite{Mangano:2012mh}.  More recently it has been argued that a very precise measurement of the top quark Yukawa can be obtained from the ratio of $t\bar t h$ to $t\bar t Z$ production \cite{Plehn:2015cta}.

\section{Executive summary}
\label{sec:ExecSumm}

The restoration of $SU(2) \times U(1)$ well above $m_Z$, along with some happy accidents, leads to some interesting relations among the various diboson partonic differential cross sections.  These are obscured once the partonic processes are convolved with parton distribution functions (PDFs), and are affected by experimental realities that impact photons, $W$s and $Z$s differently.  Nevertheless, at LO we find numerous ratios of differential cross sections for LHC diboson production that have the potential to be interesting observables.

In \sec{LO} below, we investigate possible diboson variables at LO.  We show that diboson processes naturally divide up into three classes:
\begin{equation}
    (1)~\gamma\gamma,\, Z\gamma,\, ZZ, \qquad
    (2)~W^\pm \gamma,\, W^\pm Z, \qquad
    (3)~W^+ W^-.
\end{equation}
(We do not consider same-sign $W^\pm W^\pm$ processes here since extra jets must accompany them.)  Each of the first two classes is self-contained, and observables can be built by taking ratios of various differential cross sections.  The $W^+W^-$ process can be related to linear combinations of processes in the first two classes, but is more complicated theoretically.

Our observables involve differential cross sections for $V_1 V_2$ production binned in various kinematic variables, which we loosely denote $\sigma(V_1V_2)$ here for brevity.  We are interested in symmetric and antisymmetric combinations $\sigma_{S}$ and $\sigma_A$; here the asymmetry is taken with respect to reversing the relative pseudorapidity $\Delta \eta \equiv \eta_1 - \eta_2$ of the two bosons, signed relative to their longitudinal boost direction. (That is, events are weighted by $\text{sign}(y_{12} \Delta \eta)$, where $y_{12} \approx \frac12 (\eta_1 + \eta_2)$ is the diboson rapidity. See \ssec{pdfs} for more details.)
We propose that the following ratios are of interest:\footnote{Although the central values of these observables are not all independent --- for instance $R_{1c} = R_{1b}/R_{1a}$, $R_2^+/R_2^- = C_{2b}/C_{2a}$, $A_2^+/A_2^- = D_{2b}/D_{2a}$ --- the pattern of theoretical and statistical uncertainties is different for each ratio.}
\begin{align}
\displaystyle
  &\bullet~~ R_{1a}=\frac{\sigma_S(Z \gamma)}{\sigma_S(\gamma\gamma)}, \quad
             R_{1b}=\frac{\sigma_S(ZZ)}{\sigma_S(\gamma\gamma)}, \quad
             R_{1c}=\frac{\sigma_S(ZZ)}{\sigma_S(Z \gamma)}, \nonumber \\[5pt]
  &\bullet~~ C_{2a} = \frac{\sigma_S(W^+\gamma)}{\sigma_S(W^-\gamma)}, \quad
             C_{2b} = \frac{\sigma_S(W^+Z)}{\sigma_S(W^-Z)}, \quad
             D_{2a} = \frac{\sigma_A(W^+\gamma)}{\sigma_A(W^-\gamma)}, \quad
             D_{2b} = \frac{\sigma_A(W^+Z)}{\sigma_A(W^-Z)}, \nonumber \\[5pt]
  &\hspace{7mm}  R_{2}^\pm = \frac{\sigma_S(W^{\pm}Z)}{\sigma_S(W^{\pm}\gamma)}, \quad
                 A_2^\pm = \frac{\sigma_A(W^{\pm}Z)}{\sigma_A(W^{\pm}\gamma)}, \nonumber \\[5pt]
  &\bullet~~ R_3 = \frac{\sigma_S(W^+W^-)}{\sigma_S(V_1^0V_2^0)}, \quad
             A_3 = \frac{\sigma_A(W^+W^-)}{\sigma_A(WV^0)}, 
\label{eq:OurRatios}
\end{align}
where $V^0$ denotes $Z$ or $\gamma$, and $\sigma_A(WV^0)$ is some linear combination of $\sigma_A(W^+V^0)$ and $\sigma_A(W^-V^0)$. See \ssec{RatObs} for a more precise discussion of $R_3$ and $A_3$.

In figures~\ref{fig:rat-za-lo}--\ref{fig:ww-lo} of \ssec{RatObs}, these ratios, calculated at LO and binned in $\sh$, are shown.  All of the ratios are slowly varying, and each has its own special features.  Observables $R_{1a}$, $R_2^\pm$, and $A_2^{\pm}$ are, to first approximation, independent of the PDFs (and hence have very small PDF uncertainties). At LO they depend only on ratios of SM couplings and charges, from which we learn $R_{1a}$ is nearly constant, $R_2^+ \approx R_2^-$, and $A_2^\pm \approx -1$.  By contrast, observables $R_{1b}, R_{1c}, C_{2a},C_{2b}, D_{2a}, D_{2b}$ are dominated by the difference between up and down PDFs; all SM couplings cancel in the $C_2$ and $D_2$ ratios. Observables $R_3$ and $A_3$ are more complex.

These observables are simplest for $\sqrt{\sh} \gg 2 m_Z$ or $\bmT \gg m_Z$, where the difference between the massless $\gamma$ and the massive $W,Z$ is of diminished importance. But as discussed in \ssec{StatUnc}, the low production rates for diboson processes at these high scales, and the low branching fraction for $Z \to \text{leptons}$, gives our observables relatively large statistical uncertainties, potentially negating the value of their low theoretical uncertainties.  (In this paper we will only consider leptonic decays of $W$s and $Z$s, though we briefly discuss other options in \ssec{Final}.)  At 300 fb$^{-1}$, the $R_{1a}$, $C_{2a}$ and $R_3$ observables can be measured in multiple bins with 5\% statistical uncertainties.  This is comparable to the theoretical uncertainties that we will claim below. The variables $R_2^\pm$ and $D_{2a}$ can only be measured in a single bin, making them only marginally useful.  At 3000 fb$^{-1}$, it appears all the variables are potentially useful excepting only $D_{2b}$ and $A_2^-$, and with $A_2^+$ marginal.

In \sec{beyondLO}, we study the simplest of these observables, the $R_1$ ratios, beyond LO.  As described in \ssec{cuts}, we choose our cuts and our observable carefully to avoid strong jet vetoes, problematic kinematic regions with very large \Kfac factors,  etc.; see \tab{bosonJetCuts} and \tab{leptonCuts} below. We also include $gg$ production, formally NNLO but numerically important. To fix its normalization, we use the fact that the dominant correction to $gg \to \gamma\gamma$ at the next order is known \cite{Bern:2002jx}.  We also use this to normalize the other $gg \to V_1^0 V_2^0$ processes.\footnote{As this paper was nearing completion, a calculation for $gg\to ZZ$ analogous to ref.~\cite{Bern:2002jx} appeared in ref.~\cite{Caola:2015psa}.  Our normalization estimate appears to agree with their results.}

In \ssec{NLO-QCD}, we show that many NLO QCD corrections do cancel in these ratios, except for the region where a final-state jet is collinear with a vector boson.  There  the photon has a collinear singularity which must be regulated with, \eg, a fragmentation function, while the $Z$ singularity is regulated by its mass.   Although the ratios shift significantly in this region, we argue in \ssec{PhotonIso} that use of a ``staircase'' isolation method, as in ref.~\cite{Binoth:2010nha,Hance:2011ysa}, leaves small theoretical uncertainties. We also show in \ssec{gg} that $gg \to V_1^0 V_2^0$ causes shifts in the ratios as large as 5--20\%  at low $\bar m_T$, due in part to an interesting accidental cancellation in $gg \to Z\gamma$, though these effects are reduced at high \bmT. Moreover, we argue that the uncertainties on these shifts are small. We also discuss other known NNLO effects on our ratios.  Finally, we find in \ssec{pdf-scale} that certain other QCD theoretical uncertainties --- PDF uncertainties and scale uncertainties in particular --- do largely cancel, especially for $R_{1a}$.

\begin{figure}
    \begin{center}
       \includegraphics[width=0.7\linewidth]{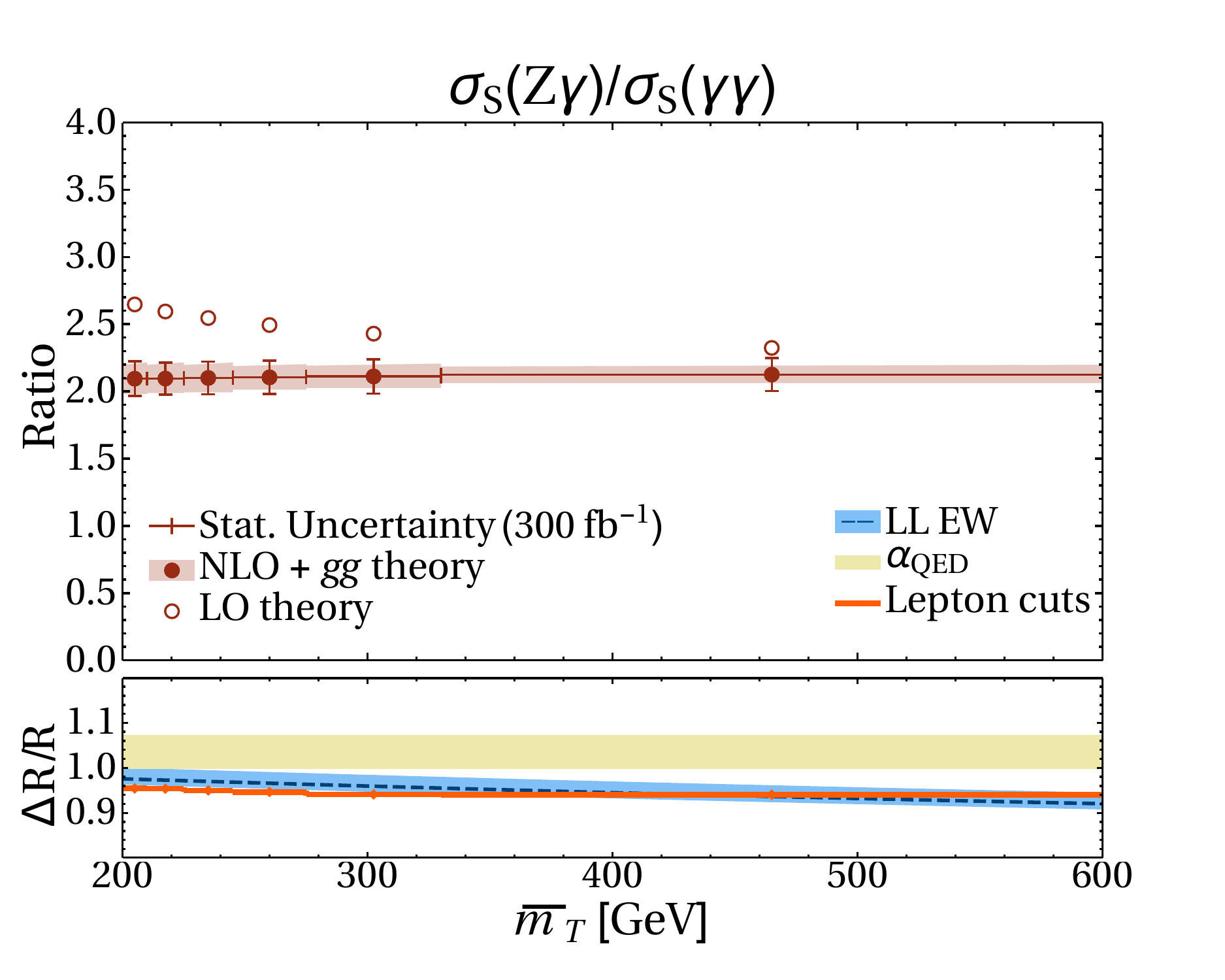}\\
       \includegraphics[width=0.49\linewidth]{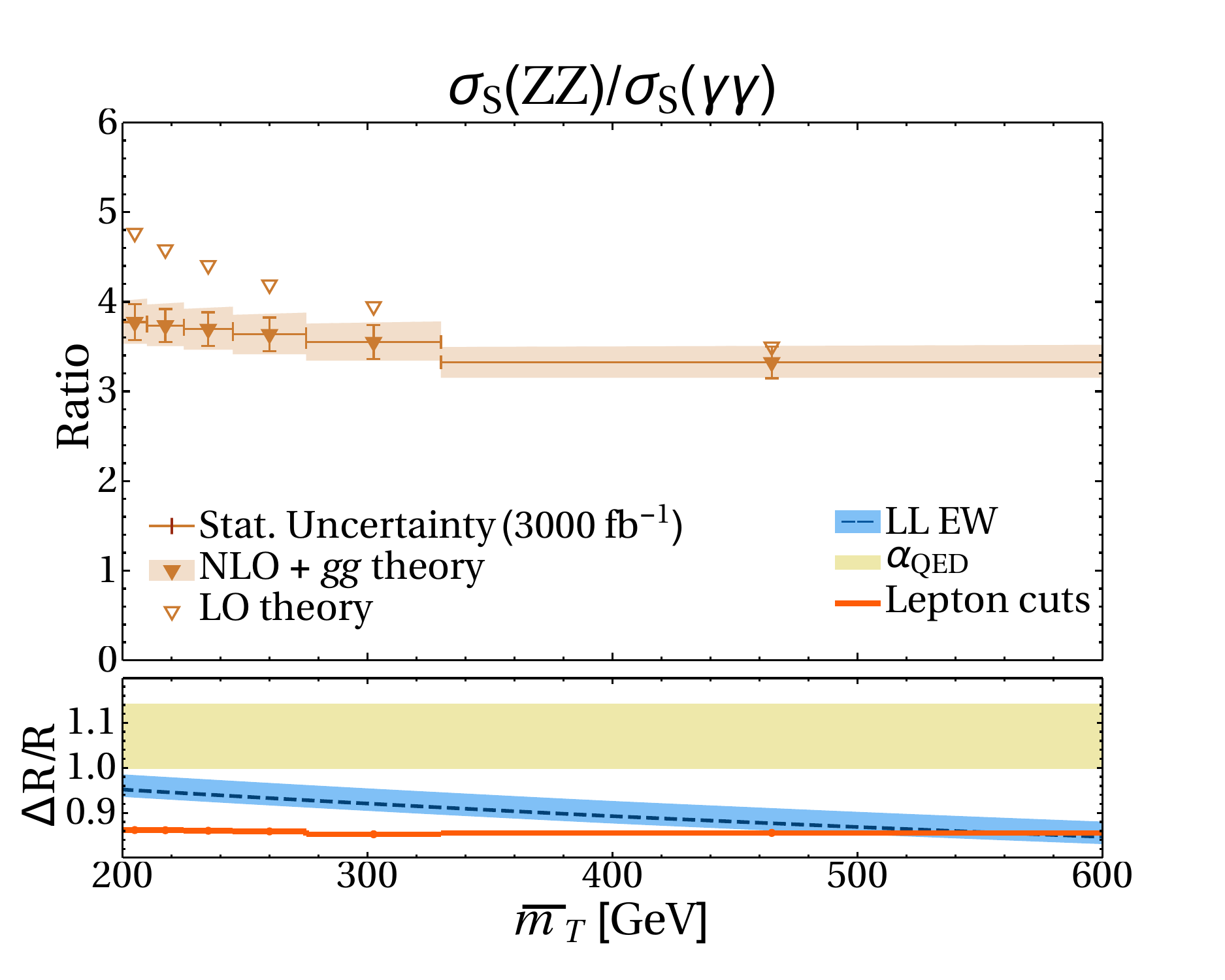}
       \includegraphics[width=0.49\linewidth]{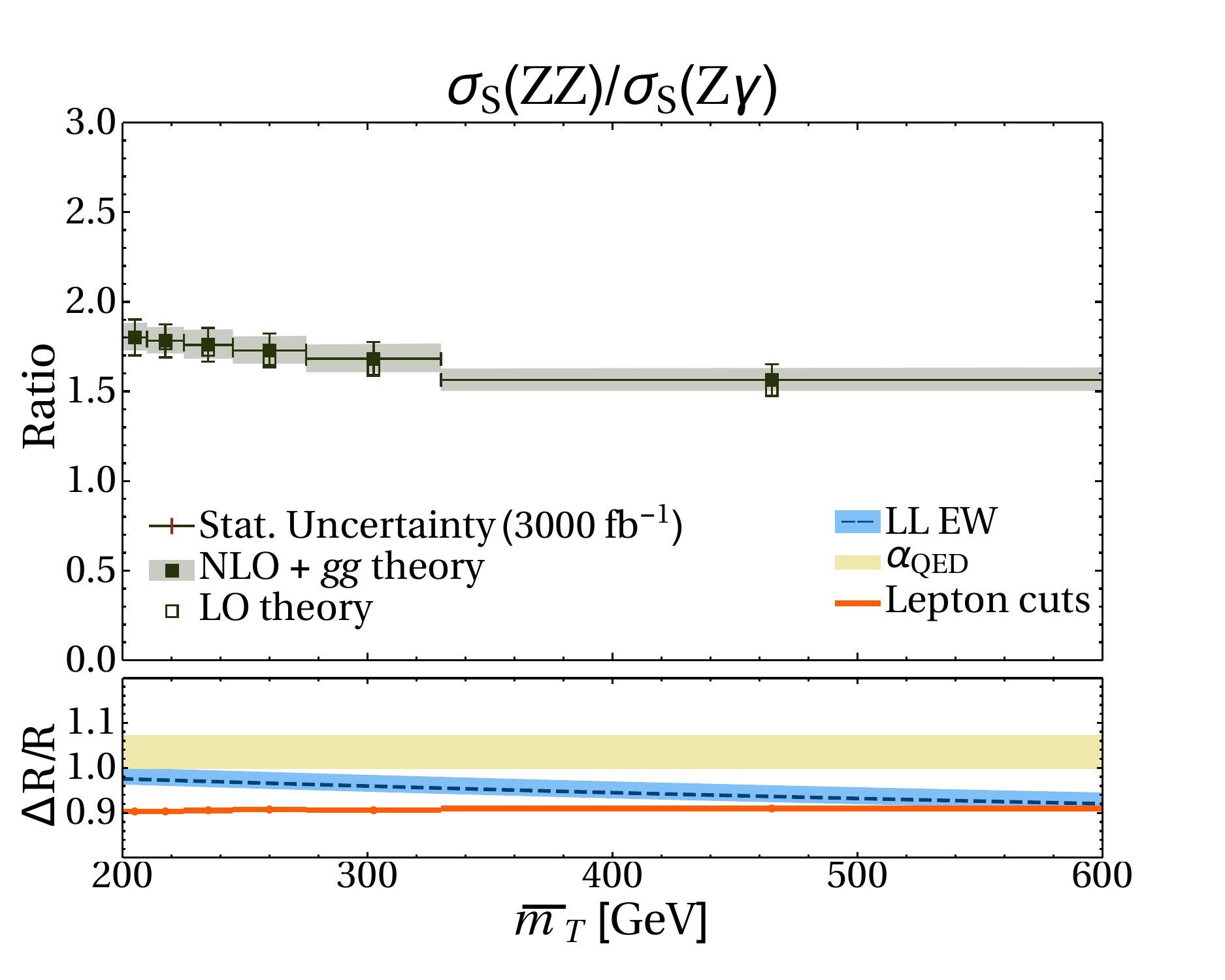}
    \end{center}
    \caption{(Top) $R_{1a} = \sigma_S(Z\gamma)/\sigma_S(\gamma\gamma)$. (Left) $R_{1b} = \sigma_S(ZZ)/\sigma_S(\gamma\gamma)$. (Right) $R_{1c} = \sigma_S(ZZ)/\sigma_S(Z\gamma)$. The solid symbols represent our NLO (+ NNLO $gg$) theoretical prediction.  Their error bars indicate the expected statistical uncertainties after 300 (3000) fb$^{-1}$ for $R_{1a}$ ($R_{1b}$ and $R_{1c}$). The shaded band around these points represents our estimate of QCD theory uncertainties; see text for important details. The corresponding LO theory prediction is given in open symbols. (By chance, higher-order corrections to $R_{1c}$ nearly cancel.) The bottom plot for each ratio shows the expected fractional correction (relative to unity) from additional non-QCD corrections: an orange solid line for the effect of $Z \to \ell\ell$ decays on the experimental measurement, a blue dashed line for an estimate of the effect of  electroweak Sudakov logarithms, with a band indicating its uncertainty, and a horizontal band for the uncertainty from the undetermined choice of $\alpha_\QED$.}
\label{fig:mainResult}
\end{figure}

These statements are summarized in \fig{mainResult}.  To explain this figure, let us focus first on the top plot, which shows results for $R_{1a}$, the ratio of $Z\gamma$ to $\gamma\gamma$ differential cross sections with respect to \bmT, obtained for the 13 TeV LHC.  The upper portion of the plot shows the ratio $R_{1a}$ as would be measured in 6 bins of 5--6\% statistical uncertainty; the last bin includes events with \bmT extending up to the kinematic limit. The open circles indicate a LO prediction, while the closed circles are our result including NLO and $gg$-initiated production.   The dominant corrections are driven by the gluon PDF, and decrease with \bmT. The error bars on the closed circles indicate the expected statistical errors at 300 fb$^{-1}$.  The shaded band indicates the theoretical uncertainties mentioned in the previous paragraphs, itemized in \tab{uncBudget} of \sec{Summ} and with all uncertainties combined linearly, except for PDF extraction uncertainties which are combined in quadrature with the others. This combination gives a conservative estimate of \emph{known} uncertainties.

We emphasize that we have not proven it impossible for additional \emph{unknown} sources at NNLO to shift the ratios' central values by larger amounts than our uncertainty estimates.  Although we believe we identified all obvious effects that do not cancel in ratios, and have either included them or estimated our uncertainties from not including them, we cannot demonstrate this directly.  Only the complete NNLO calculations, for which code is not yet public, will confirm that there are no additional subtleties.

The lower portion of the plot shows estimates of three sources of additional corrections and their uncertainties, \emph{expressed as a relative shift} of the ratio; (\ie\ 1.05 indicates an upward shift of 5\% on the ratio.) First, as discussed in \ssec{doubleLog}, leading-log EW corrections only partially cancel in the $R_{1}$ ratios.  At high \bmT Sudakov logarithmic effects will dominate and can be roughly estimated using the soft-collinear approximation, as studied in ref.~\cite{Becher:2013zua}.  The effect on $R_{1a}$ arises  as a difference between the $Z$ and $\gamma$ jet functions, and is of order $5$--$10\%$ at high \bmT, though this is probably an overestimate.
We show this estimate by plotting the effect on our ratios of the calculation of ref.~\cite{Becher:2013zua}  as a blue dashed line, along with an estimate of its uncertainty band as a shaded blue region.
At low \bmT a finite correction, still relatively small, may make the true EW shift of $R_{1a}$ somewhat larger than indicated by our blue band --- see \cite{Bierweiler:2013dja,Denner:2015fca}, although their cuts are significantly different from ours.
Nevertheless, and more importantly, our uncertainty band is conservative.
The band correctly shows the dominant uncertainty at high \bmT, from matching the resummed and fixed-order calculations.  At small \bmT the leading uncertainty, from scale variation of the EW couplings, is smaller than the band.

Second, the tan horizontal shaded bar represents an unresolved disagreement in
the community, discussed in \ssec{alphaQED}, regarding the choice of scale
$\mu$ for evaluating $\alpha_{\QED}$ when an on-shell photon is emitted in a
hadronic setting.  The difference between using $\mu = 0$ and $\mu = m_Z$
--- for each observable, an overall shift of all the bins by a nearly equal
amount --- is indicated by this bar. This issue is temporary; the
uncertainty will be eliminated once the controversy is settled.

Third, we have chosen to show our results in the upper portion of the figure without including effects from $Z$ decays to leptons.  That is, in the figure we applied cuts on the vector bosons but ignored the finite $Z$ width and the kinematic and isolation cuts that must be imposed on the leptons.  As we study in \ssec{Zdecay}, these effects, shown as an orange solid line in the lower portion of the figure, do materially change the ratios at the $\sim 5$--$15\%$ level, but with very low uncertainty.

In the other two plots of \fig{mainResult}, we show similar results for $R_{1b}$ and $R_{1c}$, but at 3000 fb$^{-1}$.  The increased integrated luminosity is required in order to obtain small statistical errors, because of the small branching fraction of $ZZ$ to four leptons.  Both QCD and EW corrections to $R_{1b}$ are larger because the differences between $Z$ and $\gamma$ contribute twice.

We see from \fig{mainResult} that the variables $R_{1a}$, $R_{1b}$ and $R_{1c}$ are nearly flat in \bmT, are potentially predictable at better than 5\%, and are measurable in several bins (using only leptonic $Z$ decays) at the $\sim 5$--$6\%$ level with 300, 3000 and 3000 fb$^{-1}$ respectively.  Corrections to the LO prediction are moderate at low \bmT and decrease with \bmT.  (In $R_{1c}$ the prediction at higher-order is nearly the same as at LO, due to an accidental cancellation between the $gg$ contribution and other corrections.)   Moreover, at 3000 fb$^{-1}$ the $R_{1a}$ ratio can be measured using tens of bins (the precise number depending on \bmT resolution) with the highest bin starting above 600 GeV, nearly double what is possible at 300 fb$^{-1}$.

  At this level of precision, these ratios are potentially sensitive both to interesting soft-collinear EW corrections and to BSM phenomena.   We are optimistic that other variables in our list will prove comparably useful, though this remains to be shown in future work.

\section{The story at leading order}
\label{sec:LO}

We begin with a study of diboson processes at tree level, which  were first computed at this order almost four decades ago~\cite{Brown:1978mq,Brown:1979ux,Mikaelian:1979nr}. In the form originally presented, the underlying broken gauge and custodial symmetries were not manifest. Making these more explicit, we identify ratios of particular interest. As we will see, each ratio has its own unique features, strengths and weaknesses, even at leading order. We will study these features first at the partonic level, where the $SU(2) \times U(1)$ structure of the rates is most clear. We then use this structure as a guide to construct our ratio observables.  Finally we show and explain the behavior of these ratios in proton-proton collisions at 13 TeV.  We conclude this section with a short discussion of the statistical uncertainties on these variables at 300 and 3000 fb$^{-1}$ at 13 TeV.

\subsection{High energy limit}
\label{subsec:LO-intro}

Well above the scale of EW symmetry breaking,
we may rewrite the SM EW bosons $W^\pm, Z, \gamma$ as the triplet $w^\pm, w^3$ and singlet $x$ of massless gauge bosons of $SU(2) \times U(1)$, along with the Goldstone scalars $\phi^\pm, \phi^3$. (We use lowercase letters for massless gauge bosons and capital letters for the mass eigenstates.) One basis for the massless diboson states consists, up to normalizations, of $SU(2) \times U(1)$ singlets and triplets:
\begin{align}
\label{eq:xx}
  xx_{\bf 1} \equiv xx &: \quad \ket{xx}\,, \\
\label{eq:wx-3}
  wx_{\bf 3} \equiv wx &: \quad \ket{w^+x}, \quad \ket{w^3x}, \quad \ket{w^-x}\,, \\
\label{eq:ww-1}
  \wwSing              &: \quad \ket{w^+w^-} + \ket{w^-w^+} - \ket{w^3w^3}\,, \\
\label{eq:ww-3}
  \wwTrip\,            &: \quad \ket{w^+w^3}-\ket{w^3w^+}, \quad \ket{w^+w^-}-\ket{w^-w^+},
                          \quad \ket{w^3w^-}-\ket{w^-w^3}\,.
\end{align}
There are also quintet $ww$ states, such as $W^+W^+$, but they require two final-state jets at LO, whereas we will focus on production with no jets at LO.  This means we only deal at LO with three $SU(2)$-singlet $q\bar{q}$ initial states
\begin{equation}
 {\ket{u_R \bar u_R}\,, \quad \ket{d_R \bar d_R}}\,, \quad \ket{u_L\bar u_L} - \ket{d_L\bar d_L}\,,
\end{equation}
and the triplet of states
\begin{equation}
  \left\{ \ket{u_L \bar d_L}\,, \quad \ket{u_L\bar u_L} + \ket{d_L \bar d_L}\,, \quad
          \ket{d_L \bar u_L} \right\}.
\end{equation}

\begin{figure}[tb!]
\begin{center}
    \setlength{\unitlength}{1mm}
    \begin{picture}(100,10)
        \put(0,0){\includegraphics[width=0.6\linewidth]{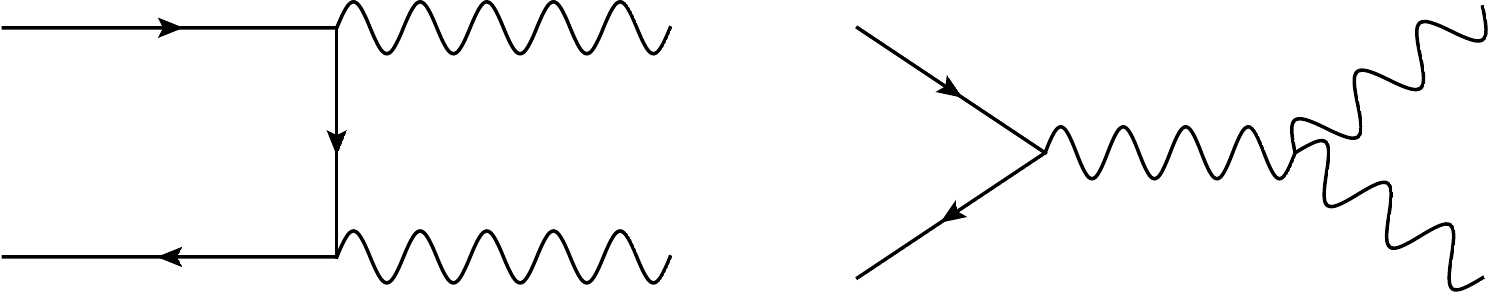}}
    \end{picture}
\end{center}
\caption{At leading order, diboson processes proceed from $q\bar q$ initial states.  The $t,u$ channels (left) and the $s$ channel (right) contribute only to particular amplitudes under $SU(2)\times U(1)$.  }
\label{fig:LOgraphs}
\end{figure}
Production rates at LO involve $s$-, $t$-, $u$-channel Feynman diagrams; see \fig{LOgraphs}.
The $s$-channel diagram, with an $f^{abc}$ symbol, only contributes for $\wwTrip$ states. Because of this, the LO production rates for $xx$, $wx$, and $\wwSing$ are proportional, differing only in the coupling constants.

This suggests that symmetries should exist among the observable cross sections of interest $\sigma(pp \to V_1 V_2)$. To determine the implications more precisely, we must take into account the production of scalars (\eg, the $\phi^3$ inside $Z$), the interference between different channels (\eg, since $W^-\gamma$ is a superposition of $wx$ and $\wwTrip$), and the convolution with PDFs.

Since the quark-scalar couplings are proportional to quark masses, we can neglect scalar production in the $t$- and $u$-channel diagrams, so the scalars contribute only to triplet processes. When final-state scalars do contribute, they do so in the spin-sum of squared helicity-amplitudes, so there are no associated interference effects.

\subsection{Squared amplitudes}
\label{subsec:SquAmp}

The production of dibosons in the limit in which their masses can be neglected can be written in a simple form.  We will denote the coupling-stripped LO singlet-, triplet- and scalar amplitudes by
\begin{align}
\label{eq:a1}
  a_1    &\propto \mathcal{M}(xx) \propto \mathcal{M}(wx) \propto \mathcal{M}(\wwSing)\,, \\
\label{eq:a3}
  a_3    &\propto \mathcal{M}(\wwTrip)\,, \\
\label{eq:aphi}
  a_\phi &\propto \mathcal{M}(\phi\phi)\,,
\end{align}
in a notation which corresponds to eqs.~\eqref{eq:xx}--\eqref{eq:ww-3}. In these schematic definitions, we leave polarizations implicit since we will always compute spin-averaged cross sections. The three amplitudes in the first line are all proportional, and this continues to hold when one includes NLO QCD corrections but not NLO EW corrections.\footnote{For instance, a virtual $w$ can attach to the final-state lines in $\mathcal{M}(\wwSing)$ but not in $\mathcal{M}(xx)$.}

In the high energy limit, the partonic cross sections of interest $d\hat{\sigma}(q\bar{q} \to V_1 V_2)$ are quadratic in the $a_i$s. The products of $a_i$s that are relevant for diboson production include\footnote{These expressions can be extracted from the high-energy limit of the partonic rates in Eqs.~\eqref{eq:firstXsec}--\eqref{eq:lastXsec} below, which were computed in refs.~\cite{Brown:1978mq,Brown:1979ux,Mikaelian:1979nr}.}
\begin{align}
\label{eq:a1sq}
  |a_1|^2    &= \frac{\th}{\uh} + \frac{\uh}{\th} \,, \\
  (a_1 a_3)  &= \left(\frac{\th - \uh}{2\sh}\right)
                 + \frac14 \left(\frac{\th}{\uh} - \frac{\uh}{\th}\right) \,, \\
\label{eq:a3sq}
  |a_3|^2    &= \frac{\th\uh}{4\sh^2} - \frac18
                 + \frac1{32} \left(\frac{\th}{\uh} + \frac{\uh}{\th}\right) \,, \\
\label{eq:aLsq}
  |a_\phi|^2 &= \frac{\th\uh}{4\sh^2}\,.
\end{align}
Here, $(a_1a_3)$ is shorthand for Re$(a_1^\star a_3)$. The $a_i$ amplitudes transform simply under $\th \leftrightarrow \uh$ exchange:
\begin{equation}
\label{eq:fbprops}
  a_1(\th,\uh) = a_1(\uh,\th), \qquad a_3(\th,\uh) = -a_3(\uh,\th), \qquad
  |a_\phi(\th,\uh)| = |a_\phi(\uh,\th)|.
\end{equation}
These properties of $a_1$ and $a_3$, required by Bose statistics and by the fact that $\wwSing$ ($\wwTrip$) is symmetric (antisymmetric) in the two $w$s,\footnote{Notice that NLO EW corrections break the $\th \leftrightarrow \uh$ symmetry of $\mathcal{M}(wx)$ since a virtual $w$ can attach to the final-state $w$ line but not to the $x$ line.} explain why in eqs.~\eqref{eq:a1sq}--\eqref{eq:aLsq} only $(a_1a_3)$ is antisymmetric under $\th \leftrightarrow \uh$.

The $\th \leftrightarrow \uh$ symmetry properties of the $a_i$s play an important role in what follows. These are forward-backward symmetries, since swapping $\th \leftrightarrow \uh$ in a $q\bar{q} \to V_1 V_2$ event reverses the sign of $\eta_1 - \eta_2$, with $\eta$ defined relative to the $q$'s momentum direction. In what follows, we will use $d \hat{\sigma}_S$ ($d \hat{\sigma}_A$) to denote $\th \leftrightarrow \uh$ symmetrized (antisymmetrized) \emph{partonic} differential cross sections. We will discuss symmetric and antisymmetric \emph{hadronic} cross sections $\sigma_S, \sigma_A$ in \ssec{pdfs}.

One important consequence of \eq{fbprops} is that $a_3$ vanishes at $\th = \uh$, that is, at center-of-mass-frame (CM) scattering angle $\theta = \pi/2$.  This ``radiation zero'' has an important impact on the diboson processes.

\subsection{Partonic cross sections at high energies}
\label{subsec:partonic}

Next we write the partonic cross sections for the production of physical dibosons $V_1V_2$, ignoring mass corrections of order $m_Z^2/p_T^2$. Our formulas are written in terms of the $a_i$s given in eqs.~\eqref{eq:a1sq}--\eqref{eq:aLsq}, making various relations among the cross sections manifest and motivating the ratio observables mentioned in \sec{ExecSumm}.

The full formulas including $\ord(m_Z^2/p_T^2)$ terms are given in \app{MassCorr}. There we define $\A_i$s as straightforward generalizations of the $a_i$s including mass corrections. These corrections are subleading in the region of phase space we study in this paper compared to certain QCD corrections, and they introduce no uncertainties.  We include them in our numerical results, but have no need to discuss them further. In fact a few useful relations, such as eqs.~\eqref{eq:WV-rel,f+b}--\eqref{eq:WV-rel,f-b}, are unaffected by the boson masses.

\subsubsection{\texorpdfstring{$\gamma\gamma, \,Z\gamma, \,ZZ$}{Diphoton, Z-photon, ZZ}}
\label{subsec:za}

Writing $c_W=\cos\theta_W$ and $s_W=\sin\theta_W$, we have
\begin{align}
\label{eq:A=wx}
  \gamma &= c_W \, x + s_W \, w^3 \,, \\
\label{eq:Z=wx}
  Z      &= c_W \, w^3 - s_W \, x \,,
\end{align}
and $Z$ also contains the scalar $\phi^3$. Pairs of photons and $Z$s can be produced in $xx$, $w^3x$, and $w^3w^3$ channels. Since $w^3w^3$ is orthogonal to the $\wwTrip$ states, the production rates in this sector are all proportional to $|a_1|^2$; see \eq{a1}. Inserting the appropriate coupling constants and writing $V^0=\gamma,Z$,
we have
\begin{equation}
\label{eq:za}
  \diff{\hat{\sigma}}{\th}(q\bar{q} \to V^0_1 V^0_2) = \frac{C^q_{12}}{\sh^2} |a_1|^2 \, ,
\end{equation}
where
\begin{align}
\label{eq:Cqaa}
  C_{\gamma\gamma}^q &= \frac12 \,\frac{\pi \alpha_2^2 s_W^4}{N_c}\, 2Q^4\,, \\
  C_{Z\gamma}^q      &= \frac{\pi \alpha_2^2 s_W^2 c_W^2}{N_c}\, \left( L^2 Q^2 + R^2 Q^2 \right)\,, \\
\label{eq:Cqzz}
  C_{ZZ}^q           &= \frac12 \,\frac{\pi \alpha_2^2 c_W^4}{N_c}\, \left( L^4 + R^4 \right)\,.
\end{align}
Here, a symmetry factor of $1/2$ has been included for identical particles, $\alpha_2$ is the $SU(2)$ coupling of the SM, $Q=T_3+Y$ is the electric charge of quark $q$, and
\begin{equation}
\label{eq:LR}
  L = T_3 - Y_L\, t_W^2, \qquad R = -Y_R\,t_W^2,
\end{equation}
with $t_W = s_W/c_W$. The $\ord(m_Z^2/p_T^2)$ corrections to \eq{za} are given in \app{MassCorr}. Each partonic rate in this sector is forward-backward symmetric, so $d\hat{\sigma}_A(V^0_1 V^0_2) = 0$  (though NLO EW corrections give a non-zero $d\hat\sigma_A(Z\gamma)$.)

\subsubsection{\texorpdfstring{$W^{\pm}\gamma, \,W^{\pm} Z$}{W-photon, WZ}}
\label{subsec:wza}

We begin this section by discussing relations among $W^+V^0$ and $W^-V^0$ rates. Since $W^+V^0$ and $W^-V^0$ production are related by $CP$, which takes $u\bar{d} \to W^+V^0$ into $d\bar{u} \to V^0 W^-$, we have (in the notation of \ssec{SquAmp})
\begin{align}
\label{eq:WV-rel,f+b}
  d\hat{\sigma}_S(u\bar{d} \to W^+V^0) &= d\hat{\sigma}_S(d\bar{u} \to W^-V^0)\,, \\
\label{eq:WV-rel,f-b}
  d\hat{\sigma}_A(u\bar{d} \to W^+V^0) &= -d\hat{\sigma}_A(d\bar{u} \to W^-V^0)\,.
\end{align}

Next we write down the partonic cross sections for producing $W^\pm V^0$.  These arise from $w^\pm w^3$ and $w^\pm x$ and involve both $a_1$ and $a_3$, as seen from eqs.~\eqref{eq:wx-3}--\eqref{eq:ww-3} and \eqref{eq:a1}--\eqref{eq:aphi}.  Scalar production $a_\phi$ also appears in $W^\pm Z$. In particular,
\begin{align}
\label{eq:wa}
  \frac{d\hat{\sigma}}{d\th}(q\bar{q}' \to W^\pm\gamma)
    &= \frac{\pi|V_{ud}|^2 \alpha_2^2 s_W^2}{N_c\,\sh^2}
         \left[\frac{Y_L^2}{2} |a_1|^2 \pm 2Y_L (a_1a_3) + 4|a_3|^2 \right]\,, \\
\label{eq:wz}
  \frac{d\hat{\sigma}}{d\th}(q\bar{q}' \to W^\pm Z)
    &= \frac{\pi|V_{ud}|^2 \alpha_2^2}{N_c\,\sh^2}  \left[\frac{s_W^2 t_W^2 Y_L^2}{2} |a_1|^2
         \mp 2s_W^2 Y_L(a_1a_3) + 4c_W^2|a_3|^2 + \frac12|a_\phi|^2\right] ,
\end{align}
where $q\bar{q}'$ is $u\bar{d}$ ($d\bar{u}$) for $W^+V^0$ ($W^-V^0$). The $\ord(m_Z^2/p_T^2)$ terms in these rates are given in \app{MassCorr}. As seen from \eq{fbprops}, these formulas obey eqs.~\eqref{eq:WV-rel,f+b}--\eqref{eq:WV-rel,f-b}.

Next we compare $W^\pm \gamma$ to $W^\pm Z$. Notice that the forward-backward antisymmetric terms in these two rates, those proportional to $Y_L(a_1a_3)$, are equal but opposite:
\begin{equation}
\label{eq:WAZ-rel,f-b}
  d\hat{\sigma}_A(W^\pm\gamma) = -d\hat{\sigma}_A(W^\pm Z)\,.
\end{equation}
These asymmetries arise from the interference between $w^\pm w^3$ and $w^\pm x$ production, a cross term that carries opposite sign for the photon versus the $Z$; see eqs.~\eqref{eq:A=wx}--\eqref{eq:Z=wx}. Alternatively, completeness requires that in the high energy limit,
\begin{equation}
  d\hat{\sigma}(W^\pm \gamma) + d\hat{\sigma}(W^\pm Z)
    = d\hat{\sigma}(w^\pm x) + d\hat{\sigma}(w^\pm w^3) + d\hat{\sigma}(\phi^\pm \phi^3) \ .
\end{equation}
Since the three terms on the right hand side are respectively proportional to $|a_1|^2$, $|a_3|^2$ and $|a_\phi|^2$, which are forward-backward symmetric, \eq{WAZ-rel,f-b} follows.

The forward-backward symmetric rates in this sector can be read from \eqs{wa}{wz} by omitting the $(a_1a_3)$ terms. Because of the smallness of $Y_L^2 = 1/36$ and the relative factor of $(8\,c_W)^{-1}$ suppressing $|a_\phi|^2$, the $|a_3|^2$ terms naively dominate the cross sections, leading to a ratio $d\hat{\sigma}_S(W^+\gamma)/d\hat{\sigma}_S(W^+Z)$ of $t_W^2 \approx 0.29$.

However, there is a small subtlety with this estimate. We noted earlier that $a_3$, antisymmetric under $\th \leftrightarrow \uh$, has a radiation zero.\footnote{This radiation zero of $a_3$ combines with $a_1$ to give the famous tree-level $f\bar f'\to W\gamma$ radiation zero \cite{Mikaelian:1979nr}, at an angle that depends on the electric charge of $f$.} Nonetheless, the coefficients of $|a_1|^2$ and $|a_\phi|^2$ are small, so this zero is only important very close to $\theta \sim \pi/2$. Moreover, by chance, the ratio of $d\hat{\sigma}_S(W^+\gamma)$ to $d\hat{\sigma}_S(W^+Z)$ is 0.19 at $\theta=\pi/2$, protecting the naive estimate of $t_W^2$ from a large correction. We will say more about this in \ssec{RatObs}.

\subsubsection{\texorpdfstring{$W^- W^+$}{WW}}
\label{subsec:ww}

The partonic amplitude for producing transversely-polarized $W^-W^+$ is a linear combination of $a_1$ and $a_3$ in the high-energy limit. One must also include the contribution $a_\phi$ from scalars $\phi^-\phi^+$, which are produced through an $\sh$-channel $w^3$ or $x$ in $q_L\bar{q}_L$-initiated processes, or through an $\sh$-channel $x$ from $q_R\bar{q}_R$.

In the high energy limit, the partonic cross sections are
\begin{multline}
\label{eq:uuww}
  \diff{\hat{\sigma}}{\th}(q\bar{q} \to W^-W^+)
    = \frac{\pi\alpha_2^2}{N_c \sh^2} \, \left\{\frac1{16} |a_1|^2 \pm \frac12(a_1a_3) + 2|a_3|^2 \right. \\
       + \left[(t_W^2\,Y_R)^2 + (t_W^2\,Y_L + T_3)^2 \right] |a_\phi|^2 \Big\}\,,
\end{multline}
where the upper (lower) sign holds for $u$-type ($d$-type) quarks. Here $T_3,Y_L,Y_R$ are the quantum numbers of quark $q$.
Note that the forward-backward symmetric rates for transversely polarized $W^-W^+$ are the same in $u\bar{u}$  and $d\bar{d}$ channels, while the forward-backward antisymmetric rates are equal and opposite; that is,
\begin{align}
\label{eq:ww,f+b}
  d\hat{\sigma}_S(u\bar{u} \to W_T^- W_T^+) &= d\hat{\sigma}_S(d\bar{d} \to W_T^- W_T^+)\,, \\
\label{eq:ww,f-b}
  d\hat{\sigma}_A(u\bar{u} \to W_T^- W_T^+) &= -d\hat{\sigma}_A(d\bar{d} \to W_T^- W_T^+)\,.
\end{align}
These relations are a consequence of $G$-parity (charge conjugation $C$ followed by a rotation by $\pi$ around the second isospin axis) which takes $u\bar{u} \to w^-w^+$ into $d\bar{d} \to w^+w^-$. Indeed, high energy production of $W_T^- W_T^+$ (which in our notation is equivalent to $w^-w^+$) proceeds at LO only through $SU(2)$ interactions, which respect $G$-parity. Alternatively one can derive \eqs{ww,f+b}{ww,f-b} using Clebsch-Gordan coefficients:
\begin{align}
\label{eq:ww|uu}
    \mathcal{M}(u\bar{u} \to w^-w^+) &= \frac12 \mathcal{M}(\qqTrip \to \wwTrip)
                                        + \frac{1}{\sqrt{6}} \mathcal{M}(\qqSing \to \wwSing) \,, \\
\label{eq:ww|dd}
    \mathcal{M}(d\bar{d} \to w^-w^+) &= \frac12 \mathcal{M}(\qqTrip \to \wwTrip)
                                        - \frac{1}{\sqrt{6}} \mathcal{M}(\qqSing \to \wwSing)\,.
\end{align}
Squaring these equations and referring to relations \eq{fbprops}, one finds that \eq{ww,f+b} must hold, with $d\hat{\sigma}_S$ given by a linear combination of $|a_1|^2$ and $|a_3|^2$. And since the cross terms have opposite signs, \eq{ww,f-b} follows, with $d\hat{\sigma}_A$ proportional to $(a_1a_3)$.

On the other hand, note that the $Y_L T_3$ terms in $d\hat{\sigma}(u\bar{u} \to \phi^-\phi^+)$ and $d\hat{\sigma}(d\bar{d} \to \phi^-\phi^+)$ are not equal even though they are forward-backward symmetric. These terms arise from an $\sh$-channel $x$ boson, which interacts with the initial-state quarks with couplings that violate $G$-parity.  However, these terms are numerically small.

Since $d\hat{\sigma}_A(W^-W^+) \propto (a_1a_3)$, the partonic asymmetry of $W^-W^+$ is proportional to\footnote{But note $d\hat{\sigma}_A(W^-W^+)$ arises as interference between $\mathcal{M}(\wwTrip)$ and $\mathcal{M}(\wwSing)$, while $d\hat{\sigma}_A(W^\pm V^0)$ is an interference between $\mathcal{M}(\wwTrip)$ and $\mathcal{M}(wx)$. Since NLO EW corrections break the LO relation $\mathcal{M}(\wwSing) \propto \mathcal{M}(wx)$, they also violate $d\hat{\sigma}_A(W^-W^+) \propto d\hat{\sigma}_A(W^\pm V^0)$. \label{foot:WW-asym}} that of $W^\pm\gamma$ and $W^\pm Z$. Meanwhile the radiation zero of $a_3$ is quite important for $d\hat{\sigma}_S(W^-W^+)$.  Later we will see that $|a_1|^2$ actually dominates the $W^-W^+$ cross section, though not overwhelmingly.  This motivates comparing $d\hat{\sigma}_S(W^+W^-)$ to $d\hat{\sigma}_S(V_1^0 V_2^0) \propto |a_1|^2$, or perhaps to a linear combination of $d\hat{\sigma}_S(V_1^0 V_2^0)$ and $d\hat{\sigma}_S(WV^0)$.

\subsection{Convolution with PDFs}
\label{subsec:pdfs}

Having discussed the partonic cross sections in detail, we now turn to the observable hadronic cross sections
\begin{eqnarray}
  d\sigma(pp \to V_1V_2)
  &=& \sum_{q,q'} dx_1 dx_2\, f_{q}(x_1)\,f_{q'}(x_2)\, d\hat{\sigma}(qq' \to V_1 V_2) 
  \cr
     &=& \sum_{q,q'} \frac{d\sh}{s}\, dy\, f_{q}(x_1)\,f_{q'}(x_2)\, d\hat{\sigma}(qq' \to V_1 V_2)\,.
\end{eqnarray}
Here $f_i(x)$ is the PDF of parton $i$, $\sh = x_1 x_2 s$ is the CM energy, and $y = \frac12 \log(x_1/x_2)$ is the rapidity of the partonic collision.

To fully specify an event, kinematic variables describing the final state must be chosen.  Since our purpose is to study ratios of different diboson processes, we want variables that keep the different processes on equal footing to the extent possible. One useful variable is $m_{VV}$, the invariant mass of the two bosons; this equals $\sqrt{\sh}$ at LO.  Considerations at LO might also suggest the use of the transverse momentum \pT of either boson.  However, the threshold value of $\sh$ required to produce the $V_1 V_2$ pair with a given \pT differs among the processes:
\begin{equation}
\label{eq:whymT}
  \sh_\text{thresh} = \left(\sqrt{p_T^2+m_1^2} + \sqrt{p_T^2+m_2^2}\right)^2 = 4\bmT^2\,,
\end{equation}
where \bmT is the average transverse mass of the two final-state bosons. Since our ratios are simpler if partonic kinematics span the same range in numerator and denominator, the above relation suggests that \bmT is a more useful kinematic variable than \pT.

The partonic cross sections $d\hat{\sigma}/d\th$ given in \ssec{partonic} can be rewritten in terms of \bmT as
\begin{equation}
  \diff{\hat\sigma}{\bmT}(qq' \to V_1 V_2)
    = \left|\diff{\th}{\bmT}\right| \, \diff{\hat{\sigma}}{\th}(qq' \to V_1 V_2) \,,
\end{equation}
where, if $m_1 = m_2$ or if both $m_1$ and $m_2$ are negligible,\footnote{The Jacobian is considerably more complicated when $m_1 \neq m_2$.}
\begin{equation}
   \left|\diff{\th}{\bmT}\right| = 2\bmT \left(1-\frac{4\bmT^2}{\sh}\right)^{-1/2}.
\end{equation}
The corresponding observable cross section takes the form
\begin{equation}
\label{eq:obsXsec}
  \sigma(pp \to V_1 V_2) = \sum_{q,q'} \int \frac{d\sh}{s} \int d\bmT
    \diff{\hat{\sigma}}{\bmT}(qq' \to V_1 V_2) \int dy\, f_{q}(x_1)\,f_{q'}(x_2)\,,
\end{equation}
where the domain of integration depends on the observable being computed and the kinematic cuts imposed.

The observables we propose in this paper involve the quantities $\sigma_S$ and $\sigma_A$ which we now define. We have already introduced $d\hat{\sigma}_S$ ($d\hat{\sigma}_A$) as the $\th \leftrightarrow \uh$ symmetric (antisymmetric) part of the differential partonic cross section. That is, $d\hat{\sigma}_A(q\bar q \to V_1 V_2)$ weights events by $\text{sign}(\eta_1 - \eta_2)$, while $d\hat{\sigma}_S$ weights events symmetrically with $+1$. At $pp$ colliders, the $q$ direction is unobservable but is typically aligned with the longitudinal boost $y_{12}$ of the diboson system, which at LO is the same as the boost $y$ of the $q\bar{q}$ center-of-mass frame. We may thus define $\sigma_A$ at LO by assigning to events the weight $\text{sign}[y(\eta_1-\eta_2)]$, as in
\begin{equation}
  \sigma_X^{\LO}(pp \to V_1 V_2) = \sum_{q_i, \bar{q}_j} \int \frac{d\sh}{s} \int d\bmT\,
    \diff{\hat{\sigma}_X^{\LO}}{\bmT} (q_i\bar{q}_j \to V_1V_2)\, \mathscr{L}^X_{q_i\bar{q}_j}\,,
\end{equation}
where $X=S,A$ and we have introduced
\begin{equation}
    \mathscr{L}^{\{{S,A}\}}_{q_i\bar{q}_j}
      = \int dy\, \{{1,\text{sign}(y)}\} \, 2f_{q_i}(x_1) f_{\bar q_j}(x_2)
\end{equation}
as symmetric and antisymmetric parton luminosities. The limits of integration on $y$ depend on $\hat s$ and $\bar m_T$ once cuts are imposed on the pseudorapidity of the bosons.

Triply-differential cross sections would show the relations among the diboson processes most directly, since the PDFs would be evaluated in small $x_1,x_2$ ranges.  However, the statistical samples required for binning in all three variables would be far larger than are available at the LHC.  To obtain measurements with small statistical errors we must integrate over two variables, namely $y$ and either $\sh$ or \bmT, and bin in the third variable.  Fortunately, even though this involves convolution with the PDFs, many of the good qualities of the partonic relations discussed above survive to $d\sigma/d\bmT$ and $d\sigma/d\sh$.

In our study of $pp \to V^0_1 V^0_2$ beyond LO in \sec{beyondLO}, we will focus on $d\sigma/d\bmT$.  However, our immediate goal in the remainder of \sec{LO} is to explain heuristically how the ratios of \eq{OurRatios} behave, and to point out their most striking features.  In this regard it is most useful to work with the variable $\sh = m_{VV}^2$.  The \bmT and $y$ integrals split cleanly as separate functions of $\sh$; see \eq{sigma,za} below. This feature makes formulas look simpler and permits simple heuristic arguments. Typically the features seen in $d\sigma/d\bmT$ are nearly the same as those seen in $d\sigma/dm_{VV}$, and moreover survive largely intact to NLO. We will see this for neutral diboson production later.

Of course the above-mentioned separation of \bmT and $y$ integrals is only formal; it ceases to hold, even at LO, when realistic kinematic cuts are included.  Such cuts are always necessary when photons are involved, since production rates diverge as $\pT^\gamma \to 0$. Thus we must introduce a lower bound $(\bmT)_\text{min}$ when integrating over \bmT in \eq{obsXsec} to compute an observable rate. In \sec{beyondLO} below we bin with respect to \bmT, beginning at 200 GeV, so this requirement is automatically satisfied there. But in our heuristic LO discussion, where we bin with respect to $m_{VV}$, we achieve this goal by imposing a cut on pseudorapidity
\begin{equation}
\label{eq:etacut}
    |\eta(V)| < 1.5
\end{equation}
for each final state boson $V$; this cut renders the LO cross sections finite.  This will not impact our heuristic reasoning but does play a role in the plots shown.

\subsection{Ratio observables}
\label{subsec:RatObs}

We now discuss the ratio observables of \eq{OurRatios}, already mentioned in \sec{ExecSumm}. We will present precise LO results in figures, and we will use schematic or approximate equations to understand the results.  In this and following sections, all results are for a 13 TeV $pp$ collider, and are obtained using \textsc{MCFM 6.8}~\cite{Campbell:1999ah,Campbell:2011bn}.  The plots of our ratios are given for diboson cross sections without decays and do not include $Z$ or $W$ branching fractions to leptons.

For $V^0_1 V^0_2 = \gamma\gamma,\,Z\gamma,\,ZZ$ we found that all the partonic cross sections are forward-backward symmetric and proportional to the kinematic function $|a_1|^2$. For each of these processes, schematically,\footnote{The lower limit of integration over \bmT depends on the pseudorapidity cut imposed at $\eta_\text{cut}=1.5$\,. In the $m_Z \to 0$ limit, $(\bmT)_\text{min} = \sqrt{\sh}/(2\cosh\eta_\text{cut})$. The limits of integration over $y$ in $\mathscr{L}^S_{q\bar{q}}$ also depend on \bmT, a point we can ignore for the heuristic arguments presented here.}
\begin{equation}
\label{eq:sigma,za}
  \diff{\sigma_S}{\sh}(pp \to V^0_1V^0_2)
    \sim \frac {\sum_q C^q_{12} \mathscr{L}^S_{q\bar{q}}(\sh)}{s\,\sh^2}
         \int^{\sqrt{\sh}/2} d\bmT \left|\diff{\th}{\bmT}\right| \, |a_1|^2\,,
\end{equation}
where the $C^q_{12}$s were defined in eqs.~\eqref{eq:Cqaa}--\eqref{eq:Cqzz}. Note the numerator of the prefactor is a weighted parton luminosity, with the PDFs weighted by process-dependent couplings and charges.  Our observable $R_{1a}$ then satisfies
\begin{equation}
\label{eq:R1a}
  R_{1a}(\sh)
    \equiv \left[\frac{\sigma_S(pp \to Z\gamma)}{\sigma_S(pp \to \gamma\gamma)}\right]_{\sh}
    \sim \frac{\sum_q C^q_{Z\gamma} \mathscr{L}^S_{q\bar{q}}(\sh)}
              {\sum_q C^q_{\gamma\gamma} \mathscr{L}^S_{q\bar{q}}(\sh)} \,,
\end{equation}
with similar relations for $R_{1b} = \sigma_S(ZZ)/\sigma_S(\gamma\gamma)$ and $R_{1c} = \sigma_S(ZZ)/\sigma_S(Z\gamma)$.

\begin{table}[tb!]
    \begin{center}
      \begin{tabular}{|c|c|c|}\hline
        $V_1^0V_2^0$ & $C^u_{12} \cdot 10^5$ & $C^d_{12} \cdot 10^5$ \\
        \hline\hline
        $\gamma\gamma$ & 1.2 & 0.07 \\ \hline
        $Z\gamma$ & 2.2 & 0.7 \\ \hline
        $ZZ$ & 1.6 & 3.3 \\ \hline
        \end{tabular}
    \end{center}
\caption{The values of $C^q_{12}$ relevant for the $R_1$ ratios.}
\label{tab:Cq12}
\end{table}

\begin{figure}[tb!]
    \begin{center}
        \includegraphics[width=0.5\linewidth]{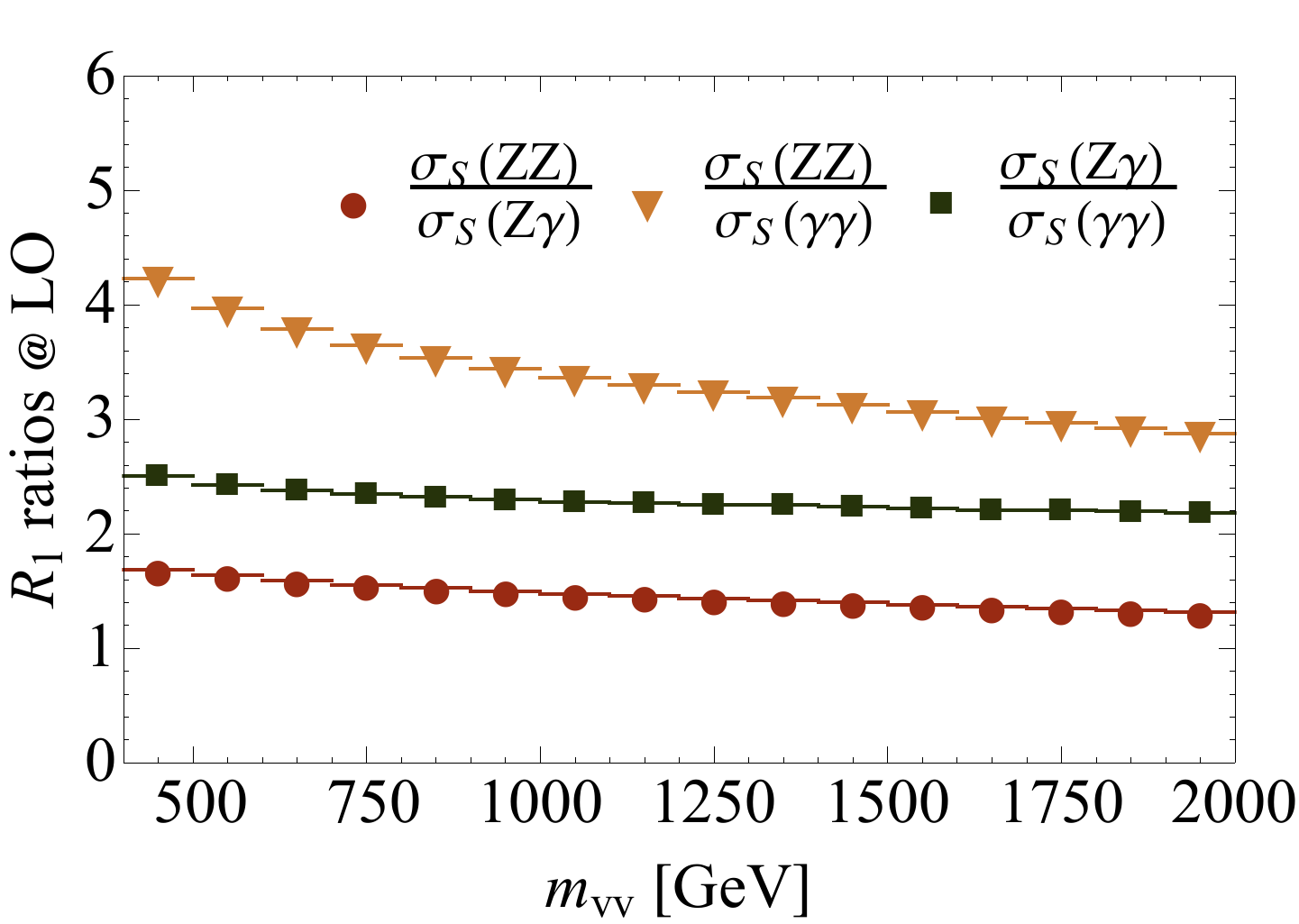}
    \end{center}
\caption{The $R_1$ ratios of $V_1^0V_2^0$ cross sections at LO, computed in MCFM at a $pp$ collider with $\sqrt{s}=13\TeV$. A pseudorapidity cut of $|\eta(V)|<1.5$ is imposed. These curves are determined almost entirely by ratios of parton luminosities, weighted by SM couplings.}
\label{fig:rat-za-lo}
\end{figure}

One can then get a rough estimate for the $R_1$ ratios by using \tab{Cq12} and applying the very crude relation $\mathscr{L}^S_{u\bar{u}} \sim 2\mathscr{L}^S_{d\bar{d}}$.  The small values of $C^d_{\gamma\gamma}, C^d_{Z\gamma}$ imply that $u\bar{u}$ initial states matter most for $R_{1a}$, and the parton luminosities largely cancel. We may therefore estimate $R_{1a} \sim C^u_{Z\gamma}/C^u_{\gamma\gamma} \sim 1.8$. Including $C^d_{12}$ and the crude relation among parton luminosities, the estimate increases to 2.1.  This estimate is very good, as we can see by looking at the actual LO $R_{1a}$ ratio in \fig{rat-za-lo}.  For $ZZ$, however, both $u\bar{u}$ and $d\bar{d}$ initial states are important. Although the similarly crude estimates $R_{1b} \sim 2.6$ and $R_{1c} \sim 1.3$ work quite well in the 1--2 TeV range, they are somewhat too small at low $\sh$ because\footnote{Effects from the $Z$ mass, neglected in these estimates, are indeed small, reaching only 3--6\% for $\sqrt{\hat s} \sim 500$ GeV.} $\mathscr{L}^S_{u\bar{u}} < 2\mathscr{L}^S_{d\bar{d}}$ for $\sqrt{\sh} \ll 1\TeV$. We will see later that NLO QCD makes only minor corrections to these ratios, especially at high energy.

\begin{figure}[tb!]
  \begin{center}
    \includegraphics[width=0.48\linewidth]{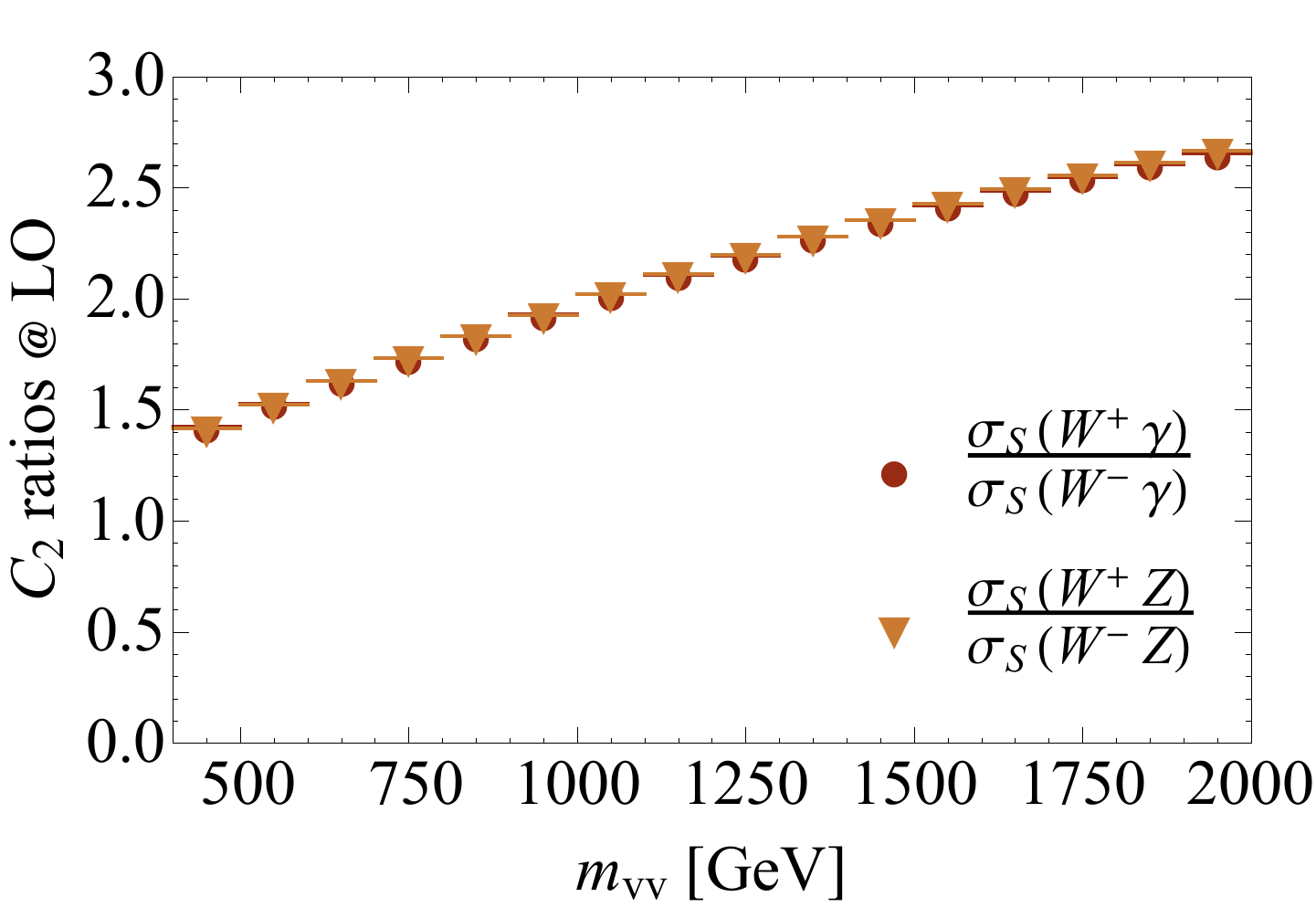}
    \includegraphics[width=0.48\linewidth]{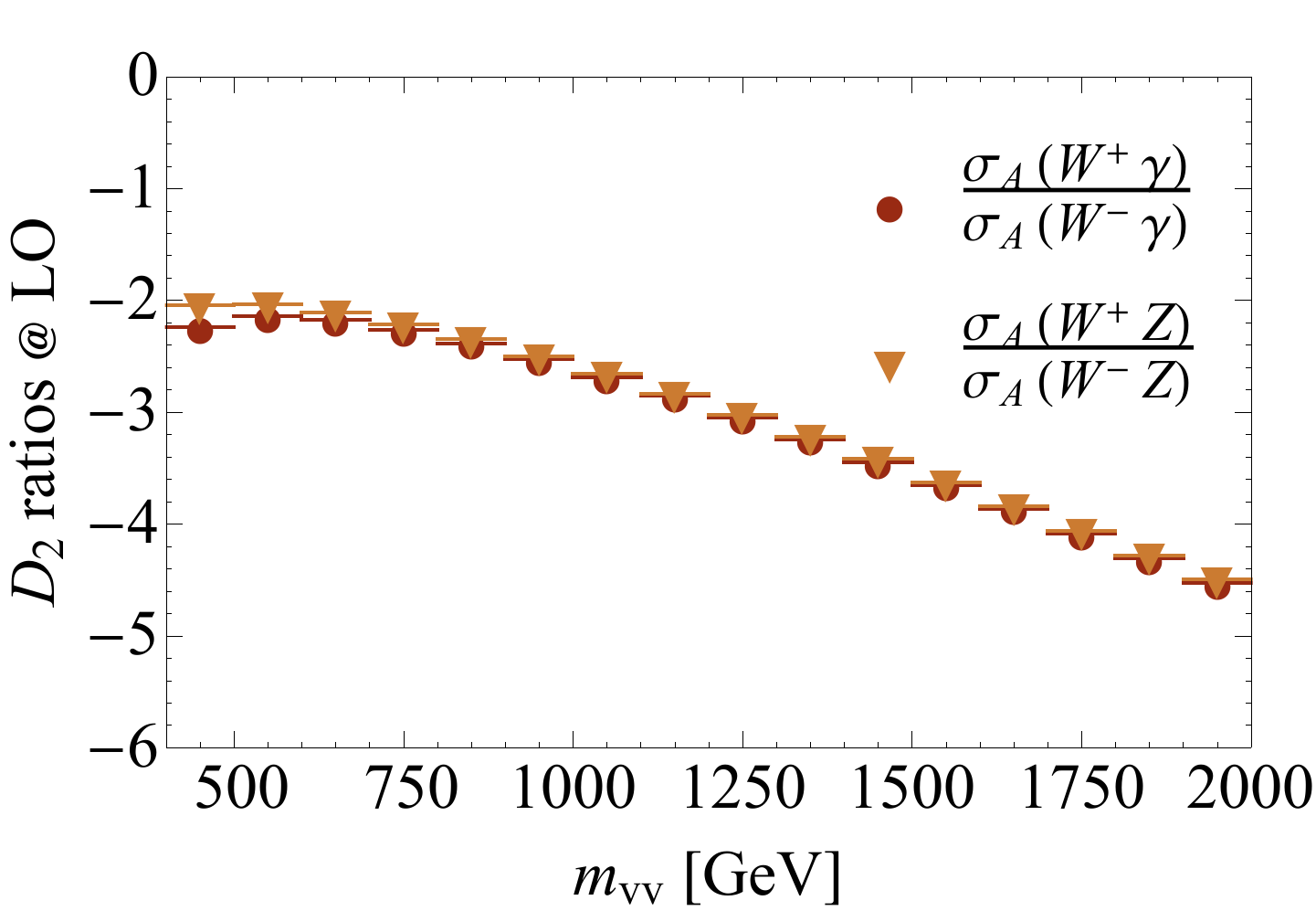}
  \end{center}
  \caption{(Left) The $C_2$ charge ratios at LO, which go roughly like $f_u/f_d$ and are identical for $W\gamma$ and $WZ$. (Right) The $D_2$ variables, also identical for $W\gamma,WZ$. These forward-backward asymmetric charge ratios have a similar dependence on the PDFs, complicated by $\text{sign}(y)$ in the asymmetric parton luminosity which results in $|D_2|>C_2$. }
\label{fig:Qrat-wza-lo}
\end{figure}

Next, we turn to the observables relating $W^+V^0$ and $W^-V^0$. We know from \eq{WV-rel,f+b} that the partonic cross sections $d\hat{\sigma}_S(W^+V^0)$ and $d\hat{\sigma}_S(W^-V^0)$ are identical. This leads to the following formula for the observable ``charge asymmetry'',
\begin{equation}
  C_{2a}(\sh)
    \equiv \left[\frac{\sigma_S(W^+\gamma)}{\sigma_S(W^-\gamma)}\right]_{\sh}
    \sim \frac{\sum_{q_u,q_d} |V_{q_u q_d}|^2 \mathscr{L}^S_{q_u\bar{q}_d}}
              {\sum_{q_u,q_d} |V_{q_u q_d}|^2 \mathscr{L}^S_{q_d\bar{q}_u}}\,,
\end{equation}
written as a ratio of weighted parton luminosities, with $V_{ij}$ the CKM matrix. The same result holds for $C_{2b} = \sigma_S(W^+Z)/\sigma_S(W^-Z)$. To derive an expectation for the magnitude and slope of these $C_2$ observables, we use the fact that $W^+V^0$ and $W^-V^0$ are produced predominantly at LO by $u\bar{d}$ and $d\bar{u}$, respectively. Then we have roughly that $C_2 \sim \mathscr{L}^S_{u\bar{d}}/\mathscr{L}^S_{d\bar{u}} \sim f_u/f_d$, which has a magnitude of order 2, grows with energy, and is identical for $W\gamma$ and $WZ$ with negligible mass corrections. These expectations are confirmed in \fig{Qrat-wza-lo}.

Similarly, because $d\hat\sigma_A(W^+V^0)$ and $d\hat \sigma_A(W^-V^0)$ are equal in magnitude and opposite in sign (see \eq{WV-rel,f-b}), we define
\begin{align}
\label{eq:D2}
D_{2a}(\hat s) ~&\equiv~
\left[\sigma_A(W^+\gamma) \over \sigma_A(W^-\gamma)\right]_{\hat s}
~\sim~ {\sum_{q_u,\,q_d} \, |V_{q_uq_d}|^2 {\mathscr L}^A_{q_u\bar q_d}\,
    \over - \sum_{q_u,\,q_d} \, |V_{q_uq_d}|^2 {\mathscr L}^A_{q_d\bar q_u}\, }
 ~\sim~- {{\mathscr L}^A_{u\bar d}\over {\mathscr L}^A_{d\bar u}}\,.
\end{align}
An identical result, with negligible mass corrections, holds for the $WZ$ processes in $D_{2b}$.
As we can see in \fig{Qrat-wza-lo}, $D_2$ has a similar shape to $C_2$, but with opposite sign and somewhat larger magnitude.  This can be understood by recalling ${\mathscr L}^A_{q\bar q} = \int dy \ \text{sign}(y) \,2\,f_q(x_1) f_{\bar q}(x_2)$. If the $y<0$ portion of the integral were zero, then we would have $|D_2|= C_2$. Instead, this portion is small, negative, and nearly identical for ${\mathscr L}^A_{u\bar d}$ and ${\mathscr L}^A_{d\bar u}$.  The fact that $|D_2|$ is fractionally larger than $C_2$ is merely a consequence of the inequality $(a-\epsilon)/(b-\epsilon)>a/b$ for $a>b>\epsilon>0$.

\begin{figure}[tb!]
    \begin{center}
      \includegraphics[width=0.48\linewidth]{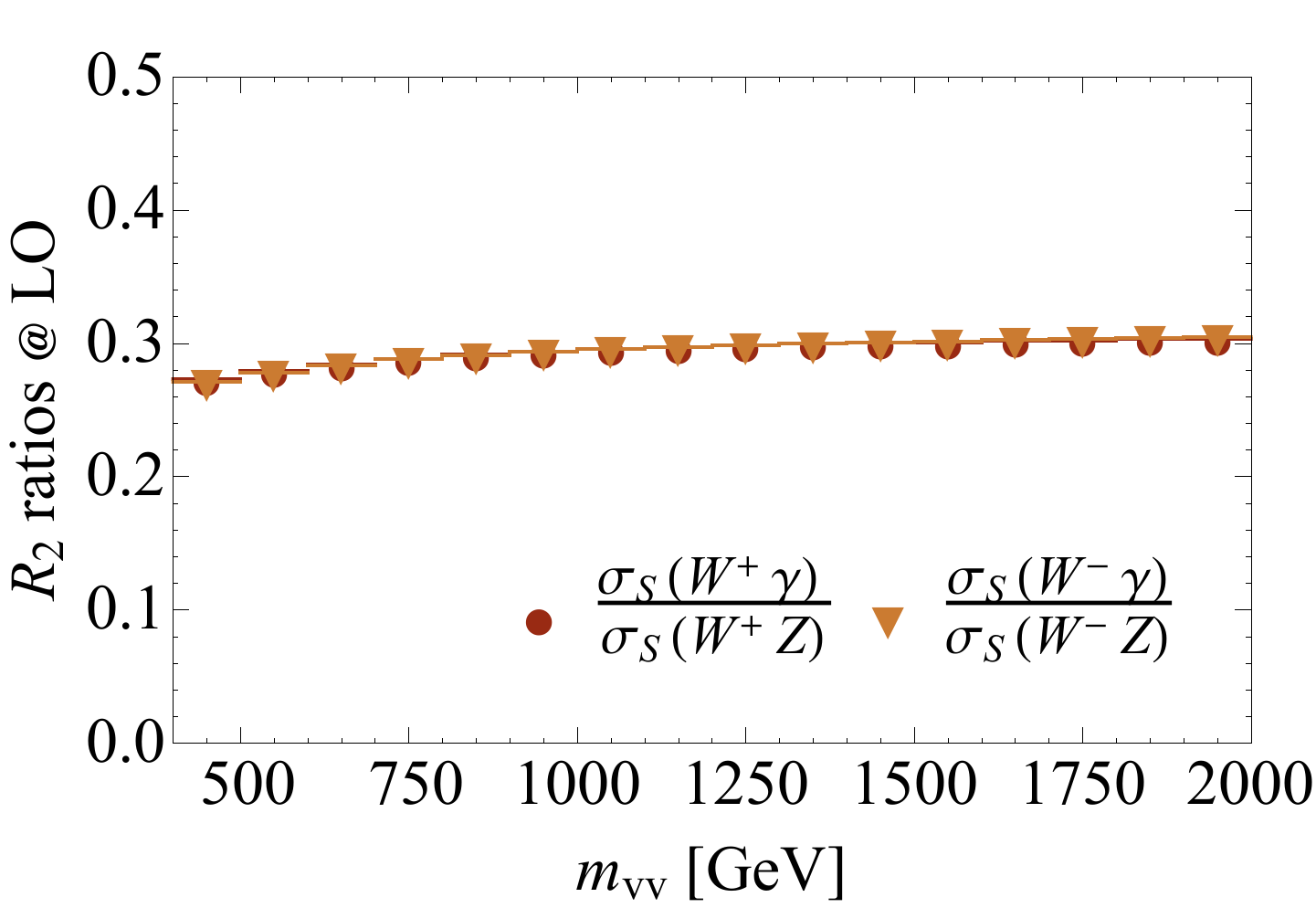}
      \includegraphics[width=0.48\linewidth]{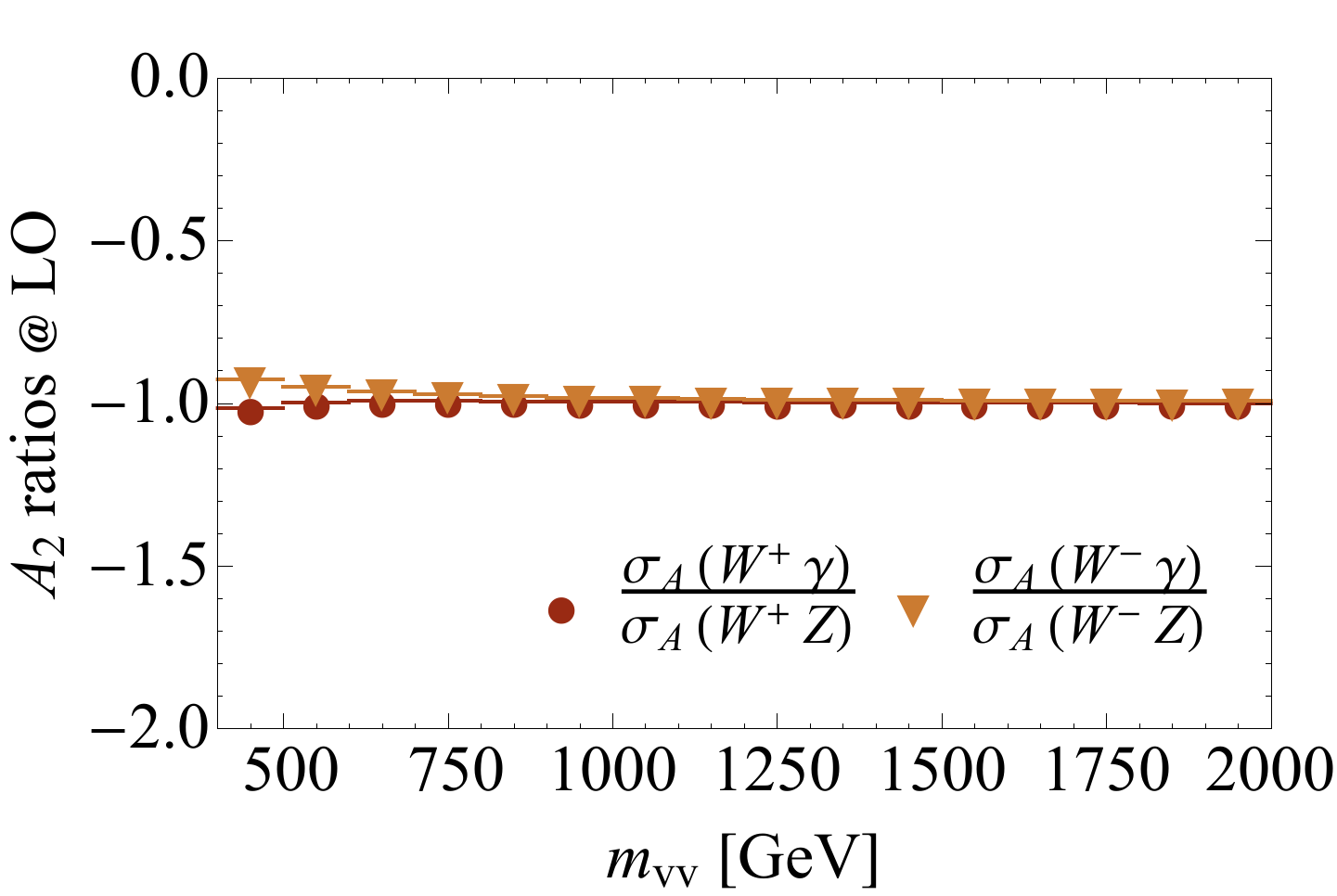}
    \end{center}
    \caption{
(Left) The $R_2$ ratios,  identical for $W^+,W^-$. These ratios are nearly $\tan^2(\theta_W)$, due to the coefficients of $|a_3|^2$ in the partonic rates.
(Right) The $A_2$ ratios at LO, also identical for $W^+$ and $W^-$. These equal $-1$ because partonic forward-backward asymmetries are equal and opposite for $W\gamma$ and $WZ$, which depend on the same PDFs.          }
\label{fig:Arat-wza-lo}
\end{figure}

Now we consider the observables that compare $W^\pm\gamma$ to $W^\pm Z$.
Both $\sigma_A(W^+\gamma)$ and $\sigma_A(W^+Z)$ depend on the same weighted parton luminosity, which appears as the numerator of \eq{D2}.  The antisymmetric partonic cross sections are equal in magnitude, opposite in sign, and proportional to $(a_1a_3)$. Everything thus cancels out of their ratio, leaving
\begin{equation}
  A_2^+(\hat s) ~\equiv~ \left[{\sigma_A(W^+\gamma) \over \sigma_A(W^+Z)}\right]_{\hat s}
  \approx - 1 \ .
\end{equation}
As seen in \fig{Arat-wza-lo}, this ratio differs from $-1$ at low $\hat s$ due to few-percent $m_Z^2/\hat s$ effects.\footnote{In addition to the mass corrections to $(a_1a_3)$ given in \app{MassCorr}, the Jacobian $|d\hat t/d\bar m_T^2|$ and the limits of integration also have mass dependence that differs in numerator and denominator.} The same holds for the $W^-V^0$ processes in $A_2^-$. Since the PDFs are absent, these ratio observables can be computed with relatively low theoretical uncertainty. It is most unfortunate that these ratios have the largest statistical errors, as we will see in \ssec{StatUnc}.

As we discussed at the end of \ssec{wza}, we naively expect
\begin{equation}
  R_2^\pm(\hat s) \equiv \left[{\sigma_S(W^\pm\gamma) \over \sigma_S(W^\pm Z)} \right]_{\hat s}
  \sim \tan^2 \theta_W \approx 0.29\,.
\end{equation}
The one subtlety is the radiation zero in $a_3$ at $\theta=\pi/2$, which is potentially important because this is the region of phase space where $d\hat\sigma_S/d\bar m_T$ peaks (due to the Jacobian $|dt/d\bar m_T|$). However, as seen in \fig{Arat-wza-lo}, the above estimate is a good one.  The reason is a combination of two pieces of good fortune.  The first is that the ratio of the partonic amplitudes everywhere lies between 0.29 and 0.19\,. Since $|a_1|^2=2$ and $|a_\phi|^2=1/16$ at $\theta=\pi/2$, we see from \eqs{wa}{wz} that
\begin{align}
    {d\hat\sigma_S}(u \bar d \to W^+\gamma) \propto {s_W^2\,Y_L^2}\,, \qquad
    {d\hat\sigma_S}(u\bar d \to W^+Z) \propto s_W^2\,t_W^2\,Y_L^2 + {1 \over 32}\,,
\end{align}
which means $d\hat\sigma_S(W^+\gamma)/d\hat\sigma_S(W^+Z)\to  0.19$ there.  The second is that the coefficients of $|a_1|^2$ and $|a_\phi|^2$ are so small that $|a_3|^2$ is numerically very important despite its radiation zero.

\begin{figure}[tb!]
    \begin{center}
        \includegraphics[width=0.48\linewidth]{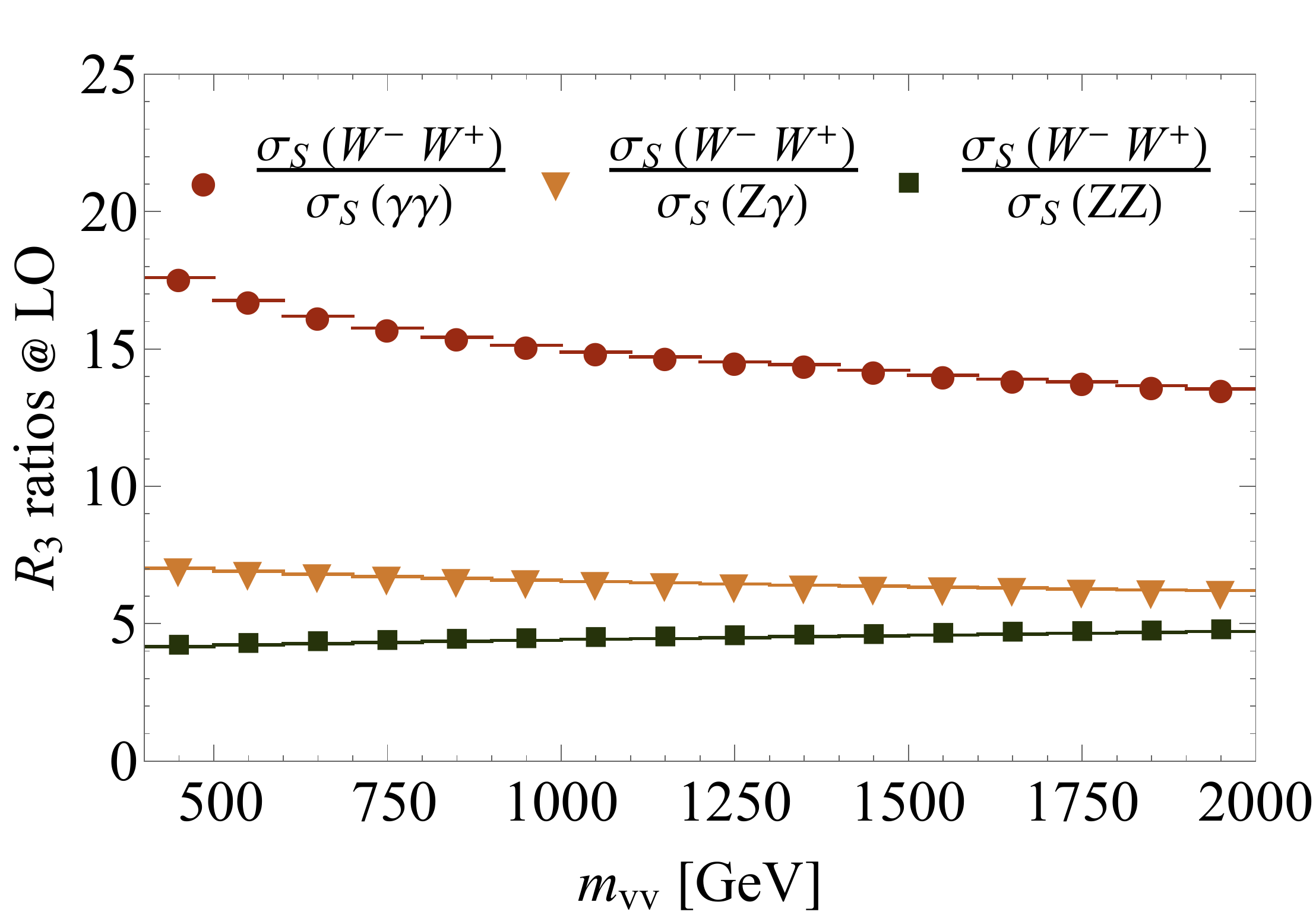}~~
        \includegraphics[width=0.48\linewidth]{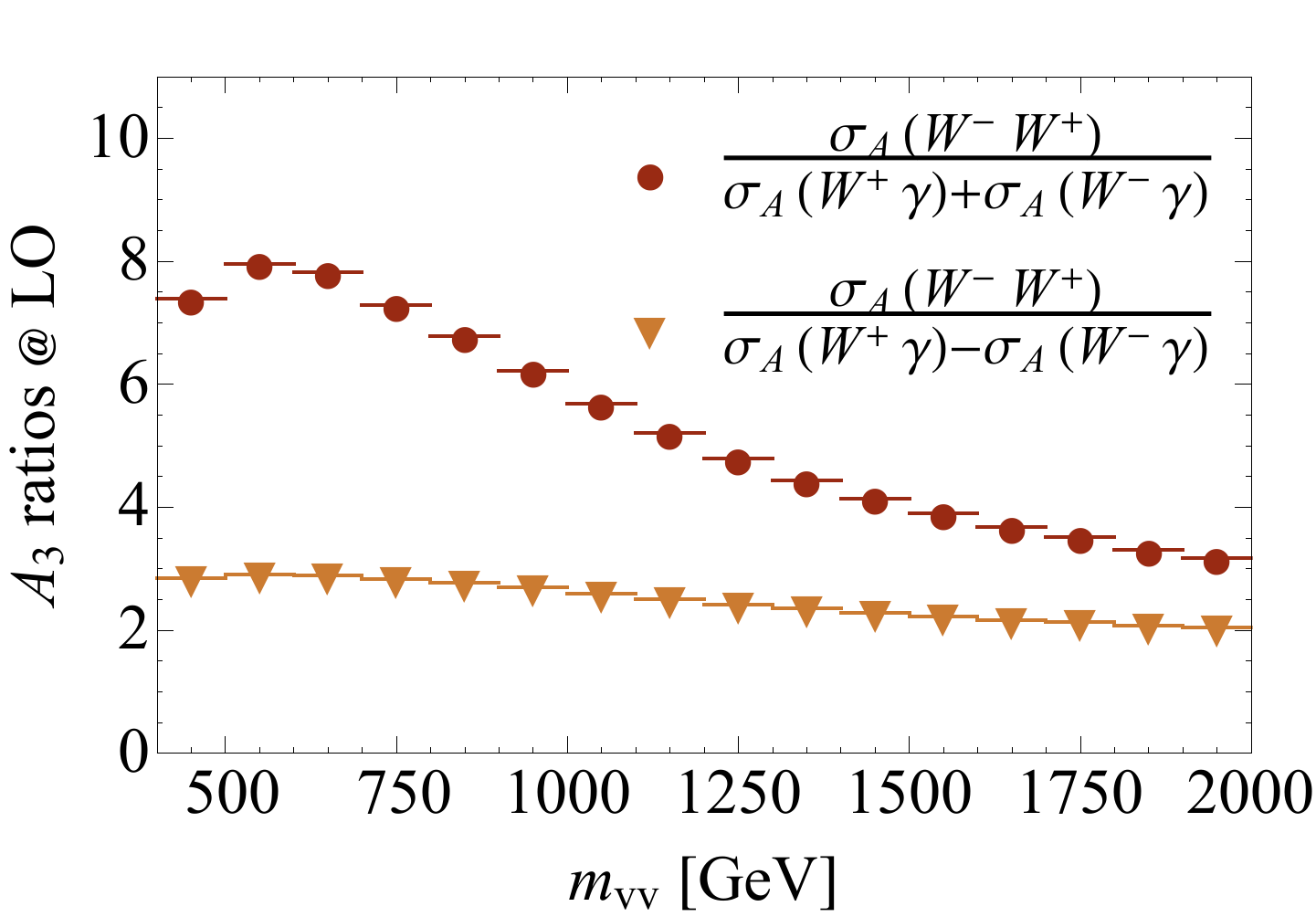}
    \end{center}
\caption{(Left) Possibly useful $R_3$ variables involving $W^-W^+$. Ratios are taken with $V_1^0V_2^0$ processes because $W^-W^+$ is dominantly produced as an $SU(2)$-singlet at LO. (Right) Possible $A_3$ variables involving forward-backward asymmetric $W^-W^+$ production. A property of the PDFs explains the flatness of the lower curve.}
\label{fig:ww-lo}
\end{figure}

This last statement is not true for $W^-W^+$; from \eq{uuww}, the relative coefficient of $|a_1|^2$ is 1/32, {\it vs.} $Y_L^2/8$ in $W\gamma$.  Consequently $d \sigma_S(W^-W^+)$ is dominated by the singlet term, making it roughly proportional to $d\sigma_S(V^0_1V^0_2)\sim |a_1|^2$.  This leads us to consider ratios such as
\begin{equation}
    R_{3a}(\hat s) \equiv \left[{\sigma_S(W^-W^+) \over \sigma_S(\gamma\gamma)} \right]_{\hat s}\,,
\end{equation}
and similarly $R_{3b} = \sigma_S(W^-W^+)/\sigma_S(Z\gamma)$ and $R_{3c} = \sigma_S(W^-W^+)/\sigma_S(ZZ)$. These possibilities are displayed in \fig{ww-lo}. We can estimate their magnitudes just as we did for the $R_{1}$ ratios above.  Comparing the coefficients of $|a_1|^2$ in \eqs{za}{uuww}, referring to \tab{Cq12}, and using the crude relation $\mathscr L^S_{u\bar u} \sim 2\,\mathscr L^S_{d \bar d}$, we get an estimate
\begin{equation}
    R_{3a} \sim {{1 \over 16}\left(\mathscr L^S_{u\bar u} + \mathscr L^S_{d\bar d}\right) \over s_W^4\,Q_u^4\,\mathscr L^S_{u\bar u}} \sim 10\,.
\end{equation}
Similar estimates for $R_{3b}$ and $R_{3c}$ then follow from the $R_1$ ratios in \fig{rat-za-lo}.

Although these estimates are not wildly off, they do come up somewhat short, even after allowing for $\mathscr L^S_{u\bar u} < 2 \, \mathscr  L^S_{d \bar d}$ at low $\hat s$.  This is  because we cannot actually ignore the $|a_3|^2$ contribution to $\sigma(W^-W^+)$, which makes up about 20\% of the total cross section. Because of this, one may be led to include some admixture of $\sigma_S(W^\pm V^0)$ in the denominators of the $R_3$ ratios. We leave it to further study to decide which admixture would have the most desirable properties at NLO.

Finally, we turn to ratios involving $\sigma_A(W^-W^+)$. As we saw earlier, the leading order partonic asymmetry in $W^-W^+$ is proportional to $(a_1a_3)$, as was the case for $W^\pm\gamma$ and $W^\pm Z$ (but see footnote \ref{foot:WW-asym}).  We therefore expect that a ratio of $d\sigma_A(W^-W^+)$ to any linear combination of the $d\sigma_A(W^\pm V^0)$ is given by a ratio of parton luminosities weighted by SM coefficients.  The asymmetries in $W^\pm Z$ suffer from low statistics, so we consider linear combinations of $d\sigma_A(W^+\gamma)$ and $d\sigma_A(W^-\gamma)$:
\begin{equation}
\label{eq:A3}
    A_3(\hat s) \equiv \left[{\sigma_A(W^-W^+) \over a\,\sigma_A(W^+\gamma) + b\,\sigma_A(W^-\gamma)} \right]_{\hat s} \sim {\mathscr L^A_{u\bar u} - \mathscr L^A_{d\bar d} \over 4\,|V_{ud}|^2\,s_W^2\,Y_L\,\left(a\,\mathscr L^A_{u\bar d} - b\,\mathscr L^A_{d\bar u}\right)}\,.
\end{equation}
It is an interesting non-obvious feature of the PDFs that, as functions of $\hat s$ in the kinematic region of interest,
\begin{equation}
    \mathscr L^A_{u\bar u} ~\propto~ \mathscr L^A_{d\bar d} ~\propto~ \mathscr L^A_{u\bar d}+ \mathscr L^A_{d\bar u}\,;
\end{equation}
the first (second) relation holds at the 2\% (15\%) level. This suggests the use of $a=-b=1$, which has the further advantage of minimizing the relative statistical uncertainty in the denominator of \eq{A3}.   Whether this is the ideal choice after NLO corrections are included remains to be seen.  We can see in \fig{ww-lo} that, at LO, this choice leads to a much flatter and smaller ratio than the choice $a=b=1$.

\subsection{Limitations of finite statistics}
\label{subsec:StatUnc}

Attractive as these ratios are, the reality of low cross sections means that many of these observables are not useful in the near term.  In \tab{numEvents} we show a rough estimate of the number of high-energy events ($m_{VV} > 400$ GeV) expected for each process. We assume 300 fb$^{-1}$ at $\sqrt{s}=13$ TeV and account for leptonic branching fractions of the $Z$ and $W$. We computed these numbers imposing a pseudorapidity cut $|\eta(V)|<1.5$ on the bosons (as in \tab{bosonJetCuts} below), and have separated events into ``Forward'' and ``Backward'' by the sign of $y(\eta_1-\eta_2$) as described in \ssec{pdfs}.

Any one of our ratios becomes interesting as a precision observable once its statistical uncertainty becomes of order 5--10\%, so that its exceptionally low theoretical errors become experimentally relevant.  If such small uncertainties are possible for a particular ratio only by combining all events together into a single bin, \eg~using ratios of total cross sections with $\bmT > 200\GeV$, then this measurement is likely to be useful only for testing methods for SM predictions, since it will be sensitive mainly to physics only up to the $\sqrt{\hat s} \sim 400$ GeV range.  However, more can be done once the events can be divided into multiple bins of varying width, each with statistical uncertainty of order 5--10\%, as in \fig{mainResult}.  In this case the lower bins serve as a test of the predictive techniques, while the higher ones are useful for other purposes, including searches for BSM phenomena and tests of important EW corrections that grow with energy and do not entirely cancel in these ratios.

\begin{table}
    \begin{center}
        \begin{tabular}{| c | c | c |}\hline
            $V_1V_2$ & $N_f+N_b$ & $N_f-N_b$  \\ \hline\hline
            $\gamma\gamma$ & 12\,000 & 0 \\
            $Z\gamma$ & 2000 & 0 \\
            $ZZ$ & 220 & 0 \\
            $W^+\gamma$ & 3300 & $-500$ \\
            $W^-\gamma$ & 2100 & 220 \\
            $W^+Z$ & 790 & 33 \\
            $W^-Z$ & 520 & $-16$ \\
            $W^-W^+$ & 9500 & $-430$ \\\hline
        \end{tabular}
    \end{center}
\caption{At LO, the number of events with $\sqrt{\hat s} = m_{VV} > (400\text{ GeV})^2$ assuming 300 fb$^{-1}$ and leptonic decays of $W$ and $Z$. $N_f$ and $N_b$ indicate forward- and backward events. These numbers increase by a factor of order 1.5--2 at NLO, but are reduced by a comparable amount when using the variable $\bar m_T$ instead of $m_{VV}$.}
\label{tab:numEvents}
\end{table}

As we saw in \fig{mainResult} of \sec{ExecSumm}, the ratio $R_{1a}$ permits 6 bins at 300 fb$^{-1}$ with 6\% statistical uncertainties.  At this integrated luminosity, the other variables that allow multiple bins with $\sim$5\% uncertainties are $C_{2a}$ and $R_3$, as one can see using \tab{numEvents}. Meanwhile $R_2^\pm$, $C_{2b}$, and $D_{2a}$ allow for a single bin.

The situation will improve at 3000 fb$^{-1}$, though the high pileup environment may lead to some loss of statistics.  If we simply assume the total rate increases by a factor of 10 without significant losses, we find that in addition to the above six variables, the variables $R_{1b}$, $R_{1c}$ and $A_3$ also permit multiple bins.  The $A_2^+$ ratio can be used in a single bin. The two variables $A_2^-$ and $D_{2b}$ involving $\sigma_A(W^-Z)$ are too small to measure.

It may prove useful to improve statistics slightly by combining observables predicted to be equal within the SM.  For instance, one could replace $R_2^+$ and $R_2^-$ with
\begin{equation}
R_2^{0}=
 \frac{\sigma_S(W^+Z)+\sigma_S(W^-Z)}{\sigma_S(W^+\gamma)+\sigma_S(W^-\gamma)}.
\end{equation}
Similar combinations would assist with $A_2^+$ and $A_2^-$ (see \fig{Arat-wza-lo}),  $C_{2a}$ and $C_{2b}$, and $D_{2a}$ and $D_{2b}$ (see \fig{Qrat-wza-lo}).

\section{Beyond leading order for \texorpdfstring{$\gamma\gamma$, $Z\gamma$, $ZZ$}{diphoton, Z-photon, ZZ}}
\label{sec:beyondLO}

In \ssec{partonic} we saw that the differential LO partonic cross sections for $V_1^0V_2^0 = \gamma\gamma$, $Z\gamma$, $ZZ$ are all proportional to the same function $|a_1|^2$, up to $m_Z^2/\pT^2$ effects (provided in \app{MassCorr}).  Consequently, at high energy, the ratios of these partonic cross sections are given by constants of the SM. Since the up quark PDF dominates $\gamma\gamma$ and largely dominates $Z\gamma$, the hadronic ratio $R_{1a}$ is approximately constant and equal to a simple  partonic ratio.  Although the PDFs have a greater effect on the hadronic cross sections for $R_{1b}$ and $R_{1c}$, these two observables still vary rather slowly with $\sqrt{\hat{s}}=m_{VV}$, with easily understandable values, as we saw in \fig{rat-za-lo}.

\begin{figure}
    \begin{center}
        \includegraphics[width=0.48\linewidth]{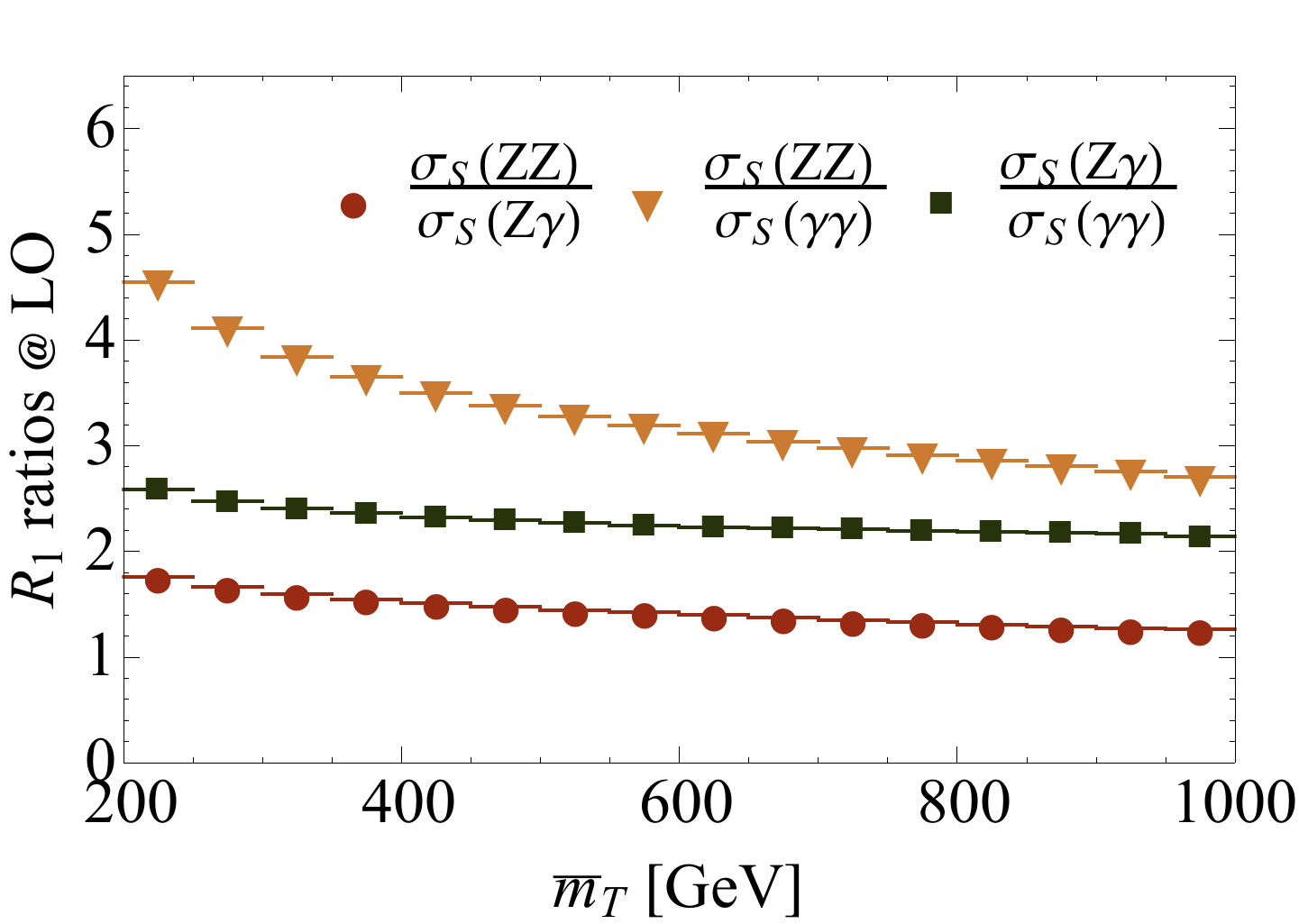}
    \end{center}
    \caption{The $R_1$ ratios at LO binned in $\bar m_T$.}
\label{fig:mtRatio}
\end{figure}

Beyond LO, we will study the $R_{1}$ ratios differentially with respect to $\bar{m}_T$, the average $m_T$ of the two vector bosons, \eq{whymT}.  The LO ratios in this variable are given in \fig{mtRatio}.
Comparing with \fig{rat-za-lo}, one can see that the LO ratios as functions of $\bar m_T$ and as functions of $m_{VV}/2$  are quite similar. This is because the hadronic cross section for a given $\bar m_T$ is dominated by the region with $\sqrt{\hat s}\sim 2\bar m_T$.

The fully-differential cross-sections for the diboson processes have been known for quite some time~\cite{Smith:1989xz,Ohnemus:1990za,Mele:1990bq,Ohnemus:1991gb,Ohnemus:1991kk,Frixione:1992pj,Bailey:1992br,Ohnemus:1992jn,Frixione:1993yp}. In this and following sections, all calculations are carried out using \textsc{MCFM 6.8}~\cite{Dixon:1998py,Campbell:1999ah,Campbell:2011bn}, except for an NNLO real emission study which used \textsc{MadGraph 2.3.0}~\cite{Alwall:2014hca}.  Renormalization and factorization scales $\mu_R,\mu_F$ are chosen at $m_{VV}$ except when otherwise specified.  We use MSTW 2008 NLO [LO] PDFs~\cite{Martin:2009iq} for all NLO [LO] calculations and for $\ord(\aS^3)$ [$\ord(\aS^2)$] $gg\to V_1^0V_2^0$ calculations.  Our cuts on the bosons are presented in \tab{bosonJetCuts}. 
See \sec{Practical} for cuts on their decay products.

\subsection{Choices of observable and of cuts}
\label{subsec:cuts}

We begin with a discussion of our cuts and our observable.  It is important to choose these carefully in order to avoid large NLO and NNLO corrections to our ratios, and associated large uncertainties.

We will discuss certain experimental realities in \sec{Practical}, but for now we neglect $Z$ decay and impose cuts on the vector bosons and on any jets,\footnote{We will refer to all final-state colored partons, for brevity only, as ``jets''.  We do not include showering and hadronization in our study, but we expect these to have small effects, since we impose cuts on our observables to avoid regions where resummation plays an important role.} as in \tab{bosonJetCuts}. (Our cuts on leptons in $Z \to \ell^+\ell^-$ are given in \tab{leptonCuts} of \ssec{Zdecay}.)   In our discussion we will have at most one jet and so for us $H_T$ is simply the \pT of that jet, but it is important that $H_T$ be the variable used at higher jet multiplicity, not maximum jet \pT.  This cut ensures that multiple jets with \pT just below our cuts cannot combine together on one side of the event and force the two bosons to be close in angle, or allow one boson to be soft relative to the QCD activity. Either of these effects would allow events that are far in phase space from the LO kinematics to enter the measurement, and potentially cause large corrections and failures of cancellations in our ratios. Note also that we choose identical kinematic cuts for $Z$ and $\gamma$, which we supplement in \sec{Practical} when being more experimentally realistic.  We discuss angular isolation of the bosons in \ssec{NLO-QCD} and \ssec{PhotonIso}.

{\renewcommand{\arraystretch}{1.5}
\begin{table}
    \begin{center}
      \begin{tabular}{| c |}\hline
        Kinematic Cuts \\ \hline \hline
          $|\eta(V_i)| < 1.5$ \\  \hline
          $p_T(V_2) > \frac12 \,p_T(V_1)$ \\ \hline
            $H_T = \sum_{\text{jets}} |p_T^{j}| < \frac12 \, p_T(V_2)$ \\ \hline
        \end{tabular}
    \end{center}
\caption{ Kinematic cuts imposed on vector bosons $V_i$ and on the jets $j$ from real emission at NLO.  In our calculations we work only to single real emission so $H_T$ is simply $\pT^j$, but the use of an $H_T$ cut is important at higher orders. We define $V_1,V_2$ by $p_T(V_1) > p_T(V_2)$.  Isolation requirements and cuts on decay products are described in \sec{Practical}.}
\label{tab:bosonJetCuts}
\end{table}
}

A variety of problems can arise that can invalidate or destabilize fixed-order calculations. Our cuts, which allow the vector bosons to have unequal \pT, but require both bosons have substantially higher \pT than any jet from real emission, are chosen to avoid them.  Note also that our cuts generally scale with the overall average \pT, and roughly with our observable \bmT.

One issue we must avoid is large logarithms.  The fairly loose cut on additional hadronic activity, $H_T < \frac12\, p_T(V_2)$,  means that logarithms of  $p_T(V)/H_{T,\text{min}}$ never become so large as to require jet veto resummation~\cite{Banfi:2012yh,Banfi:2012jm,Tackmann:2012bt}. But because our cuts scale with the average \pT, we also avoid large logarithms of $p_T(V_1)/p_T(V_2)$, which (in combination with a large $qg$ parton luminosity) could have led to very large corrections \cite{Rubin:2010xp}.  Simultaneously,\footnote{We thank  Z.~Bern for alerting us to possible subtleties with these cuts and specifically to ref.~\cite{Frixione:1997ks}.} asymmetric cuts on the bosons avoid logarithms of $p_T(V)/\Delta$, where $\Delta = \pT^\text{cut}(V_1) - \pT^\text{cut}(V_2)$; these logarithms, which arise from soft gluon emission, were first identified in ref.~\cite{Frixione:1997ks} and resummed in ref.~\cite{Banfi:2003jj}.
Meanwhile our observable itself, $\bmT$, does not appear in large logarithms and requires no resummation.

Other effects can enhance the size of fixed-order terms relative to naive expectations. For instance, if radiative corrections are allowed to populate phase space at lower $\sqrt{\sh}$ than is accessible at tree level, the formally NLO calculation carries \emph{de facto} LO scale uncertainties.  This does not happen with our cuts and observable; all bins in \bmT are dominated by the LO contribution.

Another common issue with $q\bar q$ processes is the opening of new channels with large parton luminosities at higher orders.  At NLO, we have the new channel $qg\to qV_1^0V_2^0$, but our cuts mitigate the \Kfac factors, making them of order 1.5.  Moreover, these \Kfac factors are nearly process independent and largely cancel in our ratios.  At NNLO, we have the new channel $gg\to V_1^0V_2^0$, which is substantial and process dependent; we include it in our calculation.  Also at NNLO is the new channel $qq\to qqV_1^0V_2^0$, which is process dependent and potentially large for valence quarks. We estimate that with our cuts, (which avoid any large logarithmic enhancements,) this process is subleading; we do not evaluate it but include it in our uncertainty estimates.

We must also avoid situations where higher-order matrix elements (at a particular jet multiplicity) are enhanced relative to LO matrix elements dressed with soft and collinear factors (at the same multiplicity). One way this can happen is if an additional jet emission can make a threshold or resonance accessible that was inaccessible at lower order.  This can occur in QCD corrections to the $Z\gamma$ process, via radiative $Z$ decays $Z \to \ell^+\ell^- \gamma$.  Simply because we take \bmT $>$ 200 GeV, this is irrelevant at LO, and our $H_T$ cut assures this does not arise at any order in \aS.

A further potential problem can appear if a radiative emission can significantly decrease an internal propagator's virtuality compared to the analogous propagator in the LO process, thus enhancing the amplitude.  (Strictly speaking, this way of stating things is not gauge invariant,
but the enhancement itself clearly is.) With our cuts and observable, this too does not occur.

\subsection{NLO QCD corrections}
\label{subsec:NLO-QCD}

\begin{figure}
\begin{center}
    \setlength{\unitlength}{1mm}
    \begin{picture}(100,25)
        \put(20,0){\includegraphics[width=0.4\linewidth]{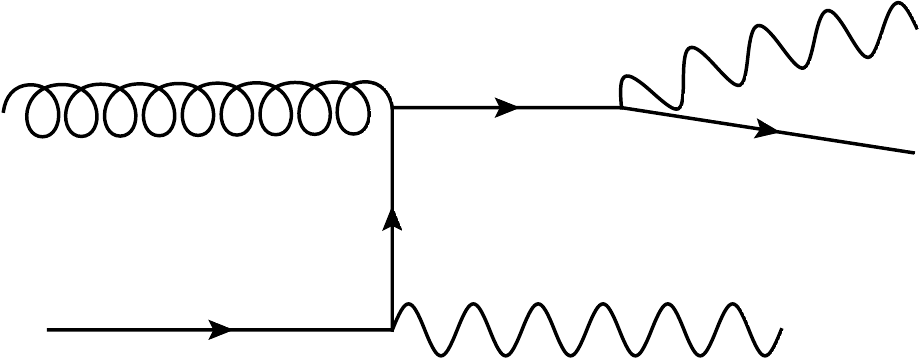}}
    \end{picture}
\end{center}
\caption{The regime in which $V$ and $q$ are nearly collinear in the final state, the source of a significant difference between photon- and $Z$-rates.}
\label{fig:CollFeyn}
\end{figure}

For our observables and with our choice of cuts, virtual and real QCD corrections to $d\hat\sigma(q\bar q\to V_1^0V_2^0)$ are largely proportional to the LO values.  Consequently the $R_1$ ratios receive only small NLO QCD corrections in most regions of phase space.  The exception is in the region where a final-state quark is nearly collinear with a vector boson;  this region is enhanced for photons by large logs from collinear emission, whereas for $Z$s the logarithmic enhancement is cut off by $m_Z$.  More specifically, for $Z$ emission the quark propagator in \fig{CollFeyn} is bounded from above by $1/m_Z^2$, while the photon's collinear singularity at low $m_{q\gamma}$  must be absorbed into a non-perturbative fragmentation function, or evaded through an angle-dependent energy isolation cut that avoids generating soft divergences at higher order.  This fundamental difference between $Z$ and $\gamma$ cannot be removed experimentally, and gives a significant NLO shift to the $R_1$ ratios at low $\bar m_T$.

The collinear-$\gamma q$ singularity can be dealt with using the smooth-cone isolation method of Frixione \cite{Frixione:1998jh}. (While theoretically elegant, this method is not practical; we will employ a more experimentally realistic version of Frixione isolation, and discuss the uncertainties inherent in its use, in \ssec{PhotonIso}.) In this method, one chooses two parameters $\delta,\epsilon$ and requires that in any cone of radius $R<\delta$, the hadronic activity is bounded by a function that goes smoothly to zero as $R \to 0$; in particular\footnote{Frixione included a third parameter $n$ as an exponent on the trigonometric function here; we have chosen $n=1$.}
\begin{eqnarray}
  \label{eq:frix}
    \sum_{h \in R} \pT^h <  p_T(V)\ {\mathcal I}(R;\epsilon,\delta) \quad \text{for all $R < \delta$} \ ,\\
  \label{eq:frixfunc}
    {\mathcal I}(R;\epsilon,\delta) = \epsilon \left({1-\cos R \over 1-\cos\delta}\right) .
\end{eqnarray}
Here the sum is over all hadrons $h$ within a cone of radius $R$ around the boson.

\begin{figure}
    \begin{center}
        \includegraphics[width=0.48\linewidth]{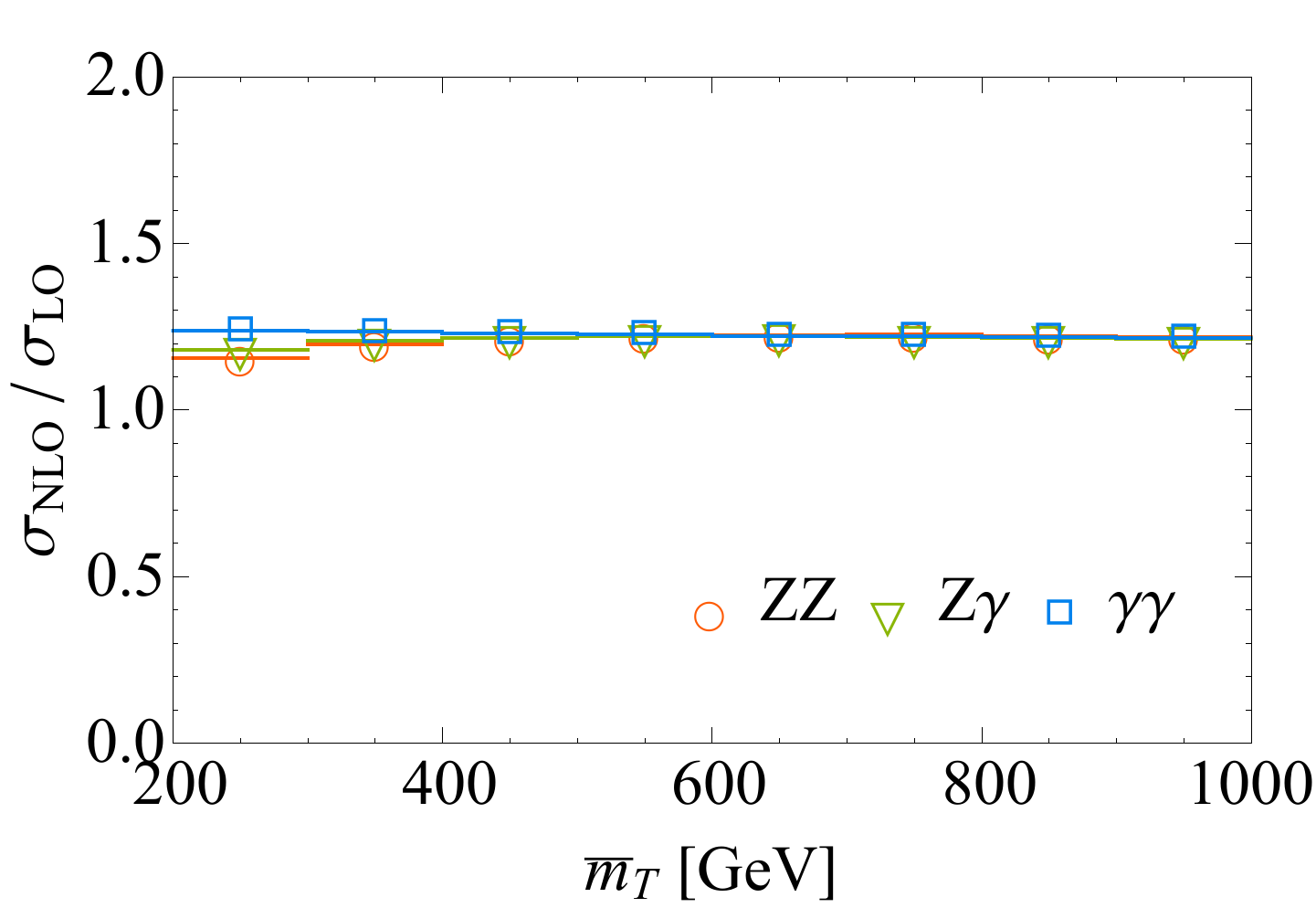}~~
        \includegraphics[width=0.48\linewidth]{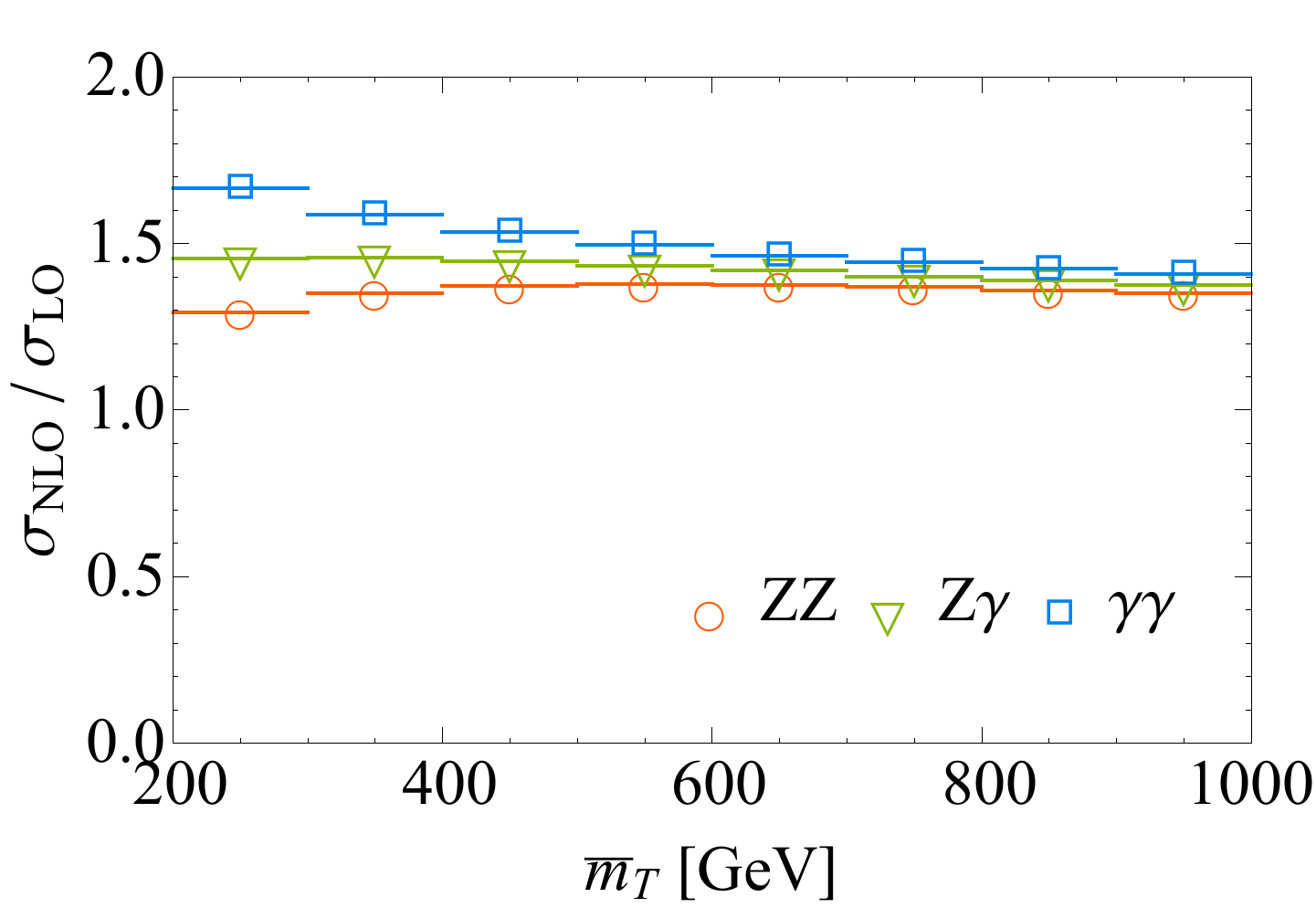}
    \end{center}
\caption{The $V_1^0V_2^0$ cross section at NLO, shown relative to the LO rates. At left, the collinear region was removed by a very strict smooth-cone isolation cut $(\delta,\epsilon)=(1.2,0.2)$ applied to both $\gamma$s and $Z$s. All 3 processes receive identical NLO corrections, thus leaving the ratios invariant. At right, with a reasonable isolation cut $(\delta,\epsilon)=(0.4,0.5)$ the NLO corrections differ significantly among the processes at low energies.}
\label{fig:dead-cone}
\end{figure}

That the $R_1$ ratios remain unchanged outside the collinear regime may be seen by applying the Frixione method with extreme parameters $(\delta,\epsilon)=(1.2,0.2)$.  This choice largely removes the collinear region.  Here (but see below) we apply isolation {\it both} to photons and $Z$s, to maintain as much congruence as possible.  At left in \fig{dead-cone}, we see that the \Kfac factors are then almost identical for the three $V^0_1V^0_2$ processes,  and so the $R_1$ ratios at NLO are the same as at LO.

However, as seen at right in the same figure, when the collinear region is restored by using more reasonable smooth-cone parameters $(\delta,\epsilon)=(0.4,0.5)$, there is a significant splitting in the \Kfac factors at low $\bar m_T$, where the $Z$ mass is particularly relevant, and thus a shift in the $R_1$ ratios away from their LO values.  Note that the splitting of $\gamma\gamma$ from $ZZ$ is roughly double that of $\gamma\gamma$ from $Z\gamma$, so the effect of the collinear regime is largest on $R_{1b}$.

In all results beyond this point we use $(\delta,\epsilon) = (0.4,0.5)$, with appropriate practical modifications discussed in \ssec{PhotonIso}. For this choice, and for the range of \bmT that is relevant for the LHC, we find it unnecessary to impose isolation on $Z$s, for the following reasons.  At low \bmT the Frixione cut removes a region where the amplitude for $Z$ emission is not enhanced. Meanwhile at larger \bmT the falling $qg$ parton luminosity  makes the collinear region less important even for photons, an effect seen at right in \fig{dead-cone}, and also tends to favor the region of low $p_T^q/p_T^Z$, which is not removed by the Frixione cut. Altogether this reduces the impact of Frixione isolation on $Z$s to the percent level, relative to the total differential cross section.
Therefore, {\it in what follows below and in our final results, we impose isolation only on photons, not on $Z$s}, and believe it is safe for the LHC experiments to do the same without negatively impacting the ratios.  At a higher-energy collider this would need to be revisited.

With these Frixione parameters, our lowest $\bar m_T$ bin sees a downward shift of $R_{1a} = \sigma(Z\gamma)/\sigma(\gamma\gamma)$ by 15\%, of $R_{1b}=\sigma(ZZ)/\sigma(\gamma\gamma)$ by 25\%, and of $R_{1c}=\sigma(ZZ)/\sigma(Z\gamma)$ by 12\% relative to the LO values.  In higher bins, the effect of the collinear regime is muted  as the $gq$ parton luminosity falls and the difference between photon and $Z$ amplitudes decreases.

It is instructive to understand why the NLO corrections to the $R_1$ ratios are so small outside of the collinear region.  The point is that most logarithmically-enhanced corrections are themselves proportional to the LO process, for reasons that even extend to many regions of phase space that are not log-enhanced. For instance, in the NLO process $q\bar q\to V_1^0V_2^0 g$, our cuts are inclusive in the initial state radiation (ISR) region of phase space, where the final-state gluon is collinear with the initial partons.  Consequently a fixed-order calculation is a reliable guide, and the NLO diagrams that appear are the same for all three processes.  Thus no large process-dependent corrections arise, and the $R_1$ ratios are hardly affected.  Meanwhile emissions of hard gluons are suppressed by our jet cuts.

Similarly, for the ISR region of $qg \to V_1^0V_2^0 q$, the ratios are little changed, for two reasons. First, the partonic cross section near this singular region displays a factorization into the tree-level cross section and a universal factor that is absorbed into the definition of the PDFs. Second, the replacement of an anti-quark PDF with a gluon PDF has a small impact, because $\bar u$ and $\bar d$ PDFs are similar.  We may see this heuristically by writing
$f_{\bar q}= \frac12(f_{\bar u}+f_{\bar d})$ and $\bar\delta= \frac12(f_{\bar u}-f_{\bar d})$, and noting the  $qg$ integrand is roughly proportional to
\begin{equation}
  \label{eq:qgpdfs}
  \left[f_{u}(x_1) d\hat\sigma^{LO}_{u\bar u} +
      f_{d}(x_1) d\hat\sigma^{LO}_{d\bar d}\right]\left[f_{g}(x_2/z)  P_{q\leftarrow g}(z) \right]
\end{equation}
while the tree-level process has integrand
\begin{equation}
  \label{eq:qqpdfs}
    \left[f_{u}(x_1) d\hat\sigma^{LO}_{u\bar u} +
      f_{d}(x_1) d\hat\sigma^{LO}_{d\bar d}\right]f_{\bar q}(x_2) + \ \ord(\bar{\delta}) \ .
\end{equation}
Here $P_{q\leftarrow g}$ is the gluon-to-quark splitting function, and we have ignored small contributions from subdominant initial states.  Since $\bar \delta\ll f_{\bar q}$ in the relevant $x$ range, these integrands are proportional, so no large correction to the LO ratios is expected from the ISR region.

\subsection{NNLO QCD corrections}
\label{subsec:gg}

Although NNLO  calculations of diboson processes have been carried out for all processes except $WZ$~\cite{Catani:2011qz,Grazzini:2013bna,Cascioli:2014yka,Gehrmann:2014fva,Grazzini:2015nwa,Grazzini:2015hta}, most of these are not yet accessible in public code. This limits our ability to refine our NLO results or to estimate the theoretical uncertainties from which they suffer.  In this context, we take the following approach.  On the one hand, we study in detail the largest known NNLO correction to our ratios, namely $gg\to V_1^0V_2^0$, which is large enough that it must be included, but fortunately is available publicly.  On the other hand, we search for additional NNLO corrections that should affect our ratios, and make rough estimates of their size to see if they are important; if so we include them as a theoretical uncertainty.

We saw in \fig{dead-cone} and eqs.~\eqref{eq:qgpdfs}--\eqref{eq:qqpdfs} that many NLO corrections are common to all three $V^0_1V^0_2$ processes and cancel in the $R_1$ ratios.  Similar logic would suggest that many NNLO QCD corrections are also common to the three processes and that, away from the collinear-$qV$ regions, new real contributions like $q\bar{q} \to V_1^0V_2^0 gg$, or $qg \to V_1^0V_2^0 qg$  are likely to cancel.   But by looking carefully at the physical origin of various effects, we can also see where such cancellations will fail.

Before we do so, let us forestall an obvious question.  Below, we will assume that many NNLO corrections cancel in ratios, and that the largest one that does not cancel comes from the $gg\to V_1^0V_2^0$ loop graph (as suggested in ref.~\cite{Bern:2002jx}), which we will include explicitly below.
One might question this assumption based on the existing NNLO and near-NNLO literature, which suggests potentially large $\Kfac_\text{NNLO/NLO}$ factors ($1.3 - 1.6$), substantial process-dependence in these \Kfac factors, and effects that can be much larger than the $gg\to V_1^0V_2^0$ loop graph.  How, then, can we possibly claim that NNLO corrections to our ratios could be brought under control, and further assume that even higher-order effects can be ignored?

Here one needs to look carefully at the details, which we do in \app{NNLO}.  The large $\Kfac_\text{NNLO/NLO}$ arise only in situations where the cuts on the bosons and jets are very different from our own, causing even the $\Kfac_\text{NLO/LO}$ factor to be much larger than the $\sim 1.5$ that we found above in \fig{dead-cone}.  The process-dependent differences among the $\Kfac_\text{NNLO/NLO}$ factors also appear much smaller when one restricts to kinematic regions and observables similar to the ones we are considering.  In those regions there is no clear indication that the $gg$ loop is not the main process-dependent effect.  Thus there is no clear evidence against our assumptions, and even some mild (though hardly decisive) evidence in their favour.   Let us note again that our choice of observable and of cuts appears to be crucial in this regard; many other observables and cuts would have larger NNLO corrections in ratios.

With that issue set aside, we now consider obvious sources of NNLO corrections that will not cancel in our ratios.  Since the dominant NLO correction to the ratios, shown in the right-hand plot of \fig{dead-cone}, was from the collinear-$qV$ region, corrections to that region of phase space will not cancel. NNLO real and virtual corrections to this single-collinear effect will impact the ratios. However, we expect these to give an order $\alpha_s$ adjustment to the splitting shown in \fig{dead-cone}, which puts them below 2\%.

Another important contribution could come from the double-collinear region in $q\bar q, gg \to V_1^0V_2^0 q \bar q$.  This too is very small, despite the large NLO single-collinear correction.  To see this, note the following.  The reason that $qg\to V_1^0V_2^0q$ is so important is that ${\mathscr L}_{qg}\gg {\mathscr L}_{q\bar q}$, partially canceling the extra $\alpha_s$ at NLO.  There is no corresponding enhancement for two independent collinear emissions.  The double-collinear region at the next order should be thought of predominantly as $q\bar q\to q\bar q$, $gg\to q\bar q$, with double emission $q\to q V_1$ and $\bar q\to \bar q V_2$.  (Our cuts remove the region where both $V_1$ and $V_2$ radiate off a single quark.)  For $q\bar q\to q\bar q$ the parton luminosity is the same as that arising at LO, so the $q\bar q$-initiated process is indeed suppressed by $\ord(\alpha_s^2)\sim 1\%$ compared to LO.  Meanwhile,  ${\mathscr L}_{gg}$ is comparable to or smaller than ${\mathscr L}_{qg}$ at the relevant energies; and  furthermore $gg \to q\bar{q}$, which lacks a $t$-channel gluon, has a smaller partonic cross section than $qq\to qq$ and $q\bar{q} \to q\bar{q}$. Altogether it appears the double-collinear regime shifts the ratios at the percent level or below.

A qualitatively new source of non-canceling corrections is from the opening of a new channel at NNLO, namely the (dominantly valence-quark) process $qq \to qqV_1^0V_2^0$.  When each of the two fermion lines emits one vector boson, the resulting contribution is generally no longer proportional to the LO $q\bar{q} \to V_1^0V_2^0$ process. Still, we estimate that the $qq$-initiated processes at NNLO correct the ratios by just a few percent.  Our argument proceeds as follows.  The process $qq \to qqV_1^0V_2^0$ has a collinear divergence near the beampipe and can only be defined by requiring both jets to have \pT greater than some minimum $\pT^{j,\text{min}}$.  However, the divergence is proportional to the LO $q\bar q\to V_1^0V_2^0$ process, and largely cancels in the ratios.  Calculating the effect on the ratios for different values of $p^j_{T,\text{min}}$ between 5 and 30 GeV, and extrapolating  $p^j_{T,\text{min}}\to 0$ by fitting to a falling exponential, we find shifts for $R_{1a}$ ($R_{1b}$) [$R_{1c}$] of 3\% (3.5\%) [2.5\%] or less. Consequently, although our estimates are crude and this source of NNLO corrections may well be one of the largest on the $R_1$ ratios, it does not seem to present issues that exceed our fiducial benchmark of 5--6\% theoretical uncertainties.

Finally, the largest known NNLO correction to the $R_1$ ratios is from $g g\to V_1^0V_2^0$.  Fortunately, much is already known about this correction, which is separately gauge-invariant and finite.  It  has been known for some time \cite{Combridge:1980sx,Glover:1988rg} and can consistently be combined with the NLO calculation on its own. As it gives the largest source of NNLO corrections in most regions of phase space and has a different dependence on EW quantum numbers than does the tree-level process, it has an important effect on our ratio observables.

\begin{figure}
    \begin{center}
        \includegraphics[width=0.48\linewidth]{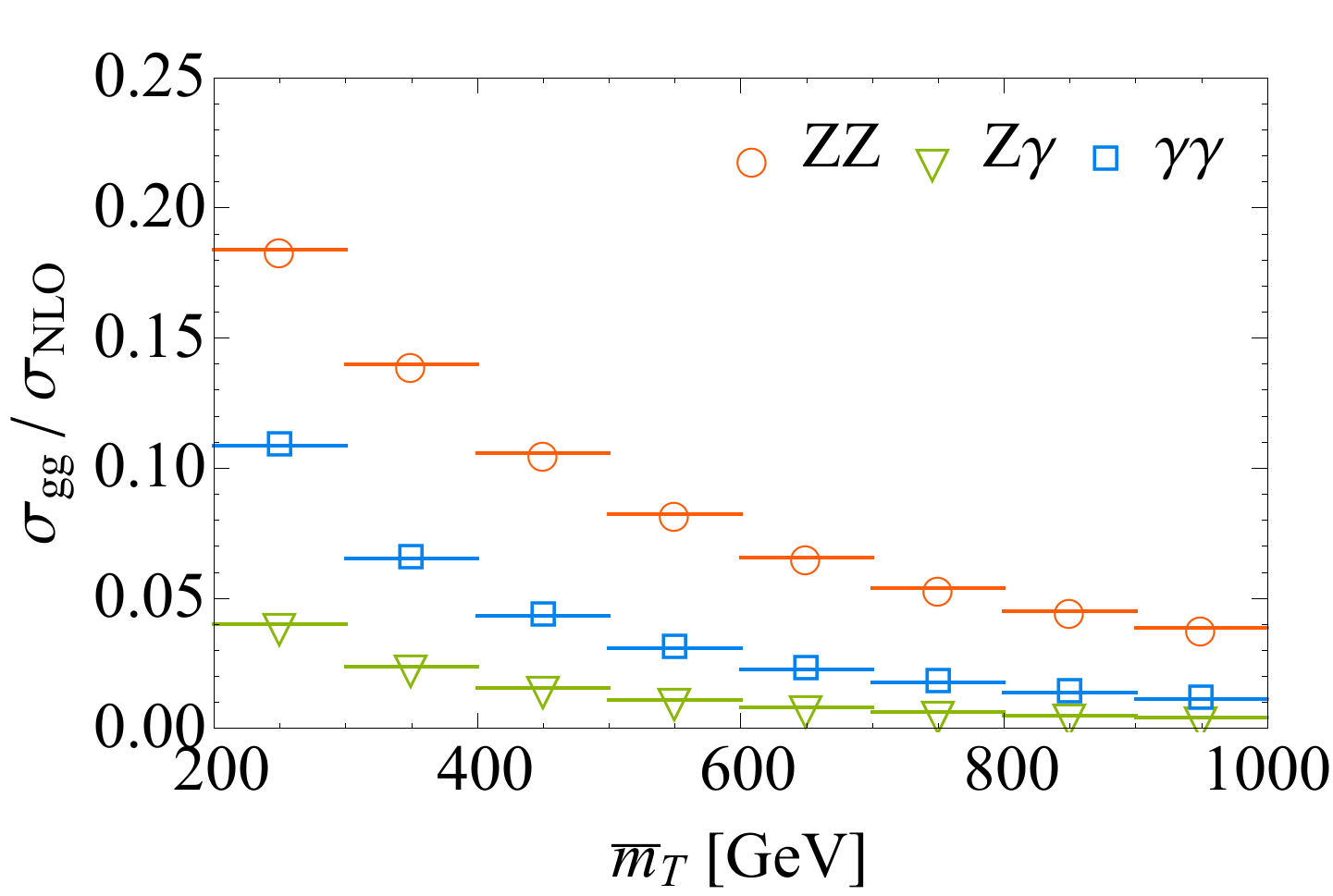}~~
        \includegraphics[width=0.48\linewidth]{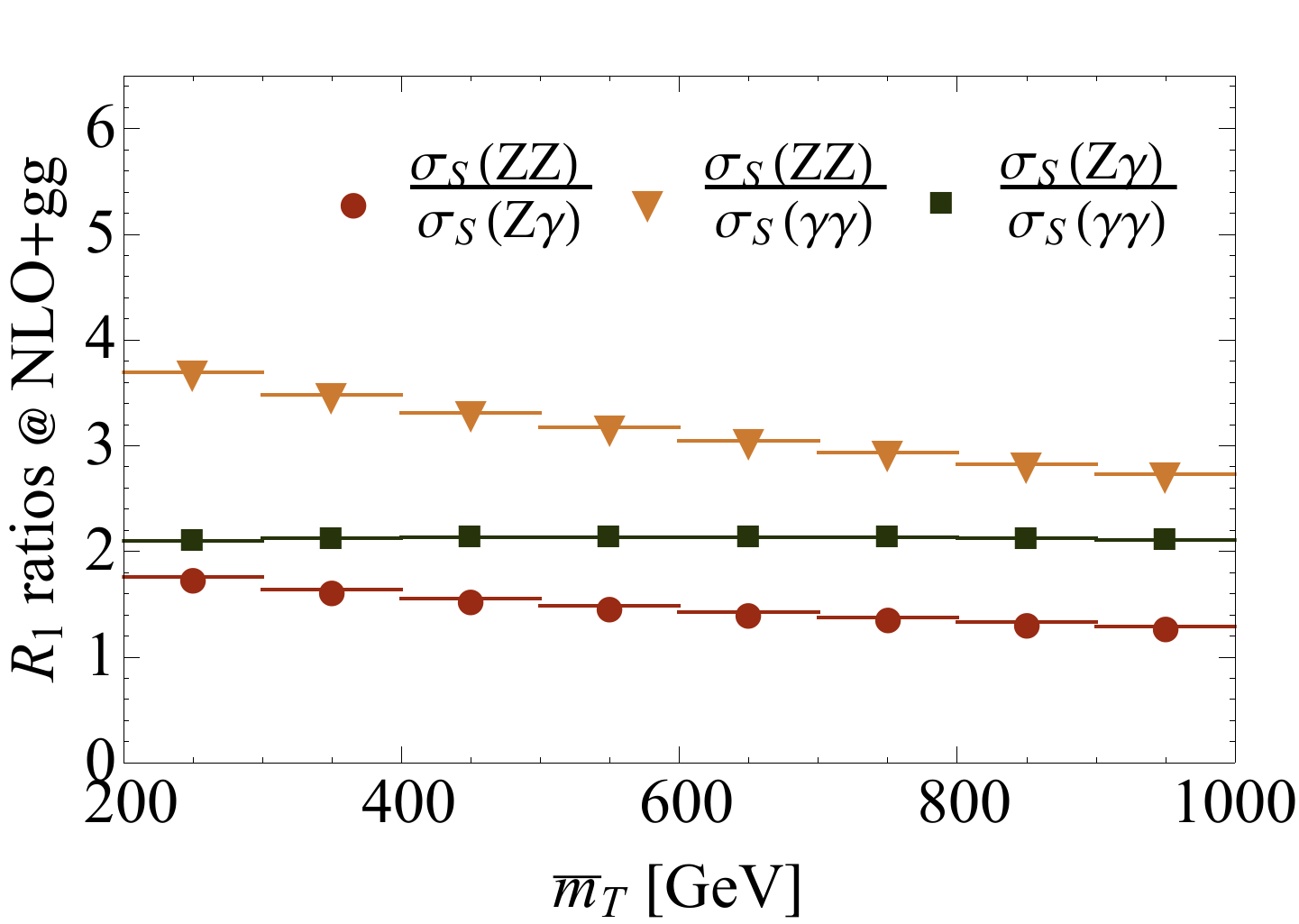}
    \end{center}
    \caption{(Left) Contribution from $gg\to V_1^0V_2^0$ to the $V_1^0V_2^0$ cross sections, expressed relative to the corresponding NLO
      cross section. We used $gg \to \gamma\gamma$ at $\ord(\alpha_S^3)$ to estimate $gg \to Z\gamma,ZZ$ at this order. (Right) The $R_1$ ratios, including the NLO and $gg\to V_1^0V_2^0$ contributions.}
\label{fig:kgg}
\end{figure}

Because $u$- and $d$-type quarks contribute coherently in the loop, the formulas for $gg\to V_1^0V_2^0$ are not proportional to the tree-level $q\bar q\to V_1^0V_2^0$ formulas. In fact $gg\to w^3 x$ is zero by $SU(2)$ conservation, and so $gg\to Z\gamma$ is relatively small compared to $gg\to ZZ, \gamma\gamma$. In \fig{kgg} the $gg$ contributions to the cross sections are shown relative to the corresponding NLO
differential cross sections; they represent  a  13\% (5\%) [20\%] correction for $\gamma\gamma$ $(Z\gamma)$ $[ZZ]$ at low $\bar m_T$,
though less at higher energies where the gluon PDFs are smaller.

Partial cancellations still take place in our ratios.
The observable $R_{1a}$ is shifted downward by as much as 7\% from its NLO value at the lowest values of $\bar m_T$ we consider; however, this $gg$-shift is reduced at higher $\bar m_T$, quickly becoming of order $3\%$. Meanwhile $R_{1b}$ $(R_{1c})$ shifts up 7\% (14\%) at low $\bar m_T$; this $gg$-shift remains at the 6\% (9\%) level for moderate $\bar m_T$ before shrinking more rapidly to 3\% (3\%) at high $\bar m_T$.

\Fig{kgg} displays the $R_1$ ratios including the $gg\to V_1^0V_2^0$ channel along with the NLO
contributions. This plot should be compared with \fig{mtRatio}, which shows the LO ratios.  Notice that $R_{1a}$ is  accidentally flatter than at LO, as a result of the above-mentioned corrections.

This plot of course depends on a choice of renormalization and factorization scales $\mu_R$ and $\mu_F$ used for the $gg\to V_1^0V_2^0$ computation.  For $gg\to\gamma\gamma$ the scale dependence can be reduced because the dominant\footnote{In \app{gg} we argue that the terms neglected in ref.~\cite{Bern:2002jx} are indeed subleading. For $gg \to ZZ$ a similar calculation appeared very recently \cite{Caola:2015psa}, as this paper was nearing completion.} part of the $\ord(\aS^3)$ correction is known \cite{Bern:2002jx}.  For $gg\to Z\gamma, ZZ$, we can use the fact that at NLO all three processes have a nearly universal $\mu_R,\mu_F$ dependence for $\hat s\gg m_Z^2$.  This is because (i) the three processes have the same $\aS$-dependence and involve the same PDFs, (ii) the SM is anomaly free and so no new non-universal diagrams appear at $\ord(\aS^3)$, and (iii) the contribution of longitudinal $Z$s to $gg \to ZZ$ is rather small \cite{Glover:1988rg}, of order 10--15\%.  Thus for reasonable values of $\mu_R$ and $\mu_F$,
\begin{equation}
\label{eq:Kgg}
    K_{gg} \equiv \frac{d\sigma_{(3)}(gg \to \gamma\gamma)}
    {d\sigma_{(2)}(gg \to \gamma\gamma)}
    \approx
    \frac{d\sigma_{(3)}(gg \to Z\gamma)}
    {d\sigma_{(2)}(gg \to Z\gamma)}
    \approx
    \frac{d\sigma_{(3)}(gg \to ZZ)}
    {d\sigma_{(2)}(gg \to ZZ)},
\end{equation}
where $d\sigma_{(n)}$ marks the cross section calculated at order $\aS^n$. We can then use MCFM to compute the known $\ord(\aS^2)$ and $\ord(\aS^3)$ cross sections for $gg\to \gamma\gamma$, thereby determining the $\ord(\aS^3)$ cross sections for the other processes to a fairly good approximation. For our central values we choose scales $\mu_R = \mu_F = m_{\gamma\gamma}$ everywhere in \eq{Kgg}.\footnote{We have observed, by direct comparison across our \bmT range, that the procedure just outlined is essentially identical to calculating the $\ord(\aS^2)$ cross sections for the three processes with scales $\mu_R \sim 0.34\,m_{VV}$ and $\mu_F\sim 0.20\, m_{VV}$. The fact that these are reasonable scales serves as a sanity check of our method.}

\begin{figure}
    \begin{center}
        \includegraphics[width=0.48\linewidth]{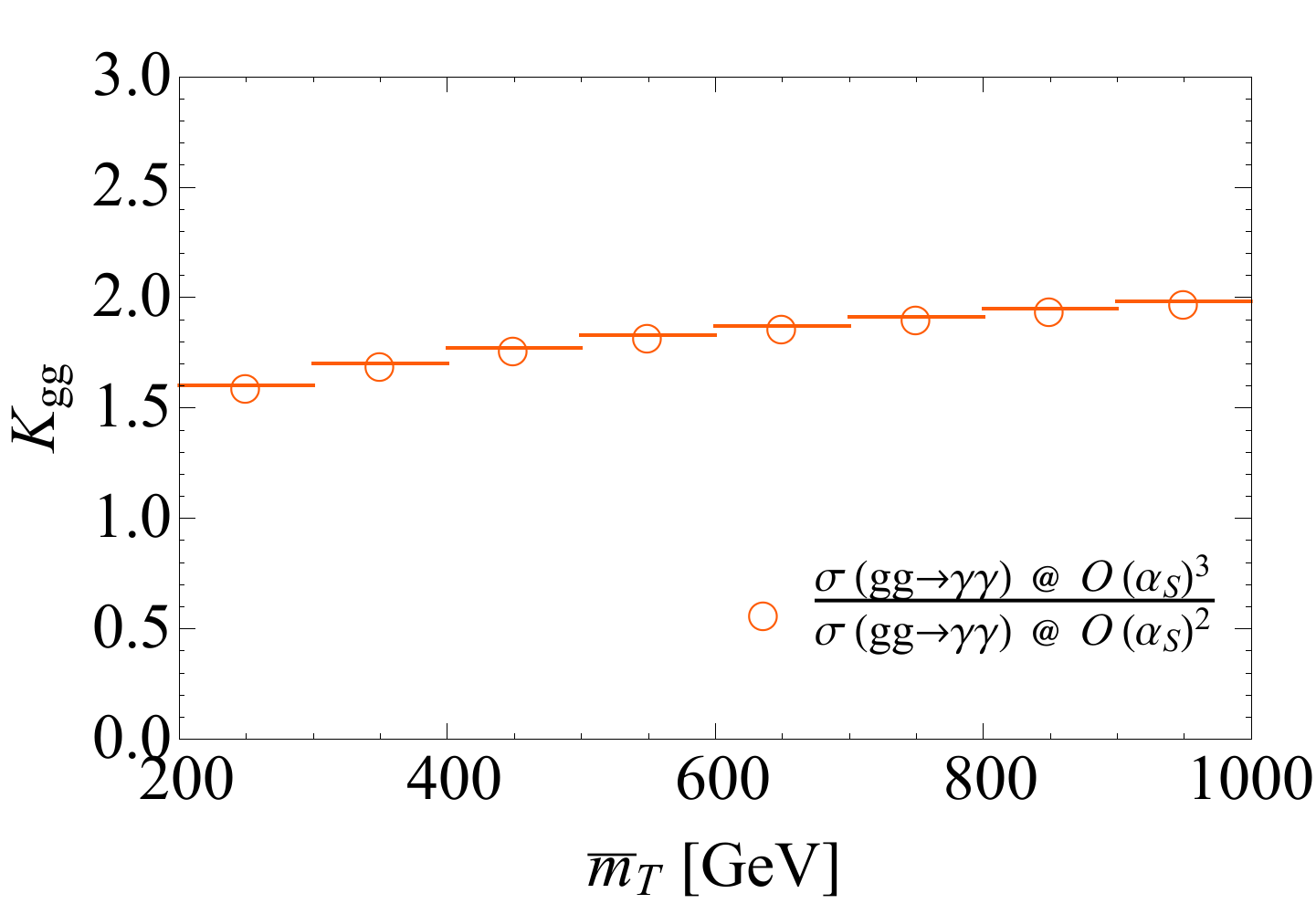}~~
        \includegraphics[width=0.48\linewidth]{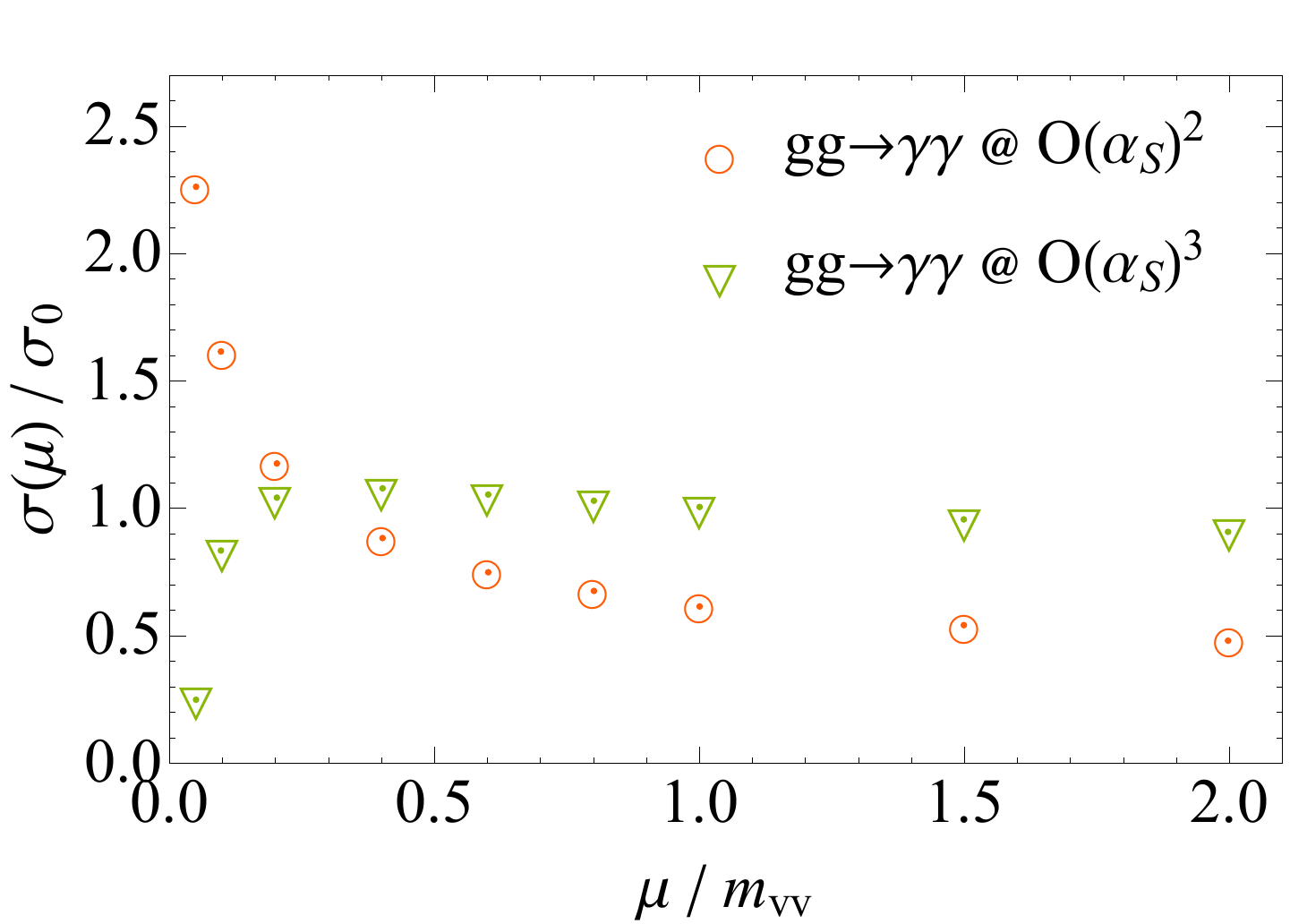}
    \end{center}
    \caption{(Left) The size of $\ord(\aS^3)$ corrections $K_{gg}$ to $gg \to \gamma\gamma$ as a function of $\bar{m}_T$, with scales set to $\mu_R=\mu_F=m_{\gamma\gamma}$ in numerator and denominator. This function allows us to estimate the $\ord(\aS^3)$ cross section for $gg \to Z\gamma$ and $ZZ$. (Right) As a function of scale $\mu_R=\mu_F=\mu$, the $gg \to \gamma\gamma$ rate in the kinematic region $\bar m_T > 200$ GeV, shown at $\ord(\alpha_s^2)$ and (with the partial calculation implemented in MCFM) at $\ord(\alpha_s^3)$. The cross sections are normalized with respect to $\sigma_0\equiv \sigma_{(3)}(gg \to \gamma\gamma, \mu=m_{VV})$.}
\label{fig:ggCorr}
\end{figure}

We show the values of $K_{gg}$ in left panel of \fig{ggCorr}. Since the values of $K_{gg}$ are large, one might wonder whether, as in $gg\to h$, the $\ord(\aS^4)$ correction to $gg \to V_1^0V_2^0$ could itself be quite large.  However, unlike $gg\to h$, where the NLO prediction exceeds the LO substantially at all $\mu$, the situation is milder here. As can be seen in the right panel of \fig{ggCorr}, which shows $gg \to \gamma\gamma$ at $\ord(\aS^2)$ and $\ord(\aS^3)$ with a variety of scale choices, the higher-order prediction turns over at small $\mu$, and above the turnover varies only slowly.  We therefore expect $\ord(\alpha_s)\sim 10-20\%$ uncertainties on $gg\to\gamma\gamma$, and $\sim 1-2\%$ uncertainties on the $R_{1}$ variables, from the unknown $\ord(\aS^4)$ terms. We will estimate uncertainties from this source in \ssec{pdf-scale} and find them consistent with this expectation.

\subsection{Partial cancellation of PDF and scale uncertainties}
\label{subsec:pdf-scale}

Now we turn to standard sources for potential theoretical uncertainties: the PDFs and the choices of renormalization and factorization scales in QCD corrections.  These show significant cancellations and become subleading compared to other uncertainties that we have already discussed.

\begin{figure}[tb]
    \begin{center}
        \includegraphics[width=0.48\linewidth]{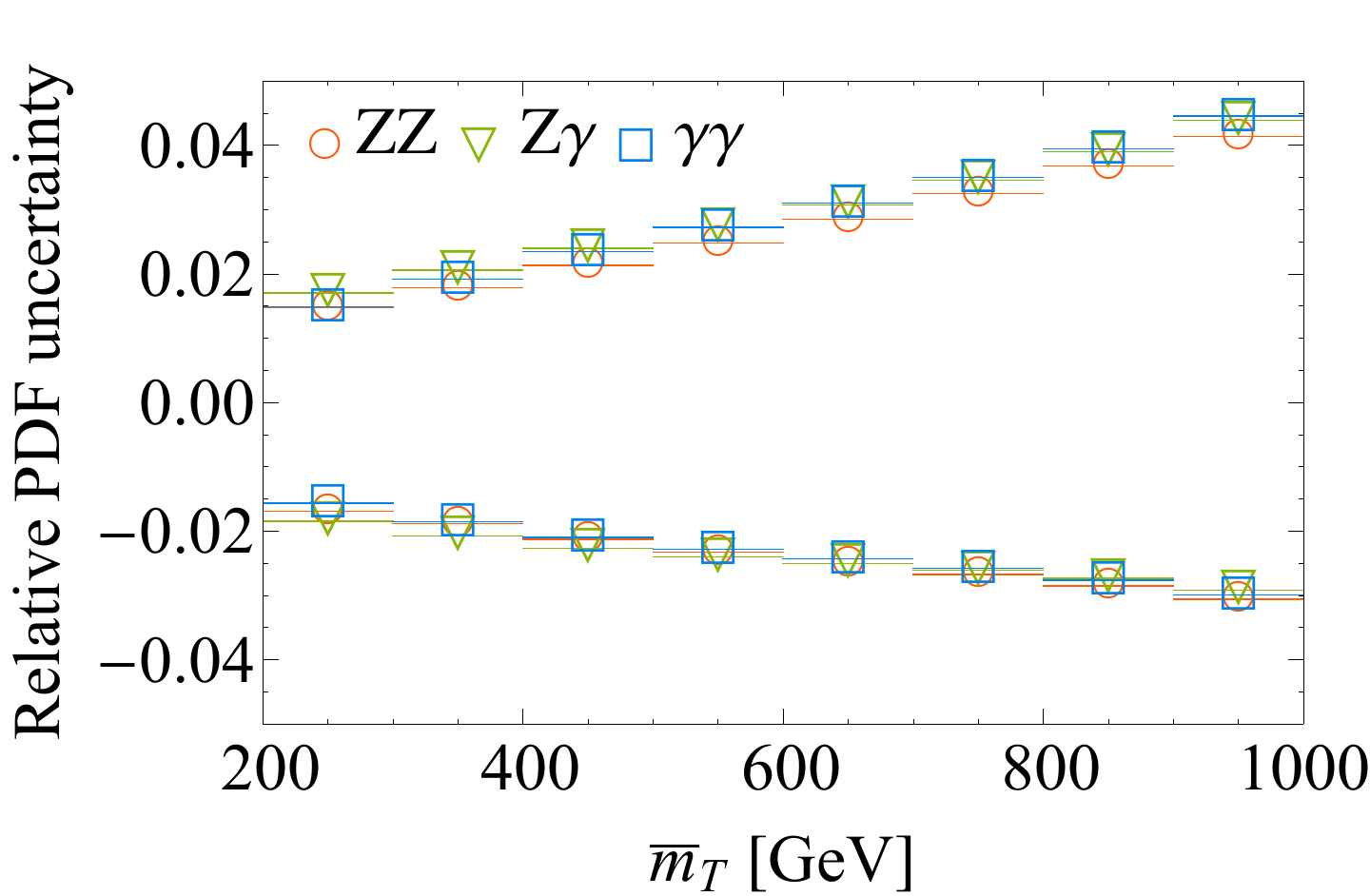}~~
        \includegraphics[width=0.48\linewidth]{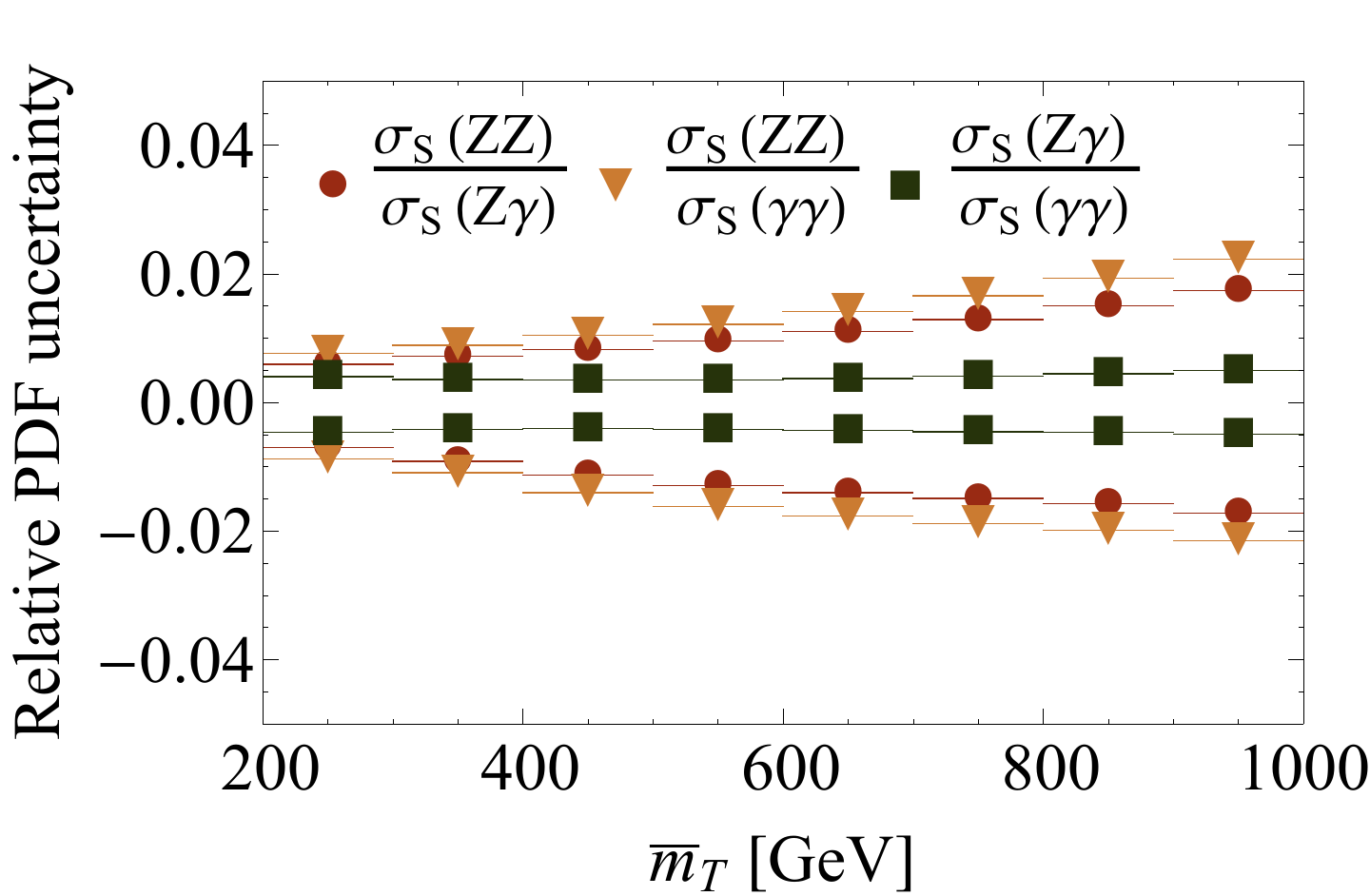}
    \end{center}
\caption{The relative PDF uncertainty bands for the individual $V^0_1V^0_2$ cross sections (left) and the $R_1$ ratios (right).  PDF variations of $gg \to V^0_1V^0_2$ are included.  See text for more details. }
\label{fig:pdfunc}
\end{figure}

The PDF uncertainties for the individual channels, and their reduced values for the ratios, are shown in \fig{pdfunc}. For the $R_{1a}$ ratio the uncertainties are of order $1\%$ and can be essentially ignored; as we saw in \ssec{RatObs}, the parton luminosity ${\mathscr{L}}^S_{u\bar u}$ dominates both numerator and denominator, so that PDF variations nearly cancel.
For the others, the uncertainties are still significantly reduced, rising only to about $2\%$ even up to $\bar m_T\sim 1$ TeV.

These uncertainties were determined using \textsc{MCFM 6.8}.  The $pp\to V_1^0V_2^0$ cross sections are evaluated for the central ($S_0$) and all 20 pairs of error sets ($S_i^\pm$) of the MSTW 2008 PDF set \cite{Martin:2009iq}.
With the cross sections $d\sigma(S_i) $, we use the prescription of ref.~\cite{Martin:2009iq} to determine the PDF uncertainties on individual channels.  The upper edge of the uncertainty band is calculated with
\begin{equation}
  \Delta_+(d\sigma) =
  \sqrt{ \sum_i \left(\max \left[0,d\sigma(S_i^+)-d\sigma(S_0),d\sigma(S_i^-)-d\sigma(S_0)\right]\right)^2},
\end{equation}
while the lower edge is the same with ``max'' replaced with ``min''.\footnote{We actually carry this out with the 90\% confidence-level NLO MSTW 2008 PDF sets, and then rescale the result, formally a $2\sigma$ variation, by 1.645 to obtain a formally $1\sigma$ variation.  This is almost the same as using the 68\%-level confidence sets, but because of non-Gaussian tails gives a slightly more conservative estimate of uncertainties.}
Because the error sets of MSTW 2008 are eigenvectors of the covariance matrix, the PDF uncertainties for the ratios can then be obtained in a similar fashion.\footnote{
For example:
\begin{equation} \nonumber
  \Delta_+\left(R_{1a}\right) =
  \sqrt{ \sum_i \left(\max \left[0,\frac{d\sigma(Z\gamma,S_i^+)}{d\sigma(\gamma\gamma,S_i^+)}-\frac{d\sigma(Z\gamma,S_0)}{d\sigma(\gamma\gamma,S_0)},\frac{d\sigma(Z\gamma,S_i^-)}{d\sigma(\gamma\gamma,S_i^-)}-\frac{d\sigma(Z\gamma,S_0)}{d\sigma(\gamma\gamma,S_0)}\right]\right)^2}.
\end{equation}}

All this is straightforward except for one subtlety.
Since we do not have access to the $\ord(\aS^3)$ calculation for $gg\to Z\gamma$ and $gg\to  ZZ$, we obtain them by rearranging \eq{Kgg} as
\begin{equation}
d\sigma_{(3)} (gg \to Z\gamma,\text{pdf}_1) \approx d\sigma_{(3)}(gg \to\gamma\gamma,\text{pdf}_1)\frac{d\sigma_{(2)} (gg \to Z\gamma,\text{pdf}_2)}{d\sigma_{(2)} (gg \to \gamma\gamma,\text{pdf}_2)} ,
\label{eq:ggpdf}
\end{equation}
where $d\sigma(\ldots;\text{pdf}_i)$ is the cross section evaluated for PDF set $S_i$. A similar expression holds for $gg \to ZZ$.  Inaccuracies in this procedure will be subleading in our uncertainties since $gg\to V_1^0V_2^0$ is itself sufficiently small.

Now we turn to uncertainties in our NLO calculation from renormalization and factorization scales $\mu_R,\mu_F$. Typically the cancellation of correlated scale variations in ratios of various processes should be viewed as accidental, since the actual structure of higher-order corrections in differing processes is uncorrelated. We wish to argue that this is not the case here. The renormalization scale is sensitive to the ultraviolet region of higher-order corrections, where EW symmetry is restored (up to longitudinal polarizations, which first appear at NNLO in $gg \to \phi^3\phi^3)$, and where we expect higher-order corrections in general to take a nearly identical form for all $V^0_1V^0_2$ processes. Meanwhile, factorization scale sensitivity primarily comes from divergences associated with emissions off the initial state. While this is not directly affected by the restoration of EW symmetry, it is sensitive to the color structure of the processes order-by-order in the perturbative expansion of QCD, which is also identical for the three $V^0_1V^0_2$ processes. For these reasons the cancellation of scale dependence we observe in our ratios is physical, since the scale choices really are probing correlated higher-order effects.

\begin{figure}[tb]
    \begin{center}
        \includegraphics[width=0.48\linewidth]{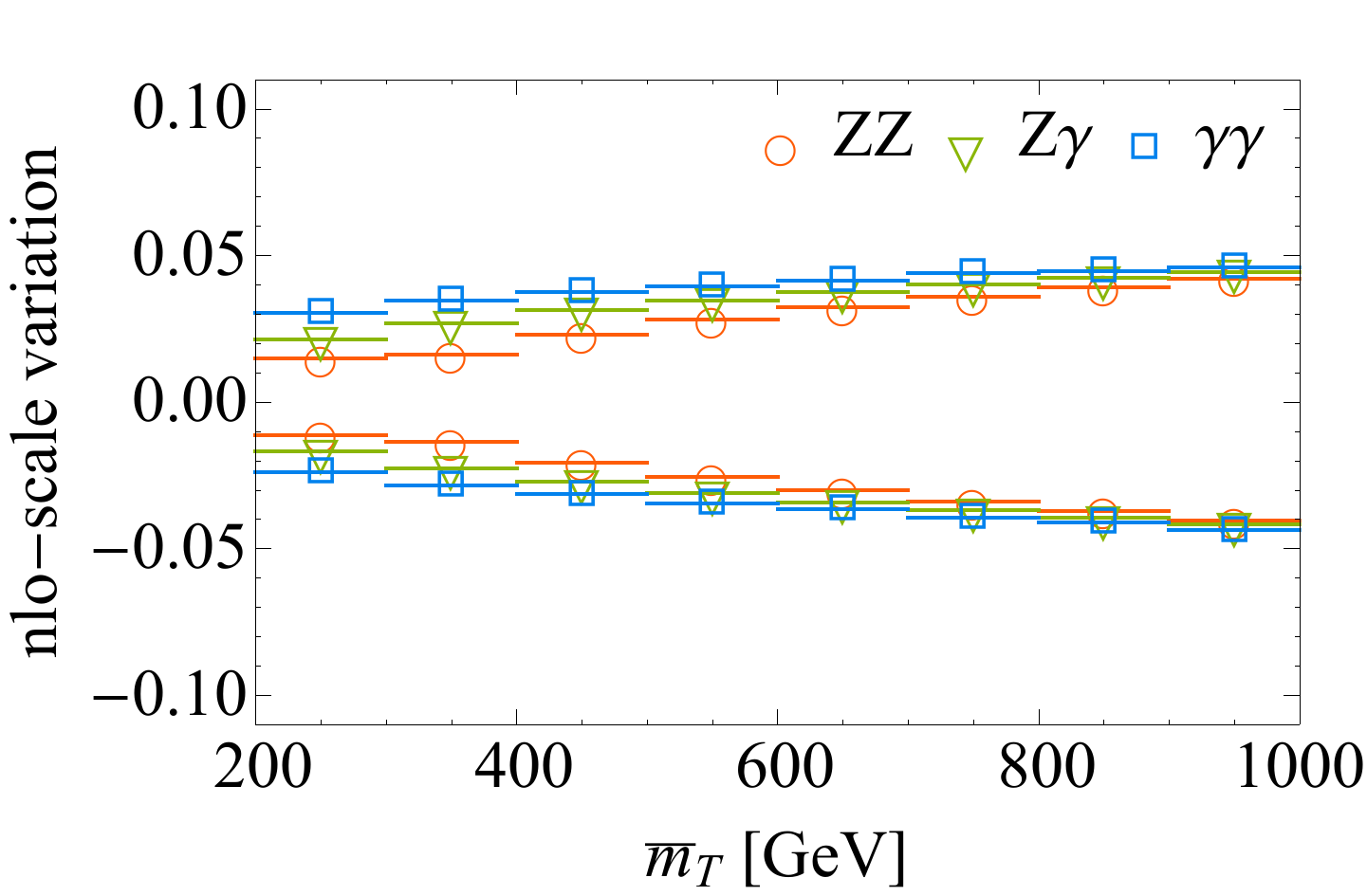}~~
        \includegraphics[width=0.48\linewidth]{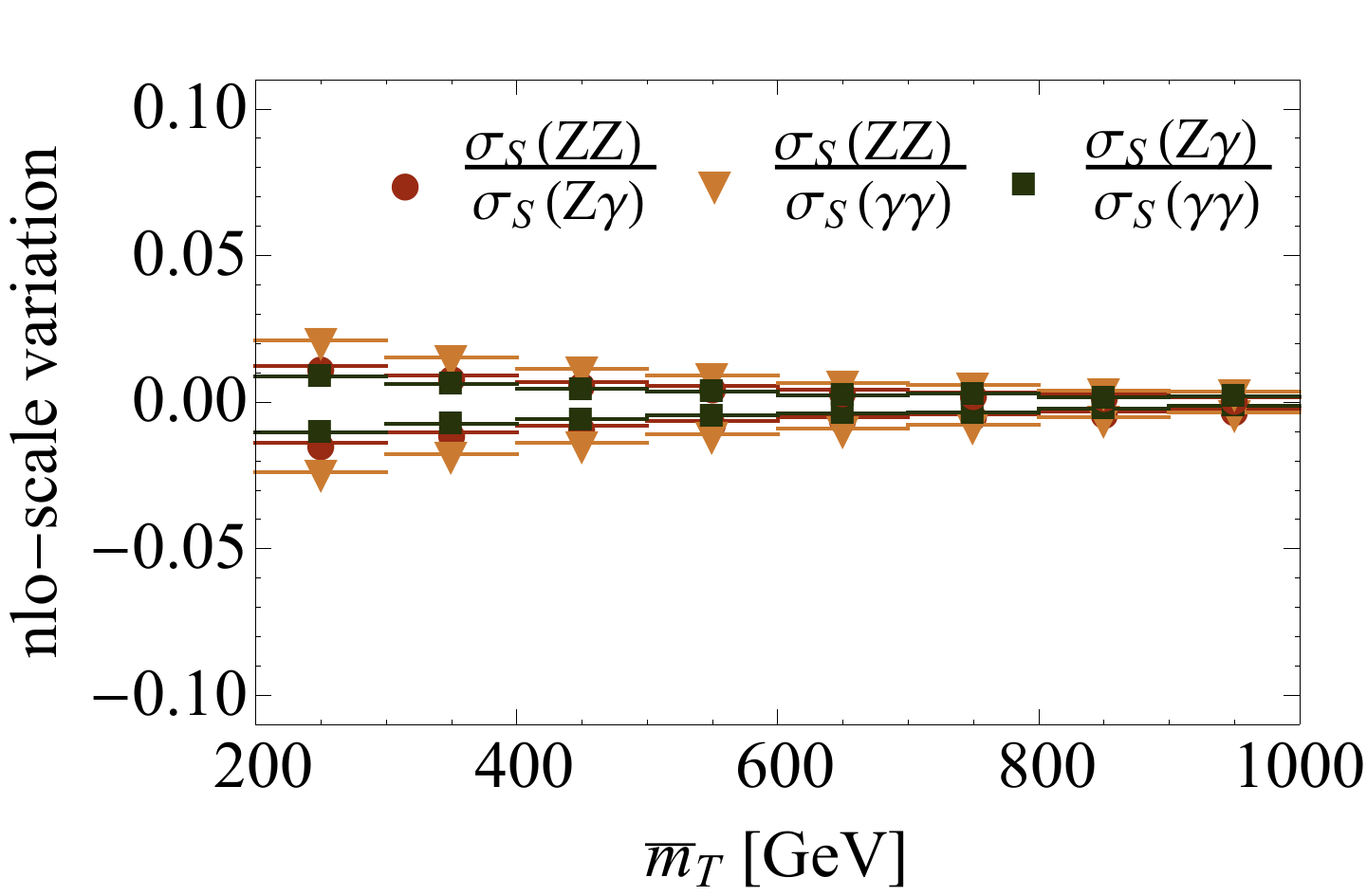}
    \end{center}
    \caption{The relative uncertainty band on the $V^0_1V^0_2$ cross sections (left) and $R_1$ ratios (right) found by varying the renormalization and factorization scales $\mu_R,\mu_F$ up and down by a factor of 2. Here the scales appearing in the $gg\to V_1^0V_2^0$ process are {\it not} varied; see \fig{scalevarGG} below.}
\label{fig:scalevar}
\end{figure}

As shown in \fig{scalevar}, scale-dependence is reduced from several percent in the cross sections to 1--2\% in the ratios, where the cancellation is significant for all three ratios and works best at high energy.  Here we have varied the scales $(\mu_R,\mu_F)$ independently from $\frac12 \,m_{VV}$ to $2\,m_{VV}$ and plotted the envelope of the relative variation in each quantity.
However, in \fig{scalevar} we have held the scales in the $gg\to V_1^0V_2^0$ processes {\it fixed}.  The calculation to NLO of $q\bar{q} \to V_1^0V_2^0$ begins at $\ord(\alpha_s^0)$, while the calculation of $gg\to V_1^0V_2^0$ begins at $\ord(\alpha_s^2$).  To the order we are working there are no terms in the former calculation which are at the same order as terms in the latter, and thus there is no sense in which the perturbative expansion of the one can affect that of the other.  Correspondingly there is no sense in which these two calculations must or should be evaluated with the same value of $\mu_R$, and so their $\mu_R$ dependence must be computed separately.  While in principle there could be correlation in the $\mu_F$-dependence through the pdfs, it turns out that $gg\to V_1^0V_2^0$ depends much more strongly on $\mu_R$, and so any such correlation is unimportant.

\begin{figure}[tb]
    \begin{center}
     \includegraphics[width=0.48\linewidth]{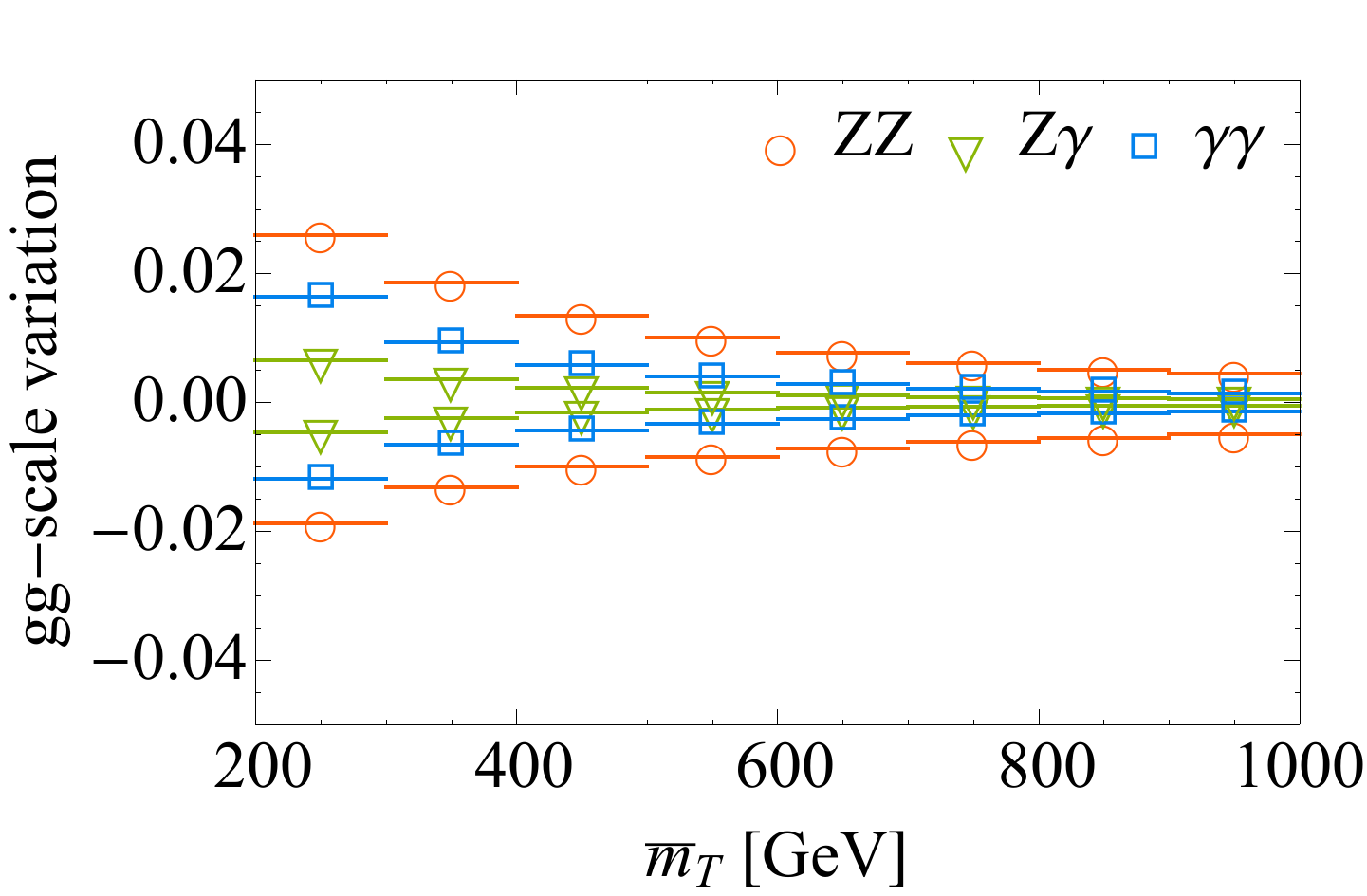}~~
     \includegraphics[width=0.48\linewidth]{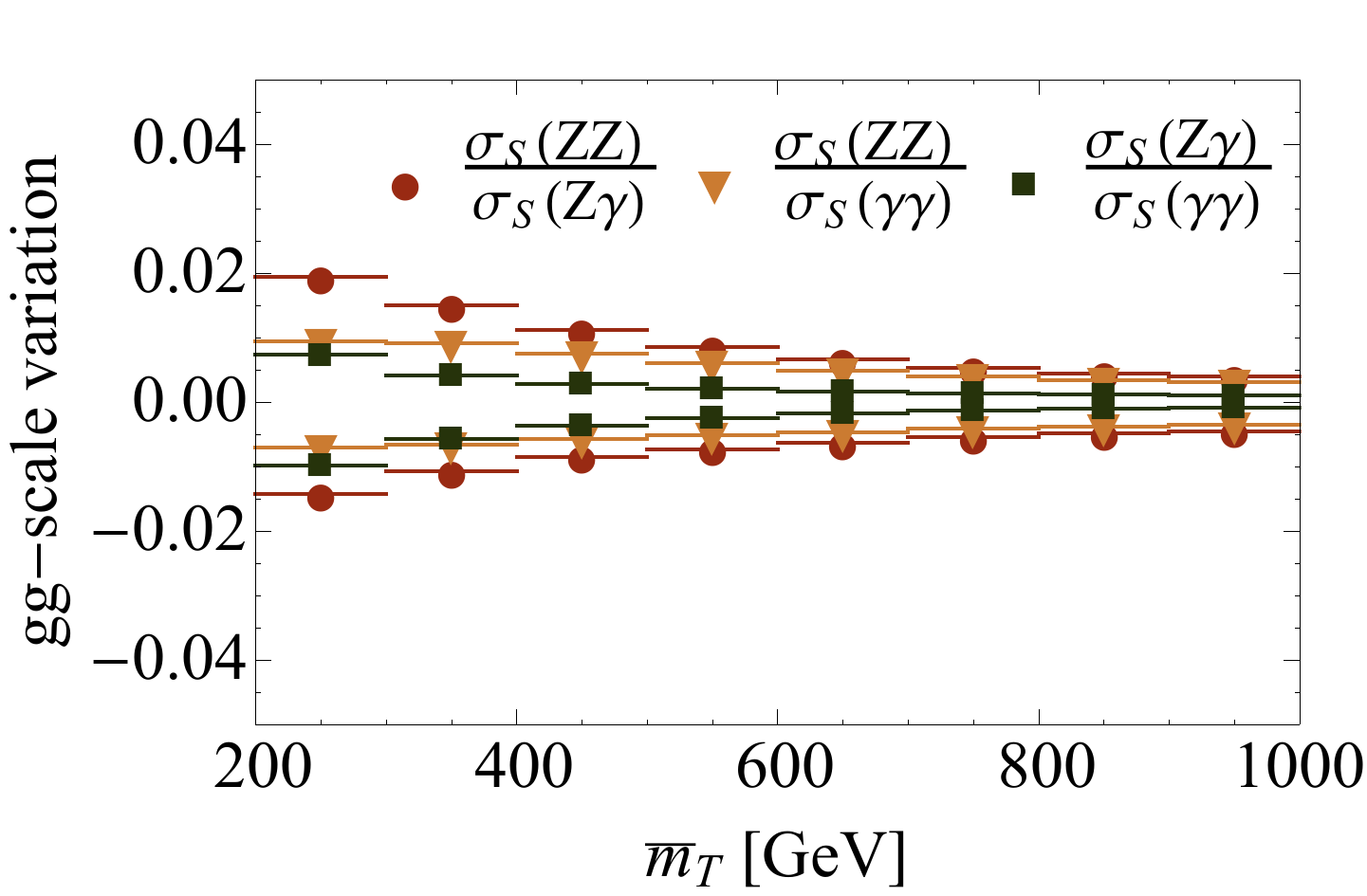}
    \end{center}
\caption{The relative error band on the $V^0_1V^0_2$ cross sections (left) and $R_1$ ratios (right) found by varying $\mu_R,\mu_F$ up and down by a factor of 2. Here {\it only} the scales appearing in $gg\to V_1^0V_2^0$ are varied.}
\label{fig:scalevarGG}
\end{figure}

Based on this reasoning, we have also computed the effects of scale variations on the $gg \to V_1^0V_2^0$ component of the cross sections, holding all other components fixed. Lacking the $\ord(\alpha_s^3)$  differential cross sections for $gg\to Z\gamma$ and $gg \to ZZ$, we again rely on another incarnation of \eq{Kgg}:
\begin{equation}
d\sigma_{(3)} (gg \to Z\gamma,\{\mu_1\}) \approx d\sigma_{(3)}(gg \to\gamma\gamma,\{\mu_1\}) \frac{d\sigma_{(2)} (gg \to Z\gamma,\{\mu_2\})}{d\sigma_{(2)} (gg \to \gamma\gamma,\{\mu_2\})},
\label{eq:ggscal}
\end{equation}
where $\{\mu_i\}$ stands for a choice of $\mu_R$ and $\mu_F$.
The resulting uncertainties due to scale variation of the $gg\to V_1^0V_2^0$ processes are shown in \fig{scalevarGG}; these are consistent with our estimate from \ssec{gg}.   Although small for each individual channel compared to the scale variation in the left-hand plot of \fig{scalevar}, cancellations are not as significant as for the NLO scale variations.  Consequently the two classes of scale variation turn out to be quite similar in size and shape for the $R_1$ observables, as can be seen in the right-hand plots of \fig{scalevar} and \fig{scalevarGG}.

Overall, we can see that while the PDF and scale uncertainties form a significant portion of the theoretical error budget for individual cross sections, these uncertainties are substantially reduced in ratios (in particular in $R_{1a}$) and become subleading.  This presumably reflects true symmetry-related cancellations in the many NNLO corrections that are common to the three  neutral diboson processes.

\subsection{EW corrections}
\label{subsec:EW-corr}

\subsubsection{Sudakov enhancements}
\label{subsec:doubleLog}

For the level of precision we pursue, higher-order EW corrections to our 
ratio observables are important. Complete calculations of NLO EW effects 
for $\gamma\gamma$, $Z\gamma$, and $ZZ$ exist, 
though public code is not yet at our disposal and the results have been 
presented with different cuts from our own. As an approximation of the EW corrections,
and to estimate the magnitude of their uncertainties, we employ a leading-log calculation 
in the threshold limit. Comparison of our results below with the full NLO calculations of 
refs.~\cite{Bierweiler:2013dja,Denner:2015fca} reassures us that our estimates are reasonable.

Because of various sources of $SU(2) \times U(1)$ breaking, large EW logarithms do not 
entirely cancel even in fairly inclusive observables such as $d\sigma/d\bmT$. At very large \bmT, 
ignoring finite NLO EW corrections and resumming the leading Sudakov logarithms, 
of the form $\alpha^n \log^{2n}([\bmT/m_{W,Z}]^2)$, is justified and should give a good approximation of the dominant effects.

An estimate of the Sudakov logarithm-enhanced corrections can be obtained from a 
calculation at threshold, where all the energy of the initial state goes into 
production of the electroweak states. The threshold limit corresponds to a strict veto 
on the real emission of EW bosons, so at high \bmT it overestimates the true EW correction.
Since we do not have such a strict veto in our observables, the large virtual corrections 
above are reduced by our partial inclusion of the real radiation of gauge bosons. 
For instance, soft $W$ and $Z$ bosons are partially included: a soft $Z$ or $W$ that decays 
hadronically typically produces soft daughters at wide angles to the hard boson, 
and thus its daughter jets will neither fail our jet cuts nor ruin isolation of the 
boson or its daughter leptons. Leptonic decays of the soft bosons are potentially more subtle, 
depending on how the extra leptons are treated experimentally. Our less extreme veto of 
soft-collinear bosons should lead to some reduction of the soft-collinear corrections.

Conversely, finite NLO corrections that we ignore in our estimates should increase the size of the
EW correction. For moderate values of \bmT, this effect may partially compensate
the above-mentioned reduction.
  Our estimates below are therefore rough guides, and the issue deserves further study.

This threshold regime was studied in the context of boson + jet 
production~\cite{Becher:2013zua}.\footnote{We thank T.~Becher for extensive discussions 
and Xavier Garcia i Tormo for providing detailed results of their calculation.} 
It was found that the EW corrections reduce the photon + jet cross section by 
$\Delta\sigma_\text{EW} = -6\%^{+3\%}_{-2\%}$ ($-11\%^{+3\%}_{-2\%}$) at 
$\pT^\gamma = 500$ (1000) GeV, while reduction of the $Z$ + jet cross section is 
roughly double this, $\Delta\sigma_\text{EW} = -13^{+4\%}_{-1\%}$ ($-22^{+4\%}_{-1\%}$). 
The difference between $Z$ and $\gamma$ arises mainly from loops involving $W$ bosons.

As these effects are primarily associated with the phase space collinear to the hard boson, 
we anticipate the effect on $\gamma\gamma$ to be roughly the square of the effect on 
$\gamma$ + jet, leading to a 12--21\% reduction in $\sigma(\gamma\gamma)$ for 
$500\GeV < \pT^\gamma < 1000\GeV$. Similarly, we expect reductions in 
$\sigma(Z\gamma)$ $[\sigma(ZZ)]$ by 18--31\% [24--39\%]. But these effects partly 
cancel in the $R_1$ ratios, reducing $R_{1a}$ $(R_{1b})$ $[R_{1c}]$ by just 
7--12\% (14--23\%) [7--12\%] in this \pT range. At high enough \pT, EW effects
become the leading correction to our ratios, dominating over QCD effects.

Importantly, the uncertainties on these EW corrections are not large and are further 
reduced in our ratios. There are several scale choices which appear in the calculation 
of ref.~\cite{Becher:2013zua}, but the scale dependence of photons and $Z$s is correlated, 
as can be seen in figure 3 of that paper. This correlation reduces the uncertainty in the 
EW corrections to our ratios. We estimate that the NLO EW uncertainty from scale choices 
that propagates into our ratios $R_{1a}$ $(R_{1b})$ $[R_{1c}]$ is no more than 
${}^{+2\%}_{-1\%}$ $\left({}^{+3\%}_{-1\%}\right)$ $\left[{}^{+2\%}_{-1\%}\right]$ 
for $\pT \sim 500$--$1000 \GeV$.    These uncertainties are comparable in size to the 
uncertainties from PDFs and unknown QCD corrections.


At lower values of \bmT, 
the finite NLO EW corrections become important, 
but our resummation approximation still serves as a rough guide to their magnitudes. 
For $\sigma(\gamma\gamma)$ and $\sigma(ZZ)$, ref.~\cite{Bierweiler:2013dja} has calculated these 
corrections as functions of $p_T$. 
The EW correction is dominated by a logarithmically growing component over
much of the \pT range relevant for our ratios,
suggesting that our approximation remains applicable in this region.
Moreover, comparison of ref.~\cite{Bierweiler:2013dja} to an 
earlier calculation of the $\alpha \log^2([p_T/m_{W,Z}]^2)$ term alone~\cite{Accomando:2004de},  
corresponding to truncation of the resummed calculation to first nontrivial order, 
found agreement at the several percent level. For similar cuts to ours, 
ref.~\cite{Bierweiler:2013dja} claims reductions in $\sigma(\gamma\gamma)$ $[\sigma(ZZ)]$ 
by 13--21\% [39--60\%] over the range $p_T \sim 500$--1000 GeV. These reductions are somewhat larger than the ones we obtained, and resummation is undoubtedly an important part of the discrepancy.
At somewhat lower $p_T$, only $\sigma(ZZ)$ shows a clear subleading $p_T$-independent correction,
which will certainly shift the EW corrections to $R_{1b},R_{1c}$ away from our leading-log predictions.
 
NLO EW results for $\sigma(Z\gamma)$ are given in ref.~\cite{Denner:2015fca}, 
  but only with a fixed and low cut
  on $p_{T,Z}$.
This makes comparison with our estimates impossible, 
because large logarithms of $p_{T,\gamma}/p_{T,Z}^\text{cut}$ arise and are 
indistinguishable from inclusive EW Sudakov 
logarithms.
Still, we have no reason to suspect 
that the behavior of the finite EW corrections should be qualitatively different 
from those of $\gamma\gamma$ and $ZZ$.
 
Most importantly for our purposes, when finite pieces numerically
dominate the NLO EW correction, its uncertainty arises mainly from
scale variation in the EW couplings.  Our earlier estimate of
the uncertainty using ref.~\cite{Becher:2013zua} is therefore an
overestimate at small \bmT.

We have summarized these statements in \fig{mainResult} of \sec{ExecSumm} by 
indicating the expected fractional shifts in the ratios due to the source of 
EW corrections derived in ref.~\cite{Becher:2013zua}, along with an estimate of 
their uncertainties. This shows that these EW effects might be observable in our 
ratios in the highest bins, where they dominate QCD effects. 
Furthermore, EW effects are under sufficient control that there will still 
be substantial sensitivity to other, non-SM contributions at high \bmT.

\subsubsection{Proper choice of EW scales for on-shell external photons}
\label{subsec:alphaQED}

Another EW issue concerns the correct choice of electromagnetic coupling
corresponding to emission of a photon.\footnote{We thank Z.~Bern for
pointing out the issue, and for conversations.}   In the literature one
finds preference for evaluating $\alpha(\mu_\QED)$ both at $\mu_\QED = 0$
and at $\mu_\QED= {\text {min}}(m_Z,\sqrt{\sh})$ (or some fraction
thereof).  Since the QED coupling runs by 7\% from 0 to $m_Z$, this
difference  affects $R_{1a}$ and $R_{1c}$ by 7\% and $R_{1b}$ by 14\%.

Typical QCD calculations may seem to suggest using $\mu_\QED \sim
\sqrt{\sh}$.
But in contrast to a quark or gluon, we can experimentally require that a
photon  is on-shell and does not shower, \ie, does not form an
electromagnetic jet of leptons and hadrons with a finite mass.  For abelian
gauge bosons, the leading effect of requiring an \emph{on-shell} photon,
rather than a photon that could be off-shell by as much as $q^2\sim \hat
s$, is given by running the coupling down from $\mu_\QED = \sqrt{\sh}$ to
$\mu_\QED = 0$. (Importantly this is not true for nonabelian gauge bosons.)
 This choice removes photons that, for instance, split to a $\mu^+\mu^-$
pair or mix with the $\rho$.  We find this argument reliable in a pure
color-singlet situation, such as Higgs decay to two photons.

Subtleties could arise, however, in a colored environment: soft ISR gluons
are present in $pp$ collisions and can be radiated into the photon
isolation cone.  On the one hand, we still want to forbid $\gamma^*\to
\mu^+\mu^-$ since this would be experimentally rejected; this tends to
suggest $\mu_\QED<2 m_\mu$. On the other hand, we should include photons
with nearby soft gluons that lie below the isolation cut
$p_{T,\text{min}}^{\text{had}}$, which could suggest\footnote{Suggested to
us by T.~Becher following ref.~\cite{Becher:2013zua}. A related suggestion was
made by M.~Schwartz.} $\mu_\QED \sim p_{T,\text{min}}^{\text{had}}$.

Faced with a lack of consensus, we have chosen not to directly address this
issue in this paper.  Instead we use MCFM 6.8
``out of the box'', for which $\mu_\QED=m_Z$ throughout.   In
\fig{mainResult} of \sec{ExecSumm}, we have indicated the potential shift
from switching to $\mu_\QED=0$  as an overall 7\% or 14\% error band that
is essentially flat and fully correlated across all bins.  (Even if this
dispute were not resolved theoretically, the measurement of the average
ratio of the lowest bins would largely fix the value of $\mu_\QED$.)  In no
sense should this be thought of as a Gaussian error band, since no
probability extends beyond the band. For now readers may adjust our results
according to their individual opinions, but clearly it is important that
consensus on the matter be reached in the near future.

\section{Additional practical considerations}
\label{sec:Practical}

\subsection{Photon isolation}
\label{subsec:PhotonIso}

In \sec{beyondLO} we used the smooth-cone photon-isolation method of Frixione,  \eq{frix}, but this is experimentally impractical.
More traditional is hard-cone isolation, simply requiring that the energy in a cone of size $R_h$ around the photon be less than $\epsilon_h \, \pT^\gamma$.
But if $\epsilon_h$ is small, a hard cone produces large logarithms due to the incomplete cancellation of virtual and soft gluon effects.  Meanwhile if $\epsilon_h$ is not small, the hard cone introduces large sensitivity to the fragmentation function $D_{q \to \gamma}(z)$ at $z\to 1$, which is dangerous to a precision calculation since
$D_{q\to\gamma}(z\to 1)$ has substantial associated uncertainties. The Frixione algorithm  avoids these issues by removing the divergent regions of phase space that require the introduction of a fragmentation function in the first place. The isolation parameters can then be set so that no large perturbatively calculable logarithms appear.
However, the smooth cone cannot be implemented experimentally since it requires the energy in a small cone around the photon to go literally to zero as that cone decreases in size.  This difficulty may be evaded by using  a discretized or ``staircase'' version of the smooth cone~\cite{Binoth:2010nha,Hance:2011ysa}.  Although sensitivity to the photon fragmentation function is thereby reintroduced, this sensitivity can be maintained small while keeping the associated logarithms of manageable size, so as to not call the accuracy of the fixed-order calculation into question.

\begin{figure}
    \begin{center}
        \includegraphics[width=0.475\linewidth]{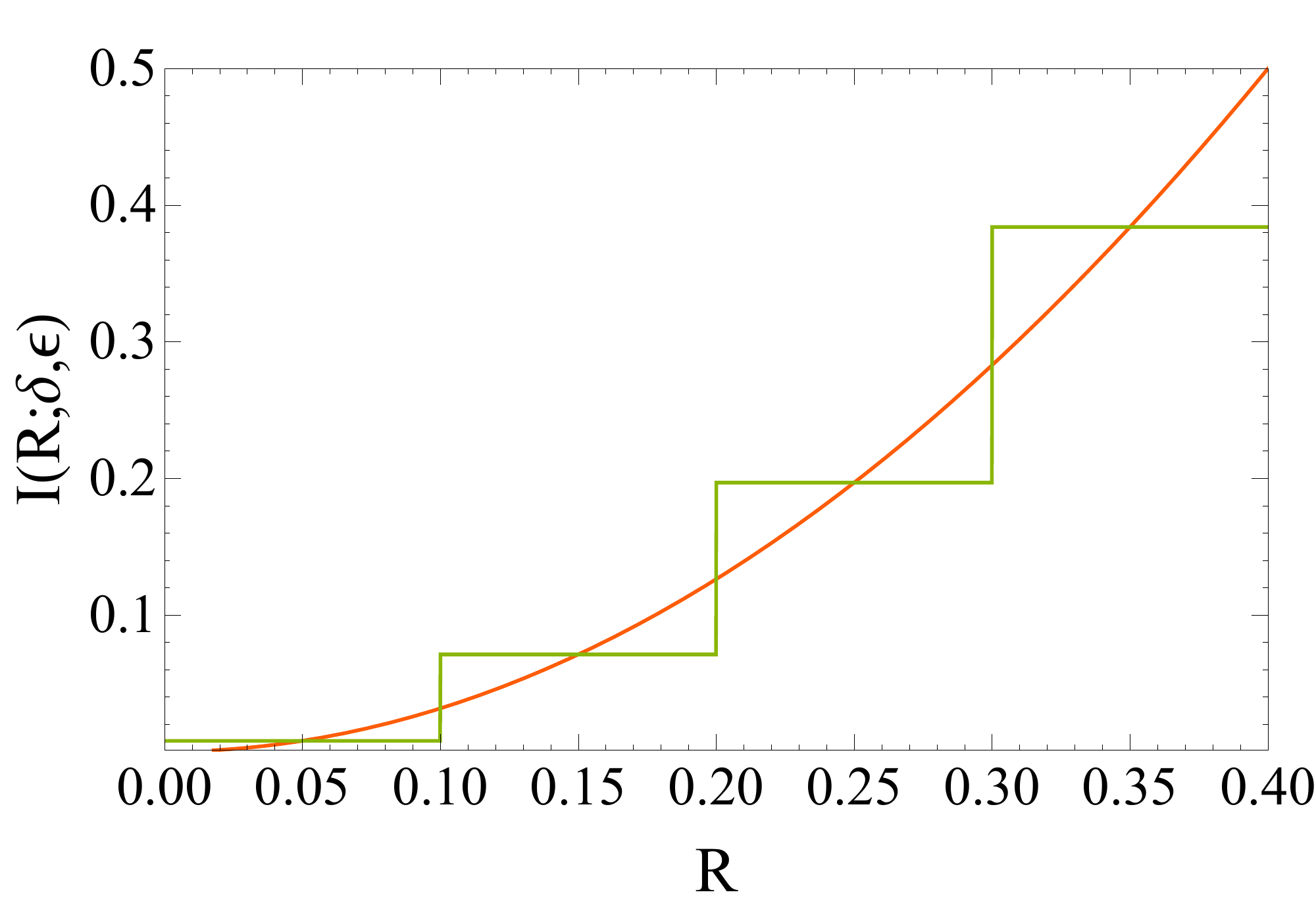}~~
        \includegraphics[width=0.5\linewidth]{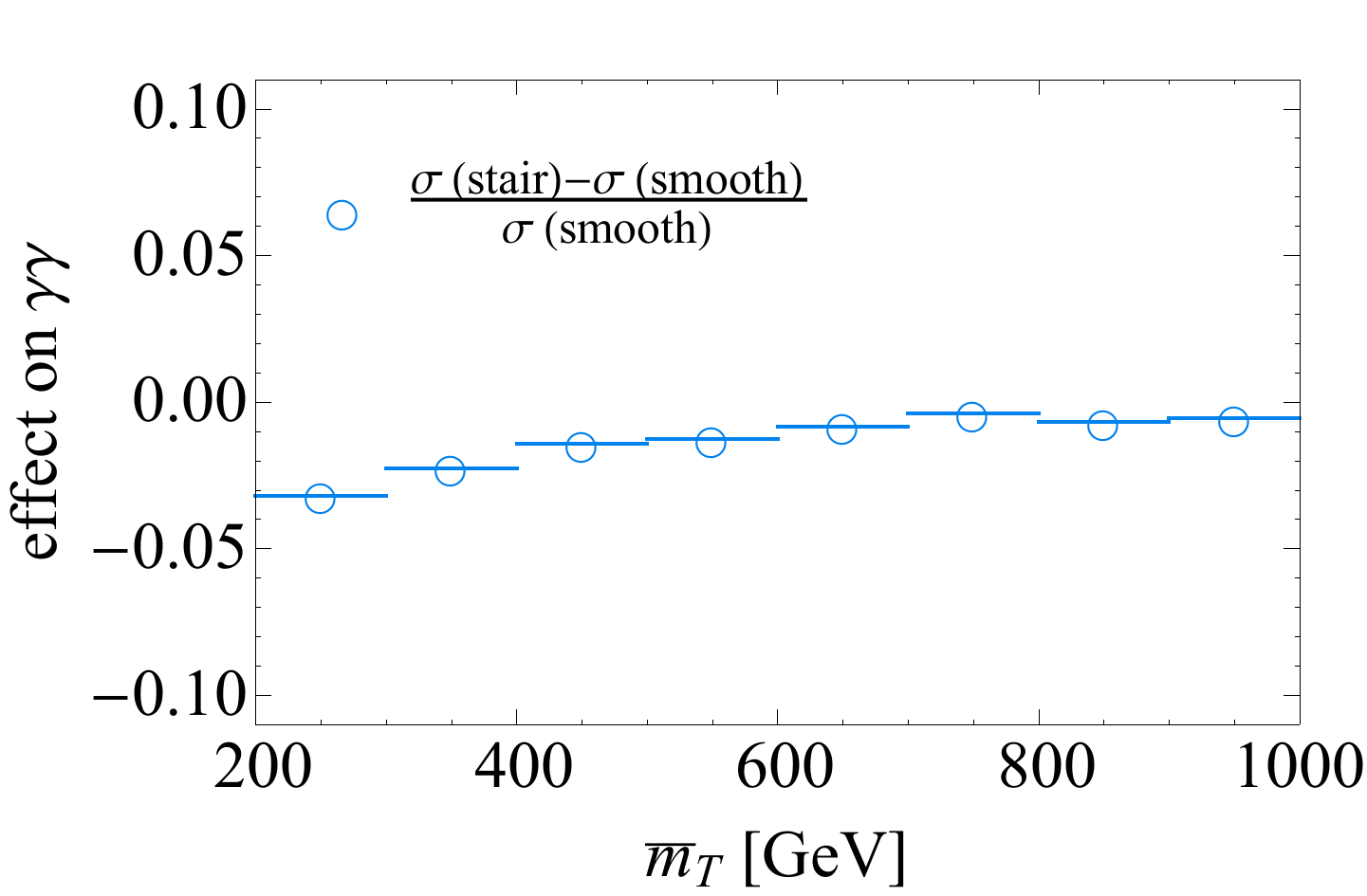}
    \end{center}
\caption{Comparing staircase isolation to the Frixione algorithm. (Left) The smooth curve is $\mathcal I(R;\epsilon,\delta)$ of \eq{frixfunc}, while the piecewise-constant curve is $\hat{\mathcal I}(R;\epsilon,\delta)$ of \eq{staircasefunc}.  (Right) The effect, on $\sigma(\gamma\gamma)$ at NLO, of changing the isolation procedure. 
Here, $\sigma$(smooth) corresponds to pure Frixione isolation, \eq{frix}, while $\sigma$(stair) is computed using \eq{staircase}. At high energies, staircase isolation is indistinguishable from the Frixione algorithm; even at low energies, the difference is slight.}
\label{fig:staircase}
\end{figure}

Our staircase isolation approximates the smooth cone of \eq{frix}, which has parameters $(\delta,\epsilon)=(0.4,0.5)$.  We choose four nested cones ($n=1,2,3,4$) with radii $R_h^{(n)}=0.1 \times n$, and approximate the function ${\mathcal I}(R;\epsilon,\delta)$ of \eq{frixfunc} by a piecewise constant function
\begin{equation}
  \label{eq:staircasefunc}
  \hat{\mathcal I}(R;\epsilon, \delta) = \epsilon
  \left[\frac{ 1 - \cos\left( \frac{1}{2} [R_h^{(n)}+R_h^{(n-1)}] \right) }{1-\cos\delta}
    \right] \equiv \epsilon_h^{(n)}, \quad  \text{for} \quad  R_h^{(n-1)}< R < R_h^{(n)},
\end{equation}
where we define $R_h^{(0)}\equiv 0$.  The constants $\epsilon_h^{(n)}$ are shown in \tab{stairs}; the functions ${\mathcal I}$ and
$\hat{\mathcal I}$ are plotted at left in \fig{staircase}. Then our staircase isolation criterion requires
\begin{equation}
\label{eq:staircase}
  \sum_{h \in R^{(n)}} \pT^h <
  \text{max}\left\{\epsilon_h^{(n)}\,\pT^\gamma\,,
  E_\text{min}^{(n)}\right\},
\end{equation}
where the energies $E_\text{min}^{(n)}$, given in \tab{stairs}, are chosen so that they lie at or above the expected level of pile-up (up to an average of  60 $pp$ collisions per crossing) over Run 2 and 3 of the LHC.
Since event-by-event pile-up subtraction techniques will remove a significant fraction of the energy deposited in the isolation cone, this choice will assure that our technique will not suffer from large efficiency losses due to pile-up.

\begin{table}
    \begin{center}
      \begin{tabular}{| c | c | c |}\hline
        $R$ & $\epsilon_h$ & $E_\text{min}$ \\ \hline \hline
        0.1 & 0.01 & 5 GeV \\  \hline
        0.2 & 0.07 & 10 GeV \\ \hline
        0.3 & 0.20 & 23 GeV \\ \hline
        0.4 & 0.38 & 40 GeV \\ \hline
      \end{tabular}
    \end{center}
\caption{Four concentric hard cones used to approximate smooth-cone isolation. $R$ is the cone angle, $\epsilon_h$ is the energy fraction, and $E_\text{min}$ is a threshold below which we do not reject events, regardless of hadronic energy fraction in the cone. Note that the value $\epsilon_h^{(1)}$ is so small that, in our kinematic regime, isolation in the innermost cone is always controlled by the energy cutoff $E_\text{min}$.}
\label{tab:stairs}
\end{table}

At right in \fig{staircase}, we compare our staircase isolation with the Frixione algorithm, by computing $\sigma(\gamma\gamma)$ with each isolation method and taking the relative difference of the results. The two methods differ by at most 4\% [2\%] in $\sigma(\gamma\gamma)$ $[\sigma(Z\gamma)$], and the difference decreases with energy.  Staircase isolation thus shifts the central value of $R_{1a}$ $(R_{1b})$ $[R_{1c}]$ up by at most 2\% (4\%) [2\%] from the values computed in \sec{beyondLO} with smooth-cone isolation.

Now, having seen that the two photon-isolation procedures are not substantially different for our ratios, let us discuss the uncertainties associated with the staircase method.   One source of uncertainties stems from the experimental extraction of the fragmentation function. We use the leading-order $q \to \gamma$ fragmentation function,
since our NLO calculations involve working only to leading order in $q\to q\gamma$ splitting. The photon fragmentation function for a quark parent has been measured most precisely at ALEPH~\cite{Buskulic:1995au}, in $Z\to \gamma$ + hadrons, in which the final state is dominated by $Z \to q\bar{q}\gamma$ and the fragmentation function contributes to the region where a quark or antiquark becomes collinear with the photon.  The function extracted at leading order by ALEPH, based on a QCD analysis proposed in ref.~\cite{Glover:1993xc}, is
\begin{align}
    D^\LO_{\gamma\leftarrow q}(z,\mu_0) &= \frac{\alpha Q_q^2}{2\pi}
    \left(P^{(0)}_{\gamma\leftarrow q} \log \frac{\mu_F^2}{\mu_0^2(1-z)^2} +C \right),\\
\label{eq:muBigUnc}
    \mu_0 &= 0.22^{+1.3}_{- 0.19} \text{ GeV}\ , \\
\label{eq:cBigUnc}
    C &= -12.1 \pm 4.3 \ ,
\end{align}
where $P^{(0)}_{\gamma \leftarrow q}$ is the tree-level perturbative splitting function.
Uncertainties on the two parameters appear large at first glance, but the parameters are highly correlated.  ALEPH suggested that one should take the relation
\begin{equation}
\label{eq:correlatedparams}
    C = \left.-1-\log\left({s \over 2\mu_0^2}\right)\right|_{s=m_Z^2}\,,
\end{equation}
and found
\begin{equation}
    \mu_0 = 0.14^{+0.21+0.22}_{-0.08-0.04}\text{ GeV}, \quad C=-13.26\,.
\end{equation}
This uncertainty in $\mu_0$ propagates into a minute (per-mil) uncertainty in our ratios. But since the correlation in \eq{correlatedparams} is not assigned an uncertainty, this approach is slightly over-optimistic.  On the other hand we can obtain an overly-conservative estimate if we ignore the correlation and vary both parameters independently by the uncertainties listed in \eqs{muBigUnc}{cBigUnc}. In this case we find uncertainties of about 1\% on our ratio $R_{1a}$. As this is surely a considerable over-estimate, we believe that this source of uncertainty is unimportant.

Several other sources might inflate the uncertainties of the isolation contribution if not handled correctly. First is the fact that, working to NLO in $V_1^0V_2^0$ production, we have done only a leading-order calculation for quark-photon splitting (and used the corresponding LO fragmentation function).  However, since the sensitivity to the fragmentation function is minimized by the staircase method, we do not think the next order correction will affect our ratios in a material way. At the same time, since none of the currently available fragmentation function fits perform any resummation of the logarithms of $\log^2(1-z)$ that appear in the perturbative fragmentation contribution, one must be careful to implement isolation in such a way that one does not weight the $z \to 1$ region of phase space too strongly. The staircase isolation that we advocate here does precisely this, in contrast to hard-cone isolation with a small radius.

\subsection{\texorpdfstring{$Z$}{Z} decay and lepton isolation}
\label{subsec:Zdecay}

Up to this point we have treated the $Z$ as though it does not decay, and imposed the same cuts on $\gamma$ and $Z$ as shown in \tab{bosonJetCuts}.  But the decay of the $Z$ forces us to impose kinematic and isolation cuts on its daughter leptons and to account for the $Z$ peak's width in defining what we mean by a $Z$.  (We consider non-leptonic decays of the $Z$ briefly in \ssec{Final}.)  This has a significant though highly predictable effect on the measurements.

To a good approximation we find that the effects of these three experimental realities factorize, meaning that the overall acceptance $\zeta$ of these three effects can be written as the product of separate acceptance factors:
\begin{equation}
\zeta = \zeta_{\Gamma,\Delta} \times \zeta_{\text{kin}} \times \zeta_{\text{iso}},
\end{equation}
where the $\zeta_i$ are defined as a relative change to the cross section due to a particular effect: $\zeta_{\Gamma,\Delta}$ is the acceptance after requiring the dilepton mass be within $\Delta$ of the $Z$ pole, $\zeta_{\text{kin}}$ is the acceptance of our lepton kinematic cuts, and $\zeta_{\text{iso}}$ is the acceptance of our lepton isolation cuts. Let us now discuss each of them in turn.

{\renewcommand{\arraystretch}{1.5}
\begin{table}
    \begin{center}

      \begin{tabular}{| c |}\hline Kinematic Cuts \\
        \hline\hline
            $| m_{\ell\ell}-m_Z| <$ 25 GeV \\ \hline
            $\pT^{\ell_1} > 20$ GeV \\ \hline
            $\pT^{\ell_{2,3,4}} > 7$ GeV \\ \hline
            $|\eta(\ell)| < 2.5$ \\  \hline
        \end{tabular}
        \hspace{0.25in}
        \begin{tabular}{| c |}\hline Isolation Cuts \\
        \hline\hline
            $\Delta R_{\ell \gamma} > 0.2$  \\ \hline
            $\Delta R_{\ell^+\ell^-} \geq 0.0$ \\ \hline
            $p_{T}(j) < 0.2 \times p_{T}(\ell)$ \\
            if $\Delta R_{\ell j} < 0.4$ \\ \hline
        \end{tabular}
    \end{center}
    \caption{Kinematic and isolation cuts imposed on daughter leptons. The leptons are \pT-ordered such that $\pT^{\ell_1} > \pT^{\ell_2} > \pT^{\ell_3} > \pT^{\ell_4}$.}
\label{tab:leptonCuts}
\end{table}
}

The $Z$'s finite width and $Z$--$\gamma^*$ interference require that we  define what we mean by a $Z$ boson.  We take a $Z$ to be an opposite-sign same-flavor dilepton pair whose mass $m_{\ell\ell}$ falls within $\Delta = 25$ GeV of $m_Z=91.2$ GeV. To quantify effects of this mass window, we define $\zeta_{\Gamma,\Delta}$ as the ratio of a finite-width cross section with mass window $\Delta$ divided by the (fictitious) zero-width cross section that we have used up to now. Note $\zeta_{\Gamma,\Delta}$ can exceed 1 if the window $\Delta$ is taken sufficiently wide.

Also at this stage, to remove a divergence in the $pp\to Z\gamma$ cross section, we apply an isolation cut between leptons and photons by requiring that $\Delta R_{\ell\gamma} > 0.2$ for any lepton-photon pair in the event. We find that if we change $\Delta R_{\gamma\ell}>0.2$ to $\Delta R_{\gamma \ell}>0.4$, the cross section changes by less than 0.5\%.  This is unsurprising since the kinematic cuts of \tab{bosonJetCuts} force the $Z$ and $\gamma$ to be well-separated. The small effect of this isolation cut is included in $\zeta_{\Gamma,\Delta}$.

We next impose realistic kinematic cuts on individual leptons, as shown in the kinematic cuts section of \tab{leptonCuts}. In order to quantify the effects of kinematic cuts on individual leptons, we define $\zeta_\text{kin}$ as the ratio of cross sections with and without these kinematic cuts.

Although the $Z$ is not itself observed, we retain the cut on $Z$s shown in \tab{bosonJetCuts}; that is, we reject $Z$s with $|\eta|>1.5$ even if the leptons pass the cuts in \tab{leptonCuts}.  This choice is somewhat arbitrary and may not be necessary, but making different $\eta$ cuts on $Z$s and $\gamma$s might inflate PDF uncertainties, which otherwise have substantial cancellations.

Finally, we impose isolation cuts between the daughter leptons of a $Z$ and all the jets in the final state. Specifically, we require that $p_{T}(j) < 0.2 \times p_{T}(\ell)$ for any jet-lepton pair with $\Delta R_{\ell j} < 0.4$. The effect of these isolation cuts is described by $\zeta_{\text{iso}}$. We define $\zeta_{\text{iso}}$ as the ratio of cross sections with and without the jet-lepton isolation cuts.

Our $Z$ bosons are often boosted.  To avoid unnecessary acceptance losses, we  do not require any lepton to be isolated from another lepton of the same flavor and opposite sign.  We do not expect this to cause a large $Z$ fake rate at the relevant $\bmT$.

\begin{figure}
    \begin{center}
        \includegraphics[width=0.45\linewidth]{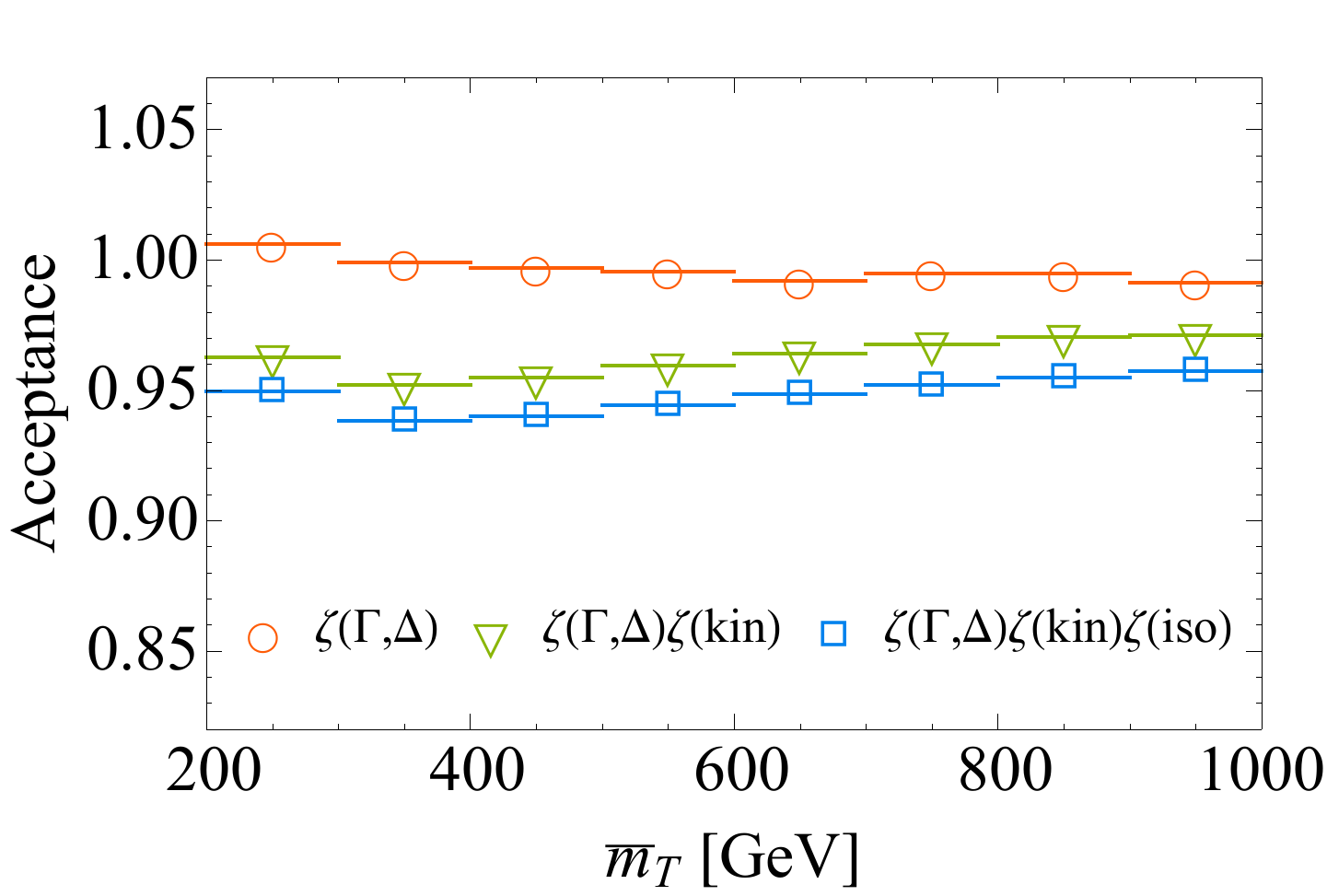}~~
        \includegraphics[width=0.45\linewidth]{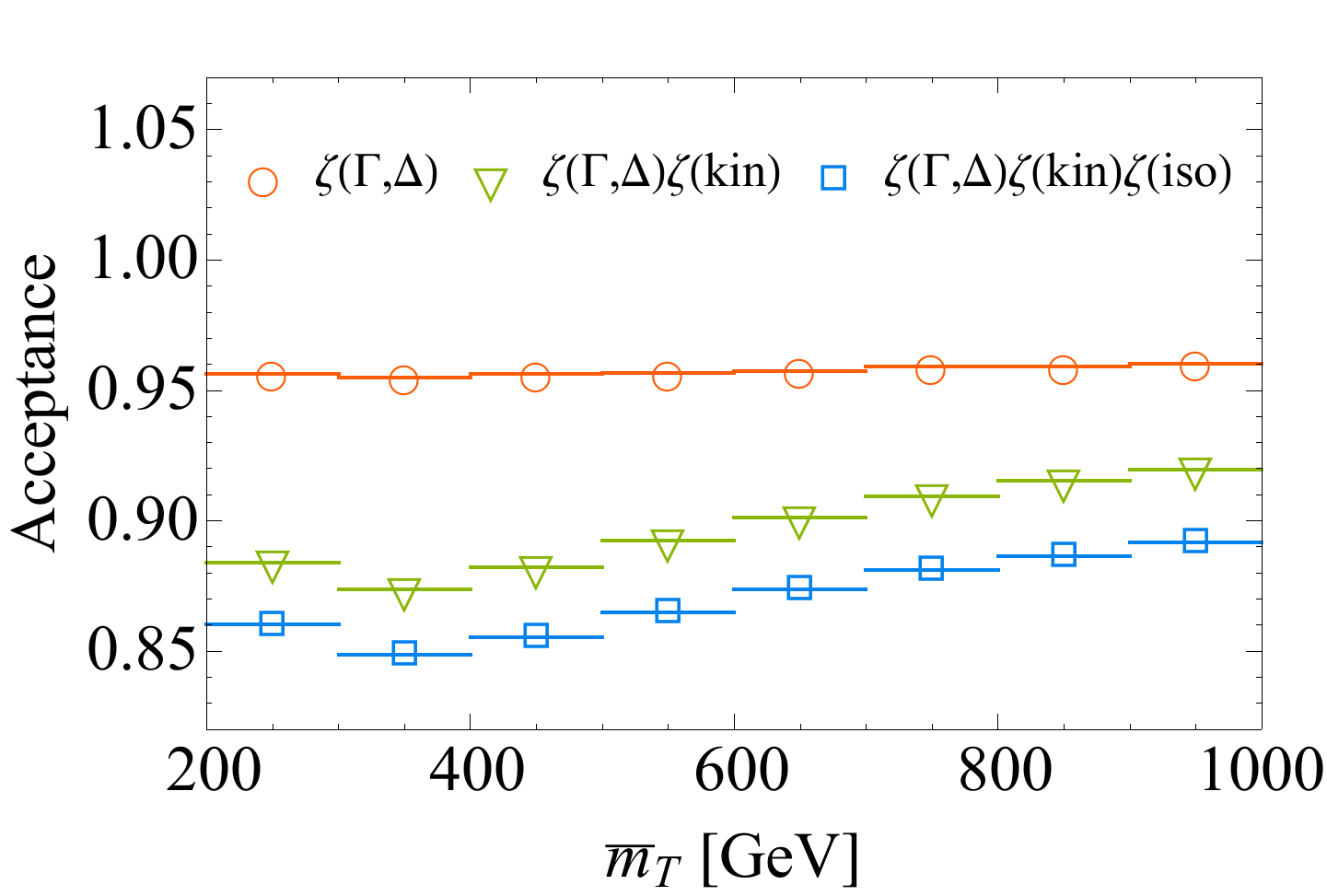}
    \end{center}
    \caption{Effects of finite $Z$ width and lepton cuts on $Z\gamma$ (left) and $ZZ$ cross sections (right). Each plot is normalized by the respective cross section with no lepton cuts and $\Gamma_Z=0$. The red circles show the effect of the $Z$ width.
      The green triangles combine the width with the kinematic cuts of \tab{leptonCuts} and photon-lepton isolation cuts. The blue squares combine these with the other lepton isolation cuts of \tab{leptonCuts}. See text for more details and notation.  }
\label{fig:Zaccept}
\end{figure}

As shown in \fig{Zaccept}, these cuts lower the $Z$ acceptance; for $Z\gamma$ ($ZZ$) events, acceptance drops to 94\% (85\%) for the lowest values of $\bar{m}_T$, rising toward 100\% (90\%) at higher values.  Losses are small at high $\bar{m}_T$ because the leptons have large \pT and have similar $\eta$ to the parent $Z$. Losses would be much greater (78\% and 58\% acceptance for $Z\gamma$ and $ZZ$) if all four leptons were required to have $\pT > 20$ GeV.

These effects thus change our ratios, reducing cross sections by 5--8\% for each $Z$.  However these effects are calculable and do not increase uncertainties.  There should be no problem to include them in the theoretical predictions or unfold them from the experimental measurements.

\section{Discussion and summary}
\label{sec:Summ}

\subsection{Uncertainty budget}
\label{subsec:UncBudget}

{\renewcommand{\arraystretch}{1.4}
\begin{table}[h!]
    \begin{center}
      \begin{tabular}{||c|c|c|c|c||}\hline
        Effect
        & $R_{1a}$
        & $R_{1b}$
        & $R_{1c}$
        & Comments \\
                          & ($Z\gamma/\gamma\gamma$) & ($ZZ/\gamma\gamma$)
                                & ($ZZ/Z\gamma$)           & \\
        \hline\hline
        $qq\to VVqq$            & 2--3\%          & 3--3.5\%         & 1.5--2.5\%
          & extrapolating $p_{T,\text{min}}^{j} \to 0$ (\ssec{NLO-QCD})\\ \hline
        $\mu_R,\mu_F$ ($gg$) & 0.5--1\%        & 1\%              & 1--2\%
          & uses NLO $gg\to\gamma\gamma$ (\ssec{pdf-scale})\\ \hline
        $\mu_R,\mu_F$ (NLO)  & 0.5--1\%        & 1.5--2.5\%       & 1--1.5\%
          & varied independently (\ssec{pdf-scale})\\ \hline
        PDF                     & 0.5\%           & 1--1.5\%         & 0.5--1\%
          & MSTW 2008 using MCFM (\ssec{pdf-scale})\\ \hline
        \hline
        EW (LL)                 & $^{+2\%}_{-1\%}$ & $^{+3\%}_{-1\%}$ & $^{+2\%}_{-1\%}$
          & EFT scale uncertainty (\ssec{doubleLog}) \\ \hline
        $\alpha_\QED$           & 7\%              & 14\%             & 7\%
          & Fully correlated (\ssec{alphaQED})\\ \hline
        \end{tabular}
    \end{center}
\caption{Summary of overall uncertainty budget. The first three entries are not independent sources of uncertainty, and combining them assuming no correlation provides a conservative estimate.}
\label{tab:uncBudget}
\end{table}
}

In \sec{ExecSumm}, we presented our claim that the three $R_1$ ratios (whose central values are related but which have different cancellations among their uncertainties) are under exceptional theoretical control. Here we present a detailed breakdown of what we include in our estimate of known theory uncertainties, as shown in \tab{uncBudget}, and justify our confidence in the small size of further higher-order effects. We now review the table line by line.

The first three lines of \tab{uncBudget} are not truly independent, as they are all striving to capture aspects of the uncertainty associated with higher-order corrections to our calculations of the ratios. Our goal in isolating them was to try to identify any particularly large effects, ones that would not show up in overall NLO scale variations, that we have not already included and would not cancel in our ratios. Although the separation we have made is both scheme and scale dependent and thus unphysical, our methods are probably sufficient to estimate the rough magnitude of the higher-order corrections that we did not include. We have also been quite conservative in our estimates and in how we combined uncertainties. 
Once NNLO calculations of all diboson processes become publicly
accessible, the uncertainties from all sources should be subsumed in the
scale variation of the analogous NNLO calculations, with the exception of
the $gg$ initial state, which only first appears at NNLO. For this last
part, two-loop results for $\gamma\gamma$ and $ZZ$ already exist \cite{Bern:2002jx,Caola:2015psa}, as do 
most components of the $Z\gamma$ calculation \cite{Gehrmann:2013vga}, allowing for a more robust
characterization of the associated uncertainties than the estimates we have
performed here.

As we noted in \ssec{NLO-QCD} above, many NNLO corrections are expected to cancel in the $R_1$ ratios. Valence quark scattering $qq\to V_1^0V_2^0 qq$, which has terms that are not proportional to the LO cross sections, gives one of the largest non-canceling terms that we cannot currently compute.  Our method for obtaining these estimates was described in \ssec{gg}.

We obtained estimates of the $\ord(\alpha_s^4)$-uncertainty in $gg\to \gamma\gamma$ production by varying the scales $\mu_R, \mu_F$ in $gg\to \gamma\gamma$, computed to $\ord(\alpha_s^3)$, up and down by a factor of two. Because of $SU(2)\times U(1)$ relations, we assumed that nearly the same relative uncertainty applies to $gg\to Z\gamma$ and $gg\to ZZ$.  See \ssec{pdf-scale} for more details.

Although it is possible that there are other large non-cancelling NNLO effects, we have not been able to identify them. In particular, although collinear effects make a large contribution at NLO, their contribution at NNLO appears to be much smaller. Moreover, there are no other new channels or new regions of phase space that open up at this order.    Consequently we naively expect other NNLO shifts to the ratios to largely cancel.
We estimated these effects in \ssec{pdf-scale} by seeing how varying the renormalization and factorization scales for the strictly NLO calculations independently affect the ratios; the $\sim 5\%$ corrections in each channel have very substantial correlations, and largely cancel in the ratios.

PDF uncertainties are extracted from the calculations that led to \fig{pdfunc}.  We find that they are very small for $R_{1a}$, and even for the other ratios are significantly smaller in percentage terms than for the individual diboson processes.

Now we turn to the EW uncertainties. The leading-log EW uncertainties, dominated by the choice of matching scales, were extracted from the threshold resummation calculation of ref.~\cite{Becher:2013zua} as described in \ssec{EW-corr}.
We also account for the differing views of how to set the scale for $\alpha(\mu_\QED)$ by varying $\mu_\QED$ between $0$ and $m_Z$.
Note this is a window and not in any sense a $1\sigma$ Gaussian uncertainty.

One item for which we do not have an error estimation is photon isolation.  An essential part of our proposal involves the use of staircase isolation, an experimentally practical approximation to the Frixone smooth-cone method, discussed in \ssec{PhotonIso}.  The use of a hard cone for isolation would introduce a substantial shift to our result and significant sensitivity to $q \to \gamma$ fragmentation.\footnote{Note that uncertainties in the fragmentation function might be reducible.  The ALEPH measurement could be repeated at the LHC, using $W$ decays arising in $t\bar t$ events.  Selecting events with a lepton, $\slashed{E}_T$, two $b$ tags and a loose photon, one could then reconstruct the tops and extract the probability that $W\to\gamma$ + hadrons.} Staircase isolation minimizes these effects.  We saw in \ssec{PhotonIso} that the difference between smooth and staircase cone, most important at low $\bar m_T$, is at most $2 \%$ for $R_{1a}$ and $R_{1c}$, and double this for $R_{1b}$.   The effect of experimental uncertainties on the fragmentation function, which we estimated by varying the parameters in ALEPH's fit, appears to be negligible.

\subsection{Final comments}
\label{subsec:Final}

\subsubsection{General reflections on our methods}

We have proposed a wide variety of ratios using LO reasoning about the $SU(2)\times U(1)$ structure of the SM.  Interestingly, the structure of $SU(2)\times U(1)$ and the radiation zero in the amplitude $a_3$ means that the naive guess for custodial-$SU(2)$ relationships among $W$ and $Z$ do not hold.
The only interesting relation between $ZZ$, $W^\pm Z$ and $W^-W^+$ production is an imperfect (and somewhat impractical) relation between $W^-W^+$ and $ZZ$, which follows only because $|a_3|^2$ is subdominant in $W^-W^+$ production.  This could be generalized to a relation between $W^-W^+$, $W^\pm Z$ and $ZZ$, but the relation is complicated as well as impractical. We also saw no interesting relation between $W^\pm \gamma$ and $Z\gamma$.  That said, charge ratios of $W^+Z$ to $W^-Z$, and of $W^+\gamma$ to $W^-\gamma$, are important tools.  Although we focused on diboson ratios at high $\bar m_T$, the nice properties of these charge ratios  do not require $SU(2)\times U(1)$, and would remain interesting even down to low $\bar m_T$.

An issue that we have not addressed is the experimental systematic uncertainty from fake photons. We have seen in the Higgs boson search that the experiments have fairly large contributions to their $\gamma\gamma$ searches from $\gamma + \text{jet}$.  Although these decrease at moderate \bmT, partially cancel in our ratios, and are typically smaller for photons in the barrel of ATLAS and CMS, they are by no means negligible, as can be seen in ref.~\cite{Khachatryan:2015qba}.  We have implicitly assumed that the systematic uncertainties from these fakes will be under very good control for $\pT^\gamma \geq 150\GeV$ by the time 300 fb$^{-1}$ has been accumulated.  If this is not true it could make the $R_1$ ratios, and others we have proposed, somewhat less useful.

We have limited our detailed study to NLO QCD effects, for practical reasons.  Many NNLO QCD and higher order EW corrections have been performed already, so it should soon be possible to improve upon our results and, most importantly, check our uncertainty estimates.  At NNLO, with two jets accompanying the two bosons, one would encounter many new issues, including vector boson scattering and potential sensitivity to new phenomena therein, as well as $SU(2)$-quintet amplitudes, including same-sign $WW$ production.  However, few of these issues may be essential in the ratios we propose, since the low rates for diboson production mean that theoretical predictions more precise than a few percent may often not be needed at the LHC.

Our results for the $R_1$ ratios involved many arbitrary choices including specific kinematic cuts, isolation requirements, binning, etc.  Although we have carefully considered these choices, we have not in any sense optimized them, and further consideration, both theoretical and experimental, should be given to them.

 Finally, in our results we have imposed an isolation cone around photons but not around $Z$s.  This appears to be a sub-percent effect (for each $Z$) with the isolation criteria that we selected.   However, at a higher energy collider this must be revisited, since at sufficiently high energy the $Z$ and photon will have to be treated on equal footing and a photon-like isolation on the reconstructed $Z$ will have to be applied; otherwise large EW logarithms will afflict our ratios.

\subsubsection{Other $Z$ decays}

The $R_1$ ratios, especially the two involving $ZZ$, suffer from low statistics, due to the small $Z\to \ell^+\ell^-$ branching fraction. One might wonder whether one can gain by looking at $Z\gamma$ events in which the $Z$ decays to neutrinos, and especially at $ZZ$ events by looking for $\ell^+\ell^-$ plus missing transverse momentum ($\slashed{E}_T$).  We have not explored this, but in the ATLAS measurement of the $ZZ$ production cross section~\cite{Aad:2012awa}, such signal events are incorporated.

An obvious downside to this approach would be an inability to put the same $\eta$ cuts on $Z$s and $\gamma$s.  Since these $SU(2)$-singlet processes are generated in the $t$ and $u$ channel, they are particularly sensitive to the $\eta$ cut, so having different cuts for $\gamma$ and $Z$ could potentially cause large NLO corrections due to imperfect cancellations.  Also, the excellent cancellation of PDF uncertainties in $R_{1a}$ could potentially fail.  There would also be backgrounds from $W\gamma$ and $WZ$ events where the $W$ decays to a hadronic tau or a soft lepton, and is mistaken for an invisibly decaying $Z$. The ideal balance between smaller statistical uncertainties and larger theoretical uncertainties will be time-dependent, and requires study by the experimental LHC groups at the time of the measurement.

Nevertheless, there might be a practical strategy using $Z\to\nu\bar\nu$ events.  One could measure $\gamma\gamma$ differentially, and use the $R_{1a}$ and $R_{1b}$ ratios to predict, in the central region, the  $Z\gamma$ and $ZZ$ distributions.  Then, assuming the SM, the full $Z\gamma$ and $ZZ$ distributions extending to larger $\eta$ could be predicted with lower uncertainties, and this prediction could then be checked against events with $\slashed{E}_T$ and a single $\gamma$ or leptonic $Z$.\footnote{A similar method is presumably needed for any ratios involving $W^+W^-$ production, where both $\eta$ and especially $\bmT$ are somewhat uncertain in each event.}

The option of using  hadronic decays of the $Z$  seems daunting. The backgrounds from $Z$- or $\gamma$-plus-jet events, where a QCD jet fakes a boosted $W$ or $Z$, are not small, and will leak into the diboson measurement. Moreover, mass- and charge resolution on hadronically decaying vector bosons is poor, so one cannot distinguish $Z\gamma$ from $W^\pm\gamma$, processes with completely different differential distributions even at LO.

A further tool that we have not explored is the use of $Z$ polarization, potentially of interest due to the parity violation in the SM. BSM physics might alter polarization ratios.

\subsubsection{Applications of the $R_1$ ratios}

The $R_1$ ratios should be useful in several ways even within the SM.  First, they allow high-precision tests of SM calculations, including Monte Carlo methods.  Second, they may serve as a place to explore higher-order EW effects. As we described in \ssec{EW-corr}, EW corrections partly cancel in these ratios, but can reach the 10--20\% level, above the level of theoretical uncertainties.  The fact that QCD corrections cancel rather completely, especially at high \bmT, means these ratios may serve as a particularly clean place to examine logarithmically-enhanced EW  effects.

In this paper we have not addressed the question of how sensitive these variables would be to BSM phenomena.  An obvious potential use of these variables is in searching for BSM interactions of the SM gauge bosons, \eg~through anomalous triple and quartic gauge couplings (aTGCs, aQCSs).  In exploring this, one should use an $SU(2)\times U(1)$ invariant classification of the various operators. We leave this for future work.  Wide resonances decaying to EW bosons might also alter the ratios without being observable in some simpler way. Most other phenomena would introduce additional hard jets, which would often be vetoed by our cuts, or large amounts of $\slashed{E}_T$, which would not impact the $R_1$ ratios but could affect ratios involving leptonic $W$s, as well as any measurements that try to use $Z \to \nu\bar{\nu}$ as we outlined above.  Strategies for events with additional jets and/or $\slashed{E}_T$ are worth further study, but it is far from clear whether our methods can be suitably generalized to such cases.

At higher collision energy, such as a 30 or 100 TeV $pp$ collider, these observables are probably still useful, but will perhaps be more complicated.   On the one hand, finite mass effects will be completely negligible, and the fragmentation contribution for $W$s and $Z$s will become similar to that for photons.  EW corrections become quite large and could easily be observed in these ratios, as would any TeV-scale new physics effects. However, other issues, such as the non-negligible rate for a hard lepton to radiate a real EW boson, as well as the general challenges of resolving and identifying gauge bosons at ultra-high boosts, will begin to have a practical impact on precise diboson measurements.  A dedicated study of this question is needed.

\subsubsection{Prospects for the other ratios}

What can we expect for the other ratios of \eq{OurRatios} at NLO?  The $R_1$ ratios are somewhat special. First, the $\gamma\gamma$, $Z\gamma$ and $ZZ$ processes are fully reconstructible, though at the cost of $Z\to \ell^+\ell^-$ branching fractions.  By contrast, $W\gamma$ and $WZ$ events with a single neutrino are only reconstructible up to a two-fold ambiguity, and leptonic $WW$ events cannot be reconstructed event-by-event.

Second, $\gamma\gamma$, $Z\gamma$ and $ZZ$ cross sections are all proportional to the singlet amplitude-squared $|a_1|^2$ at LO, and many of their NLO corrections are identical at high energy. This is not true for the other processes.  A particular complication is the fact  that the $SU(2)$-triplet amplitude-squared $|a_3|^2$ vanishes at scattering angle $\pi/2$, or in other words at $\hat s=4 \bar m_T^2$. The falling PDFs assure this is a kinematic region of particular importance for production rates at hadron colliders.  Differential cross sections for $W^\pm\gamma$, $W^\pm Z$ and $W^-W^+$ are suppressed at LO by this ``radiation zero'', but to different degrees, and what remains behind is different in each case.  The radiation zero is removed at NLO, and consequently some ratios, particularly certain asymmetries which are quite small at LO, may end up with large NLO corrections and NNLO uncertainties.  Indeed it is already well-known that the $\Kfac_\text{NLO/LO}$ factor for $W\gamma$ is much larger than that for $Z\gamma$ \cite{Ohnemus:1992jn,Grazzini:2015nwa}.

Despite these challenges, there are enough variables in our list that some may evade these concerns.  We are optimistic that a few of the remaining variables will be as precisely predictable as the $R_1$ ratios, and we plan to explore this possibility further.  In the meantime, we hope that our methods will inspire invention of other precision observables, perhaps more sophisticated and less obvious, for the LHC and for hadron colliders of the future.

\acknowledgments
We are grateful for conversations with  T.~Becher, Z.~Bern, J.~Campbell, C.~Rogan, G.~Salam, M.~Schwartz, K.~Tackmann, and C.~Williams.  The authors were supported in part by US Department of Energy grant DE-SC0013607, and National Science Foundation grants PHY-1258729, PHY-0855591, and PHY-1216270.  The computations in this paper were run on the Odyssey cluster supported by the FAS Division of Science, Research Computing Group at Harvard University.
M.J.S.~thanks Harvard University for support and hospitality during this project, as well as the Simons Foundation for support.  For hospitality and support during this research, M.F. and M.J.S. thank the Galileo Galilei Institute,  C.F. thanks the Perimeter Institute, J.S. and M.J.S. thank the CERN theory group, J.S. thanks the Munich Institute for Astro- and Particle Physics, and M.F. thanks the Aspen Center for Physics (via NSF grant PHY-1066293) and the Center for Future High Energy Physics, Beijing.

\appendix

\section{Partonic diboson cross sections with mass corrections}
\label{app:MassCorr}

In this appendix, we give the partonic diboson cross sections at LO including all dependence on $m_W, m_Z$. This is a straightforward modification of the formulas given in the high-energy limit in eqs.~\eqref{eq:za}--\eqref{eq:Cqzz}, \eqref{eq:wa}--\eqref{eq:wz}, and \eqref{eq:uuww} above. The partonic rates including all finite-mass effects can be written as
\begin{align}
\label{eq:firstXsec}
    {d\hat \sigma \over d\hat t}(q \bar q \to \gamma \gamma) &= \left({1 \over 2}\right) {\pi\,\alpha_2^2\,s_W^4 \over N_c\,\hat s^2}\left(2\,Q^4\right)|\A_1|^2,\\[5pt]
    {d\hat \sigma \over d\hat t}(q \bar q \to Z\gamma) &= {\pi\,\alpha_2^2 \, s_W^2\,c_W^2\over N_c\,\hat s^2}  \left(L^2\,Q^2+R^2\,Q^2\right)  |\A_1|^2,\\[5pt]
    {d\hat \sigma \over d\hat t}(q \bar q \to ZZ) &= \left({1 \over 2}\right) {\pi \, \alpha_2^2\,c_W^4 \over N_c\,\hat s^2}  \left({L^4 + R^4} \right) |\A_1|^2,\\[5pt]
\label{eq:WAref}
    {d\hat\sigma \over d\hat t}(q\bar q\,' \to W^\pm\gamma) &= {\pi\,|V_{ud}|^2\,\alpha_2^2\,s_W^2 \over N_c\,\hat s^2}  \left[{Y_L^2 \,|\A_1|^2 \over 2}  \pm 2\,Y_L\,(\A_1\A_3) + 4\,|\A_3^\ell|^2 \right],\\[5pt]
\nonumber
    {d\hat\sigma \over d\hat t}(q\bar q\,' \to W^\pm Z) &= {\pi\,|V_{ud}|^2\,\alpha_2^2 \over N_c\,\hat s^2}  \left[{s_W^2\,t_W^2\,Y_L^2\, |\A_1|^2 \over 2}  \mp 2\,s_W^2\,Y_L\,(\A_1\A_3) \right. \\[5pt]
\label{eq:WZref}
    &\hspace{5.24cm} \left. + 4\,c_W^2\,|\A_3^\ell|^2 + {|\A_\phi^\ell|^2 \over 2}\right],\\[5pt]
\nonumber
    {d\hat \sigma \over d\hat t}(q\bar q \to W^-W^+) &= {\pi\,\alpha_2^2 \over N_c\,\hat s^2}  \left[ {|\A_1|^2 \over 16}   \pm {(\A_1\A_3) \over 2} + {2}\,|\A_3^\ell|^2 + |\A_\phi^\ell|^2 \right.\\[5pt]
    &\hspace{5.24cm} \left. + {2}\,|\A_3^r|^2 + |\A_\phi^r|^2\right],
\label{eq:lastXsec}
\end{align}
where $\alpha_2,L,R$ were defined in \ssec{za} and $q \bar q\,'$ is $u\bar d$ ($d \bar u$) for $W^+V^0$ ($W^-V^0$). In $d\hat\sigma(W^-W^+)$, the upper (lower) sign holds for $u$-type ($d$-type) quarks. In these formulas, the superscripts $\ell,r$ refer to the handedness of the incoming quarks.

The partonic cross sections above are written in terms of $\A_i$s, generalizations of the $a_i$s, defined as
\begin{align}
    |\A_1|^2 &= (\hat t \,\hat u - m_1^2 \, m_2^2)  \left({1 \over \hat t^2} + {1 \over\hat  u^2}\right) + {2\,\hat s\,(m_1^2 + m_2^2) \over \hat t \, \hat u}\,,\\[5pt]
    (\A_1\A_3) &= (N^\ell_T\,P_s) \, (\hat T \hat U)  \left({1 \over \hat u} - {1 \over \hat t}\right) + {1 \over 4} \left(\hat t\,\hat u - m_1^2\,m_2^2\right) \left({1 \over \hat u^2}-{1 \over \hat t^2}\right)\,, \\[5pt]
\nonumber
    |\A_3^h|^2 &= (N^h_T\,P_s)^2 \, (\hat T \hat U)  +{\delta_{h\ell}}  \left[{(N^\ell_T\,P_s) \over 4} \, (\hat T \hat U)  \left({1 \over \hat t} + {1 \over \hat u}\right)\right. \\[5pt]
    &\hspace{2.9cm} \left.+ {1 \over 32}  (\hat t \, \hat u - m_1^2\,m_2^2)\left({1 \over \hat t^2}+{1 \over \hat u^2}\right) - {1 \over 16} \, {\hat s\,(m_1^2+m_2^2) \over \hat t \, \hat u}\right]\,, \\[5pt]
    |\A_\phi^h|^2 &= (N^h_\phi\,P_s)^2 \left[ \hat t \, \hat u + 2\,\hat s \, (m_1^2+m_2^2) - m_1^2 \, m_2^2\right]\,,
\end{align}
where $m_1,m_2$ are the masses of $V_1,V_2$. Here we have abbreviated
\begin{equation}
    (\hat T \hat U) \equiv \hat t \, \hat u - \hat s \, (m_1^2+m_2^2) - m_1^2\,m_2^2
\end{equation}
and defined $\hat s$-channel propagators
\begin{equation}
    P_s \equiv \left\{ \begin{array}{ll} \displaystyle {1 \over \hat s-m_W^2} & \text{for $WV^0$\,,} \\[15pt]
    \displaystyle {1 \over \hat s-m_Z^2} & \text{for $W^-W^+$}\,. \end{array} \right.
\end{equation}
Each $P_s$ appears with a coefficient:
\begin{equation}
    N^\ell_T = N^\ell_\phi = {1 \over 2} ~~\text{for}~~ WV^0\,,
\end{equation}
while for $W^-W^+$\,,
\begin{align}
    N^\ell_T &= |T_3| - {|Q|\,m_Z^2\,s_W^2 \over \hat s} \,,\\[5pt]
    N^r_T &= - {|Q|\,m_Z^2\,s_W^2 \over \hat s} \,,\\[5pt]
    N^\ell_\phi &= t_W^2\,Y_L \left({1 \over 2} - {m_Z^2\,c_W^2 \over \hat s}\right) + T_3\left({1 \over 2} - {m_Z^2\,s_W^2 \over \hat s}\right), \\[5pt]
    N^r_\phi &= t_W^2 \, Y_R  \left({1 \over 2} - {m_Z^2\,c_W^2 \over \hat s}\right).
\end{align}
One can obtain these coefficients by combining the $\gamma$- and $Z$-propagators (along with their attached coupling constants) for $\hat s$-channel production of $w^-w^+$ or $\phi^-\phi^+$.

\section{On the approximations used in $gg\to VV$ estimates}
\label{app:gg}

In all our studies of $gg\to V_1^0V_2^0$, we have been using MCFM's implementation of the dominant $\ord(\alpha_s^3)$ correction to this process, which in turn relies on the calculation of ref.~\cite{Bern:2002jx}.  This calculation computes only part of the $\ord(\alpha_s^2)$ and $\ord(\alpha_s^3)$ corrections to $pp\to V_1^0V_2^0$ production, leaving out many terms.  One may reasonably wonder whether it represents a consistent calculation, and whether we can safely use it to determine overall normalizations as in \fig{ggCorr}, to relate the contributions for different $V_1^0V_2^0$ processes as in \eq{Kgg}, and to estimate remaining scale dependence as in \ssec{pdf-scale}.

Our methods rely upon the fact that all $\ord(\alpha_s^2)$ and $\ord(\alpha_s^3)$ terms included in ref.~\cite{Bern:2002jx} contain a single fermion loop at amplitude level which is squared in the matrix element.  The loop is proportional to the number of quark generations $N_g$, so these terms are of order $N_g^2$, and are also proportional to a particular combination of $SU(2)\times U(1)$ charges that arise because the $V_1$ and $V_2$ gauge bosons must attach to the loop.

In the perturbative expansion of $pp\to V_1^0 V_2^0$, this proportionality to $N_g^2$  arises first in the leading $gg\to V_1^0 V_2^0$ box  graph, and not in any other graph at this or lower order.  Therefore, at order $\ord(\alpha_s^2)$, we may parametrically separate the loop graph from all other terms.
This justifies treating the scale choices in this part of the calculation as separate from those in the NLO calculation.

This would still be true at $\ord(\alpha_s^3)$ if all terms proportional to $N_g^2$ were included.  Only these terms can be involved in moderating the $\mu_R$ dependence of the $\ord(\alpha_s^2)$ box graph.
In turn, this would justify the methods used in \fig{ggCorr} to fix the normalization and in \ssec{pdf-scale} to estimate remaining scale dependence.  Furthermore, \eq{Kgg} would be true,
because the $N_g^2$-dependent terms for the three $V_1^0 V_2^0$ processes would differ only in the $SU(2)\times U(1)$ factors arising where the $\gamma$ or $Z$ bosons attach to the fermion loop or lines (up to small effects from the top quark and from $m_Z$).

However, ref.~\cite{Bern:2002jx} presents the two-loop corrections to $gg\to \gamma\gamma$ and the radiative correction $gg\to \gamma\gamma g$, but omits a term from  the interference  between the tree-level and one-loop-level amplitudes for the process $gg\to\gamma\gamma q\bar q$.  This interference term is also of order $N_g^2$
and needs to be included to capture the full scale dependence of the $\ord(N_g^2)$ piece at $\ord(\aS^3)$.

We now argue that the $gg\to V_1^0 V_2^0 q\bar{q}$ processes at $\ord(\alpha_s^2)$ and $\ord(\alpha_s^3)$ are subleading compared to what we have included.
The contribution of the tree-level rate for $gg \to V_1 V_2 q\bar{q}$ was studied for $V_1 V_2 = W\gamma$ and $WZ$ in ref.~\cite{Adamson:2002jb}, and for $Z\gamma$ in ref.~\cite{Adamson:2002rm}, and it turns out to be anomalously small.  In particular, even for $Z\gamma$ it turns out to be less than 30\% of the $\ord(\alpha_s^2)$ $gg\to Z\gamma$ rate, which, as discussed in \ssec{gg}, itself has a partial $SU(2)\times U(1)$ cancellation in the loop.  For $\gamma\gamma$ and $ZZ$ we expect the corresponding graph to be significantly smaller relative to the $\ord(\alpha_s^2)$ rate.

Meanwhile, at $\ord(\alpha_s^3)$, the interference of the corresponding amplitude against its one-loop correction does not affect the scale sensitivity of the resulting calculation. No additional $\mu_R$ sensitivity appears, since the only loops that arise are all in the form of subdiagrams that are themselves finite. Simultaneously, $\mu_F$ sensitivity is already taken into account by varying the PDF scale appropriately: the tree-level $gg\to V_1^0 V_2^0 q\bar{q}$ process has no QCD infrared divergences in the final state, so the interference term contributes only universal divergences, which are already included in the PDF evolution.

This (along with the recent paper \cite{Caola:2015psa}) gives us some confidence that our method for normalizing $gg\to ZZ$ and $\gamma\gamma$ has a small relative uncertainty.  The correction for $Z\gamma$, suppressed at $\ord(\alpha_s^2)$, may have a larger relative uncertainty, but this is still small in absolute terms. More quantitatively, in the $R_{1a}$ ratio the contribution of $gg\to V_1^0 V_2^0$ in the lowest \bmT bin is roughly +5\% from $Z\gamma$ and $-13\%$ from $\gamma\gamma$, and this drops off rapidly with \bmT.  Even if we took an overly conservative 30\% relative uncertainty estimate for the normalization of $gg\to Z\gamma$, and ignored any correlations with $gg\to \gamma\gamma$ that would cancel in the ratio, this would translate into at most a $1.5\%$ ($<1\%$) uncertainty in the lowest (highest) bin in \fig{mainResult} for $R_{1a}$. A similar statement applies for $R_{1c}$.

\section{Further discussion of NNLO \Kfac factors}
\label{app:NNLO}

We have assumed NNLO corrections are small compared to our $\Kfac_\text{NLO/LO}$ factors of order 1.5.  Also, as suggested in ref.~\cite{Bern:2002jx}, we assumed that the $gg$ loop contributions give the majority (or rather, more precisely and more importantly, the largest fraction that does not cancel in ratios) of the NNLO contributions.  In several cases, recently computed fully differential NNLO cross sections for the $\gamma\gamma$, $Z\gamma$, and $ZZ$ processes feature much larger corrections than what we have claimed to expect for our observables. Here we discuss how the cases where this is true are affected by one or more of the issues we discuss in \ssec{cuts}, causing the $\Kfac_\text{NNLO/NLO}$ factors to be larger than would be the case for our cuts and observable.

In figure 1 of ref.~\cite{Catani:2011qz}, which shows the fully-differential NNLO calculation of the $\gamma\gamma$ cross section, one sees \Kfac factors of $\Kfac_\text{NLO/LO} \sim 3$ and $\Kfac_\text{NNLO/NLO} \sim 1.6$.  Meanwhile the $gg$ box contribution was found to be $\approx 15\%$ of the total NNLO correction.  These results would appear to cast doubt on our assumptions. However, their calculation uses fixed asymmetric cuts for the photon of $\pT^\text{hard} \ge 40\GeV$ and $\pT^\text{soft} \ge 25\GeV$, with invariant diphoton mass $m_{\gamma\gamma}$ as the observable.  As the authors point out, this allows a large NLO contribution, at any $m_{\gamma\gamma}$, from events with $\pT^\text{soft} \sim 25\GeV$ and a large $\pT^\text{hard}$, which leads to large logarithms.  This explains both the much larger NLO corrections and (since the $gg$ process at NNLO only occurs with LO kinematics) the smallness of the $gg$ contribution compared to the total NNLO correction.  This interpretation of the origin of the large \Kfac factors is supported by the cross section presented in ref.~\cite{Cieri:2013pza}, figure 2, which, using more symmetric fixed cuts of $\pT^\text{hard} \ge 25\GeV$ and $\pT^\text{soft} \ge 22\GeV$, found $K_\text{NNLO/NLO} \sim 1.3$.  Recall that we used scaling asymmetric cuts that never allow $\pT^\text{hard}  \gg \pT^\text{soft}$, so we expect we have smaller quantum corrections than in either of these cases.

Large \Kfac factors are also reported for the NNLO $Z\gamma$ cross sections in ref.~\cite{Grazzini:2015nwa}; see figure 5 of that paper. Here the dominant issue is not \pT cuts (for the $Z$, at least, the calculation imposes none), but the treatment of resonances.  A cut on the photon \pT of 40 GeV (combined with lepton isolation cuts) ensures that at leading order, the $Z$ is kinematically forbidden from decaying to $\ell^+ \ell^- \gamma$. However once the $\ell^+ \ell^- \gamma$ system is allowed to recoil off a jet, such a configuration becomes kinematically accessible, resulting in a large \Kfac factor at small $m_{\ell^+\ell^-\gamma}$.   But one sees much smaller \Kfac factors either if the \pT cut on the photon is reduced, allowing $Z\to \ell^+\ell^-\gamma$ to arise at LO, or if $m_{\ell^+\ell^-\gamma}$ is taken much larger than $m_Z$.  In the latter case, which is more relevant for us, it is found that $K_\text{NNLO/NLO}\sim 1.2$ for their cuts; while the $\Kfac_\text{NLO/LO}$ factor is not reported, we estimate from the figure that it is $\sim 1.7-1.8$.

Meanwhile $K_\text{NNLO/NLO}$ factors reported for $ZZ$ in ref.~\cite{Grazzini:2015hta}, for cuts matching ATLAS and CMS analyses, are about $1.15$ for the 8 TeV LHC.  The figure in ref.~\cite{Cascioli:2014yka} for the inclusive $ZZ$ cross section shows a $\Kfac_\text{NNLO/NLO}$ factor growing from $1.13$ to $1.17$ between 8 and 13 TeV, over an $\Kfac_\text{NLO/LO}$ factor of about 1.5.

The same figure in ref.~\cite{Cascioli:2014yka} shows that the $gg\to ZZ$ loop contribution to the inclusive cross section is about $60\%$ of the total NNLO contribution.  Since it is evaluated at lowest order, this percentage has a large $\mu_R$ uncertainty.  Our discussion in \ssec{gg} supports using a low scale in evaluating this contribution, which may increase it further. (See also the recent result of ref.~\cite{Caola:2015psa}.)  Some further evidence that the $gg$ contribution is the majority, for an appropriate observable similar to ours, was given in ref.~\cite{Campanario:2015nha}, figure 6, though this involved only a partial evaluation of the NNLO contribution.

We have less evidence for the relative size of the $gg\to\gamma\gamma$ loop, except that in the case of the very large \Kfac factors of ref.~\cite{Catani:2011qz} it can be as small as $15\%$ of the full NNLO contribution.  Changing the cuts as was done in ref.~\cite{Cieri:2013pza} reduced the $\Kfac_\text{NNLO/NLO}$ factor to 1.3 but should have no effect on the lowest-order $gg\to \gamma\gamma$ loop calculation, which has LO kinematics, so this could bring the contribution to $\sim 30\%$.  If our cuts lead to even smaller NLO \Kfac factors, as we suspect, this could bring the $gg\to \gamma\gamma$ loop into the majority.

For $Z\gamma$, coherent cancellations between up and down quarks in the loop cause the $gg\to Z\gamma$ loop contribution to be a few times smaller than the other two, relative to the NLO calculation.  (See our \fig{kgg}.)   Although ref.~\cite{Grazzini:2015nwa} reports this loop is only 6--9\% of the total NNLO correction, one can see from figure 5 of that paper that this percentage increases to $\sim 20\%$ once the $Z^*\gamma$ system is well above the $Z$ pole. This is consistent with our estimate of the $Z\gamma$ contribution being quite uncertain in relative terms but of little importance in absolute terms.

In sum, we find no clear inconsistencies in the literature between our assumptions and existing calculations.

\bibliography{dibosonslhc}
\bibliographystyle{jhep}

\end{document}